%% file: 00-main.tex
\documentclass[onefignum,onetabnum]{siamart190516}

\input{96-shared}

\ifpdf
\hypersetup{
  pdftitle={A Large-Scale Benchmark for the Incompressible Navier-Stokes Equations},
  pdfauthor={}
}
\fi

\input{99-pre}

\begin{document}

\maketitle

% REQUIRED
\begin{abstract}
    \input{01-abstract.tex}

\end{abstract}

% REQUIRED

% REQUIRED

\input{02-introduction.tex}

\input{03-related.tex}
\input{05-testproblems.tex}

\input{06-largescale.tex}
\input{07-conclusions.tex}

\section*{Acknowledgments}

\bibliographystyle{siamplain}
\bibliography{98-references}

\appendix
\input{09-appendix}
\end{document}

%% file: 96-shared.tex
\usepackage{lipsum}
\usepackage{amsfonts}
\usepackage{graphicx}
\usepackage{subfigure}
\usepackage{epstopdf}
\usepackage{algorithmic}
\usepackage{array}
\usepackage{booktabs}
\ifpdf
  \DeclareGraphicsExtensions{.eps,.pdf,.png,.jpg}
\else
  \DeclareGraphicsExtensions{.eps}
\fi
\usepackage{anyfontsize}
\newsiamremark{remark}{Remark}
\newsiamremark{hypothesis}{Hypothesis}
\crefname{hypothesis}{Hypothesis}{Hypotheses}
\newsiamthm{claim}{Claim}

\headers{A Large Scale Benchmark for incompressible Navier-Stokes}{Z. Huang, T. Schneider, M. Li, C. Jiang, D. Zorin, and D. Panozzo}

\title{A Large Scale Benchmark for the incompressible Navier-Stokes Equations\thanks{Submitted to the editors
    \today.
    \funding{This work was partially supported by the NSF CAREER award with number 1652515, the NSF grant IIS-1320635, the NSF grant DMS-1436591, the NSF grant 1835712, the NSF grant IIS-1943199, CCF-1813624,  ECCS-2023780, the NSERC grant DGECR-2021-00461 and RGPIN-2021-03707, a gift from Adobe Research, and a gift from nTopology.}}}

\author{
    Zizhou Huang\thanks{NYU Courant, New York.}
    \and
    Teseo Schneider\thanks{University of Victoria, British Columbia.}
    \and
    Minchen Li\thanks{UCLA Math Dept., California}
    \and
    Chenfanfu Jiang\footnotemark[4]
    \and
    Denis Zorin\footnotemark[2]
    \and
    Daniele Panozzo\footnotemark[2]
}

\usepackage{amsopn}

%% file: 99-pre.tex
\definecolor{Green}{rgb}{0.4,1,0.4}
\definecolor{purple}{rgb}{0.5,0.5,1}

\newcommand{\RR}{\mathbb{R}}

%% file: 01-abstract.tex
Numerical methods for solving incompressible Navier-Stokes equations are widely applied and are among the most extensively studied.  A plethora of approaches have been introduced targeting different applications and regimes; these methods vary in space discretization, choice of time integration scheme, or reduction of the non-linear equations to a sequence of simpler problems. While the theoretical properties of most of these variants are known, it is still difficult to pick the best option for a given problem, as practical performance of these methods has never been systematically compared over a varied set of geometries and boundary conditions. 

We introduce a collection of benchmark problems in 2D and 3D (geometry description and boundary conditions), including simple cases with known analytic solution, classical experimental setups, and complex geometries with fabricated solutions for evaluation of numerical schemes for incompressible Navier-Stokes equations in laminar flow regime.  We compare the performance of a representative selection of most broadly used algorithms for Navier-Stokes equations on this set of problems. Where applicable, we compare the most common spatial discretization choices (unstructured triangle/tetrahedral meshes and structured or semi-structured quadrilateral/hexahedral meshes). 

The study shows that while the type of spatial discretization used has a minor impact on the accuracy of the solutions, the choice of time integration method, spatial discretization order, and the choice of solving the coupled equations or reducing them to simpler subproblems have very different properties. Methods that are directly solving the original equations tend to be more accurate than splitting approaches for the same number of degrees of freedom, but numerical or computational difficulty arise when they are scaled to larger problem sizes. Low-order splitting methods are less accurate, but scale more easily to large problems, while higher-order splitting methods are accurate but require dense time discretizations to be stable.

We release the description of the experiments and an implementation of our benchmark, which we believe will enable statistically significant comparisons with the state of the art as new approaches for solving the incompressible Navier-Stokes equations are introduced.

%% file: 02-introduction.tex
\section{Introduction}

The computation of the motion of incompressible viscous fluids is ubiquitous in scientific computing and engineering applications. The corresponding incompressible Navier-Stokes equations are challenging to solve numerically, and a large number of approaches have been proposed to compute their solution on discrete domains.

While a large literature considered the theoretical properties of these methods, such as convergence, and applied these to a broad range of fluid problems, a systematic comparison of these techniques on a representative common set of benchmark problems is not available. Our work strives to fill this gap by introducing a large collection of benchmark problems, varying from simple cases with known analytical solution, to manufactured solutions on thousands of complex geometries, a collection of reference implementations of representative solution methods, and an automated way of comparing results. The comparison is performed by matching a primary quantity of interest (e.g., running time) to a secondary quantity (e.g., accuracy): to achieve this, we automate the generation of discretizations for all our problems, thus enabling us to experimentally match the primary quantity and measure the secondary one. 

We consider two classes of problems: (1) problems with known closed-form solutions, using fabricated solutions both on simple domains and on a large collection of hundreds of complex 2D and 3D geometries, and (2) a set of standard test problems for which a closed-form solution is unknown, and the error has to be studied with respect to fine-resolution numerical solutions. 

\subsection{Mesh Types}
We consider three types of meshes:
\begin{enumerate}
\item[T] unstructured triangular mesh (2d) or tetrahedral mesh in (3d);
\item[Q] unstructured quadrilateral meshes (2d) and hexahedral meshes (3d);
\item[R] regular 2d or 3d grid.
%\item[MAC-GRID] 
\end{enumerate}

\subsection{Spatial Discretization}
We compare four approaches to spatial discretization (SD):
\begin{enumerate}
    \item[FE] A finite-element discretization using elements of possibly different order for velocity and pressure~\cite{polyfem}. %For instance, FE21 uses second order elements for velocty and first order ones for pressure.
    \item[FV] A finite-volume formulation, (we use the  Ansys FLUENT~\cite{ansys2016ansys} implementation)
    \item[FD] A finite difference scheme using the Marker-and-Cell grid~\cite{harlow1965numerical}.
\end{enumerate}

\subsection{Time Integration}
Spatial discretization methods are coupled with five time-integration methods (TD):
\begin{enumerate}
    \item[BDF3] The Backward Differentiation Formula of order 3 \cite{suli2003introduction}.
    \item[C] The coupled time integration scheme in Ansys FLUENT~\cite{ansys2016ansys}.
    \item[SL] A splitting formulation \cite{chorin1967numerical} using FE/FD for diffusion and pressure projection, and a semi-Lagrangian scheme~\cite{sawyer1963semi} for advection.
    \item[FLIP] A variant of the FLIP scheme \cite{brackbill1986flip} using FE/FD for diffusion and pressure projection, and particles for advection.
    \item[AB2AM2] An explicit predictor-corrector method \cite{2ndSplit} that consists of a second order Adam-Bashforth (AB2) predictor and a modified second order Adam-Moulton (AM2) corrector.
\end{enumerate}

\subsection{Methods}
Not all combinations of spatial and time discretizations are realizable. We selected six representative valid combinations  commonly used research papers, and added the most popular solution used in industry. For each method, we test it on all mesh types, with the exception of finite differences, which require a regular grid:\\
\begin{center}
\begin{tabular}{l|lll}
Acronym & Spatial D. & Time D. & Mesh\\
\hline
FE21-BDF3   & FE21   & BDF3              & T, Q, R \\
FV-C        & FV     & C                 & T, Q, R \\
FE11-SL     & FE11   & SL                & T, Q, R \\
FE11-FLIP   & FE11   & FLIP              & T, Q, R \\
FD-FLIP     & FD     & FLIP              & R     \\
FE11-AB2AM2 & FE11   & AM2AB2            & T, Q, R \\
FE21-AB2AM2 & FE21   & AM2AB2            & T, Q, R \\
\end{tabular}
\end{center}

% The approaches are tested, where applicable, on 3 discretizations: (1) unstructured triangle/tetrahedral meshes, (2) unstructured quadrilateral/hexahedral meshes, and (3) regular grids.

\subsection{Summary of Results}
Our study allows evaluating the combinations of time and spatial discretizations over multiple performance measures. (For a description of our experimental setup see Appendix \ref{app:setup}.)

\paragraph{Mesh Type} Different mesh types do not lead to noticeably different results on any of the settings that we tested. In particular, we did not observe measurable benefits for structured meshes (which are difficult to generate automatically) over unstructured meshes (which can be reliably generated automatically from a boundary description) for the type of solvers we have used. 

\paragraph{2D vs 3D} The relative pros and cons of the methods we tested are similar in our 2D and 3D test cases, thus the conclusions of our study hold for both two- and three-dimensional problems. Using a hexahedral mesh in 3D for FE11-FLIP, FE11-SL has a minor downside, since they require inverting the trilinear geometric map using a numerical method, which has a small negative effect on runtime (the issue is not present in 2D, as the bilinear geometric map can be inverted in closed form).

\paragraph{Runtime Limitations} In our experiments, we were able to get a solution from methods that linearly discretize space and time within our time budget for all mesh resolutions we considered, but the accuracy of these methods is relatively low even on the finest mesh. Higher-order methods have issues with higher resolutions. FE21-BDF3 leads to linear systems for which iterative solvers may not converge to target accuracy, thus precluding the solution of large systems. FE21-AB2AM2 allows to solve for large number of DOFs but, due to the explicit time integration, the time step has to be decreased quadratically with respect to the edge length, thus making it impractical. FV-C is a good compromise, its convergence rate is lower than FE21-BDF3, but higher than all the linear methods, and it still scales well to larger problem sizes.

\paragraph{Accuracy} Methods using third-order time-integration schemes and second-order bases for the velocity are the only ones that can obtain high-accuracy solutions in a reasonable time due to their cubic convergence. However, they suffer from scalability problems due to either numerical difficulties or restrictions in the maximal time step.

\paragraph{Pressure Boundary Accuracy} Only FE21-BDF3 produces reliably accurate results: standard linear splitting schemes do not converge to a solution, high-order splitting schemes converge, but due to their limitations on the time step are impractical.  This observation may be of importance for the choice of method for coupled fluid-structure problems. 

\paragraph{Time step Size} All methods have restrictions, due to the CFL condition, on their time step size. The methods that are less affected by this limitation are linear splitting schemes, which converge for large steps, but their accuracy is low. On the other end of the spectrum, higher-order explicit time splitting methods have better accuracy for a given grid resolution, but are only stable for time steps decreasing quadratically with edge length. 
\paragraph{Viscosity} All methods perform similarly for fluids with high viscosity. For low viscosity, their solutions are widely different, making it challenging to evaluate their accuracy on test problems without analytical solutions. 

We present our results and benchmarks first, and provide a more detailed analysis of suggested directions for future work in the conclusion.

%% file: 03-related.tex
\section{Review of Spatial and Time Discretization Types}

We briefly review some common existing algorithms for numerical solution of Navier-Stokes equations. We cover the discretization of space and time separately: while these need to be chosen in a compatible way, there is often some flexibility in how space and time discretization types are paired. 

\subsection{Incompressible Navier-Stokes Equations}\label{sec:bg}

Let $\Omega\subset \RR^d$, $d\in\{2,3\}$ be a domain with boundary $\partial\Omega$, divided into parts  $\partial\Omega_D\subset\partial\Omega$ and $\partial\Omega_N\subset\partial\Omega$ on which Dirichlet and Neumann boundary conditions are applied (we only consider these two types of boundary conditions). 
The Navier-Stokes (NS) equations are formulated in terms of velocity
\[
    u\colon \Omega\times(0, T) \to \RR^d
\]
of a fluid with kinematic viscosity $\nu$ and density $\rho$, with time $t\in (0, T)$  and pressure
\[
    p\colon \Omega\times(0, T) \to \RR.
\]
The equations for $u$ and $p$ are:
\begin{equation*}
    \begin{cases}
        \displaystyle\rho\frac{\partial u}{\partial t} + \rho(u \cdot \nabla) u - \nu \Delta u + \nabla p = b & \text{ on } \Omega\times(0, T)           \\
        \nabla \cdot u = 0                                                                                    & \text{ on } \Omega\times(0, T)           \\
        u  = d                                                                                                & \text{ on } \partial\Omega_D\times(0, T) \\
        \nu\frac{\partial u}{\partial n} + pn  = g                                                            & \text{ on } \partial\Omega_N\times(0, T) \\
        u(0)  = u_0                                                                                           & \text{ on } \Omega\times\{0\},
    \end{cases}
\end{equation*}
where $b$ is an external volume force, $n$ is the surface normal, $u_0$ is the initial condition, and $d$ and $g$ are the Dirichlet and Neumann conditions.

We consider discretization based on meshes: triangular and quadrilateral in 2D and tetrahedral and hexahedral in 3D. Thus for a given mesh size $h$ (i.e., the maximum edge length on the mesh) the different methods compute approximations $u_h(x, t)$ and $p_h(x,t) $ of the solutions $u$ and $p$ of the NS equations.

\subsection{Spatial Discretization}

\paragraph{Finite Elements}
The finite element method (FEM)~\cite{szabo1991finite, Ciarlet:2002:fem} is a classical method used to solve PDEs, and large number of libraries and software exist (e.g., DOLFIN (FEniCS)~\cite{code:fenics} Firedrake~\cite{code:firedrake}, libMESH~\cite{code:libmesh}, mFEM~\cite{mfem,mfem-web}, getFEM~\cite{getfem}, Deal II~\cite{dealii}).

The elements used for incompressible Navier-Stokes equations are typically the same as for the linear incompressible Stokes equation, e.g., \cite{femGirault1986,GLOWINSKI2003}.  To ensure convergence, the finite element space should satisfy the inf-sup condition~\cite{femSusanne, femFortin}. Conforming elements, such as  the Taylor-Hood elements~\cite{femSusanne, femFortin,GLOWINSKI2003}, and non-conforming elements such as Raviart-Thomas elements~\cite{Raviart1991} are commonly used.  To solve the nonlinear problem, iterative penalty methods~\cite{femCodina, femSusanne}, and augmented Lagrangian methods~\cite{femAugLag} were used.

In our comparison, we use the standard linear and quadratic Lagrange bases for tetrahedra ($P_1$ and $P_2$) and hexahedra ($Q_1$ and $Q_2$) \cite{szabo1991finite, Ciarlet:2002:fem}, which is a special case of the Taylor-Hood element~\cite{femSusanne}. We use the standard mixed Galerkin formulation~\cite{szabo1991finite, Ciarlet:2002:fem} with the full-order Gaussian quadrature for all our experiments. 

\paragraph{Finite Volume}
In the finite volume (FV) method for fluid simulation (e.g.,  \cite{Kolditz2002}), the solution domain is subdivided into a finite number of small control volumes (or cells) by a mesh, which does not change in time for transient calculations. Velocity and pressure are stored in the same cell location (cell center in Ansys Fluent), which is usually called the collocated or non-staggered variable arrangement\ \cite{Moukalled2016}.

The volume integrals in a partial differential equation that contain a divergence term are converted to surface integrals, using the divergence theorem. These terms are then evaluated as fluxes at the surfaces of each finite volume. Because the flux entering a given volume is identical to that leaving the adjacent volume, these methods are conservative.

We use the Ansys pressure-based FV solver and the Ansys laminar model \cite{ansys2016ansys} in Ansys Fluent\cite{ansys2016ansys}. We use the pressure-velocity coupled scheme. For spatial discretization, we select the least-square cell-based scheme for gradients, the second-order scheme for pressure, and the second upwind implicit scheme for momentum. We select the second-order implicit scheme provided in Fluent as the time integrator \cite{ansys2016ansys}.

\paragraph{Finite Differences}
In certain applications, high solution errors are more acceptable; in these cases, a structured and uniform spatial discretization with the lowest possible order of interpolation kernels for both velocity and pressure \cite{bridson2015fluid} is used. Many methods in this category adopt a linear velocity and constant pressure scheme. Since this discretization combination on collocated grids has spurious checkerboard or hour-glass modes and is not stable, it is common to use  the staggered Marker-And-Cell (MAC) grid\ \cite{harlow1965numerical} with velocities discretized on the cell surfaces and pressure on the cell centers, that stabilizes the scheme. We use the MAC grid in our experiments with FD approaches. In Appendix~\ref{app:setup} we describe how we use the MAC grid for complex geometries.

\subsection{Time Discretization}

\paragraph{Implicit Time Integration} 

A common choice for high-order time integration is the backward differentiation formula (BDF) of order 3, applied to the full system of equations, in contrast to splitting methods explained below. 

\paragraph{Ansys Coupled}
The widely used software Ansys Fluent \cite{ansys2016ansys} uses a coupled method to solve the pressure-velocity coupling. The fully implicit coupling is achieved through an implicit discretization of pressure gradient terms in the momentum equations and an implicit discretization of the face mass flux, including the Rhie-Chow pressure dissipation terms. Unlike the SIMPLE \cite{PATANKAR1972, JASAK1996} and SIMPLEC \cite{SIMPLEC} algorithms, it solves the momentum and pressure-based continuity equations together. 

\paragraph{Linear Operator Splitting}

Since directly solving the nonlinear time integration problem with incompressibility constraints is expensive, it is possible to use the Chorin's projection\ \cite{chorin1967numerical,stam1999stable} to split the solve at each time step into predictor-corrector sub-steps where pressure terms only participate in the corrector step that handles incompressibility. 

The predictor step is often further split into advection and diffusion for viscous fluids\ \cite{bridson2015fluid}. Diffusion step typically involves a global linear solve\ \cite{bridson2015fluid} or sometimes is computed using particles\ \cite{rivoalen2001particle,zhang2016resolving}. The advection step suffers from significant numerical dissipation in the classic semi-Lagrangian method\ \cite{sawyer1963semi,stam1999stable}; ways to reduce this dissipation were extensively studied. For purely grid based methods, while MacCormack\ \cite{maccormack2003effect,selle2008unconditionally} and BFECC\ \cite{dupont2003back} advect the solution forward and then backward in time to estimate and compensate error, Bimocq\ \cite{qu2019efficient}  resolve the flow map characteristics for periods longer than a single time step (as opposed to one step with semi-Lagrangian) to reduce dissipation.
Hybrid Lagrange/Eulerian techniques like Particle-In-Cell (PIC)\ \cite{harlow1962particle}, Fluid-Implicit-Particle (FLIP)\ \cite{brackbill1986flip}, AffinePIC\ \cite{jiang2017angular,ding2020affine}, and PolyPIC\ \cite{fu2017polynomial} explicitly track motion of particles in the fluid and transfer information between grid and particles, which is nearly dissipation-free, but can suffer from nonuniform particle distribution and also higher computational costs.
Ding et al.\ \cite{ding2020affine} presents a  quantitative study of accuracy of several different particle-based advection methods.
 Chorin projection itself causes dissipation: Zhang et al.\ \cite{zhang2015restoring} reduced numerical dissipation
caused by the corrector step by estimating lost vorticity and adding it back into the fluid. Zehnder et al.\ \cite{zehnder2018advection} propose a simple but effective modification to the splitting scheme that is similar to midpoint rule integration to reduce the corrector error.

In this study, we evaluate the two advection methods, semi-Lagrangian (SL) \cite{sawyer1963semi} and the Fluid-Implicit-Particle (FLIP) \cite{brackbill1986flip}, which are typical for the Eulerian and Lagrangian views respectively. The original SL method is unconditionally stable, and the method is widely used in numerical weather prediction\ \cite{bates1982multiply}. For a more thorough comparison among different advection methods, we refer to\ \cite{qu2019efficient} and\ \cite{ding2020affine}.

\paragraph{High-Order Operator Splitting}

The computational advantages of commonly used linear operator splitting have the downside of limiting their rate of convergence to linear. To overcome this limitation, while retaining some of their benefits, high-order splitting schemes can be used instead \cite{William1989,AB2AM2,2ndSplit,Weinan1995,CHALMERS2019}. In this study, we use the method proposed in \cite{2ndSplit} as a recent representative of this class. We briefly review this method for completeness.

\cite{2ndSplit} solves the incompressible NS equations using a split-step approach that decouples the solution of the velocity from the solution of the pressure. The split-step method \cite{AB2AM2,2ndSplit} is an explicit predictor-corrector method that consists of a second-order Adam-Bashforth (AB2) predictor and a modified second-order Adam-Moulton (AM2) predictor.
We implemented this method with the TN boundary condition and divergence damping in \cite{2ndSplit}. 
% \DZ{commented out the mention of WABE - one sentence left did not explain anything; perhaps best to leave this discussion out, and if the reviewers ask, then we can explain the whole story.}
%Note that we do not use the WABE boundary condition \cite{2ndSplit} as the TN boundary condition obtains the same convergence rate as the WABE boundary condition for both velocity and pressure. %, and the 3D version of the WABE boundary condition is unclear \DP{in what sense? not explained in the paper?} \ZH{They don't have a 3D version in the paper, their derivation of WABE based on 2D}.

\subsection{Convergence Rates} 
\begin{figure}
    \centering\scriptsize
    \includegraphics[width=.45\linewidth]{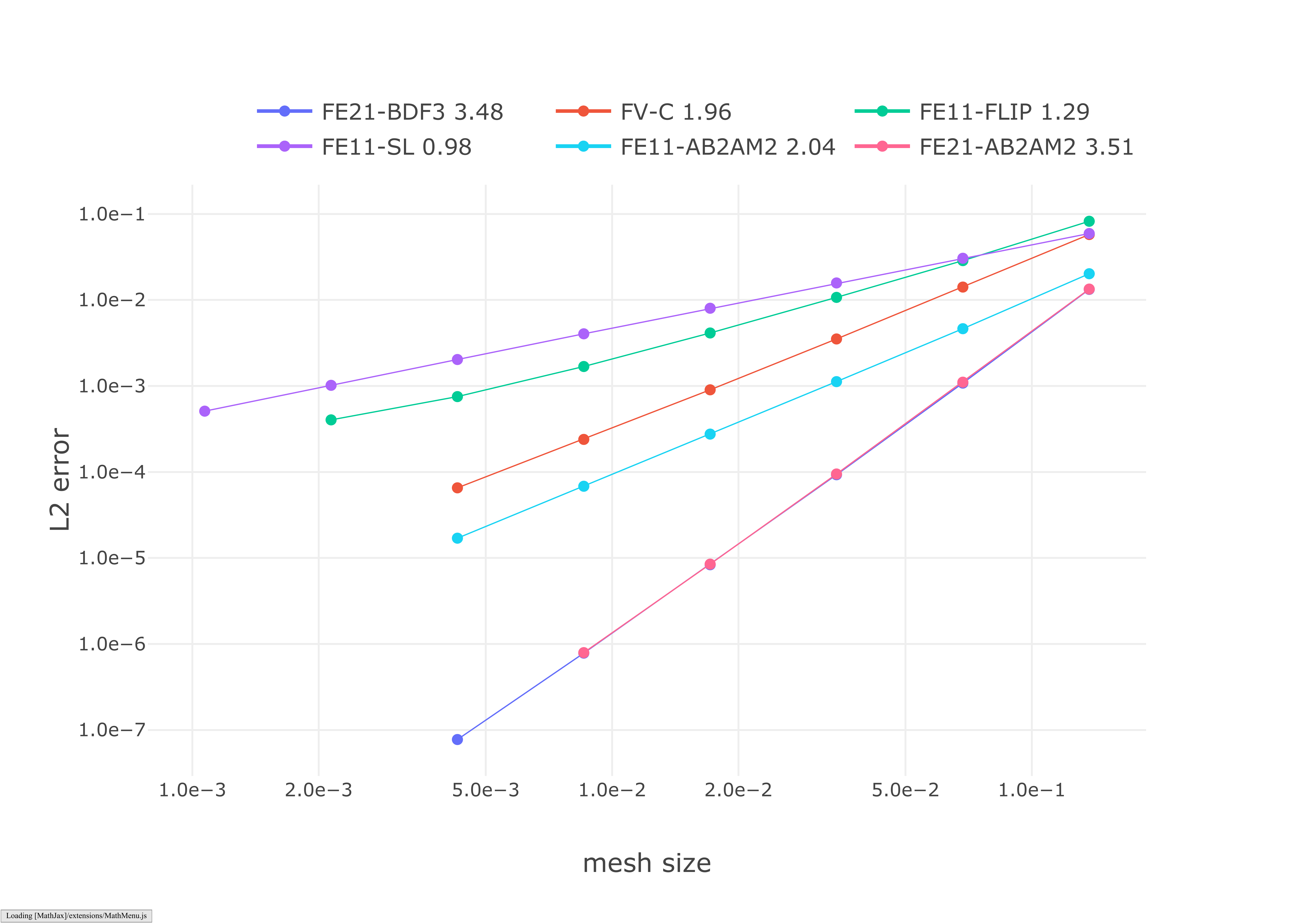}\hfill
    \includegraphics[width=.45\linewidth]{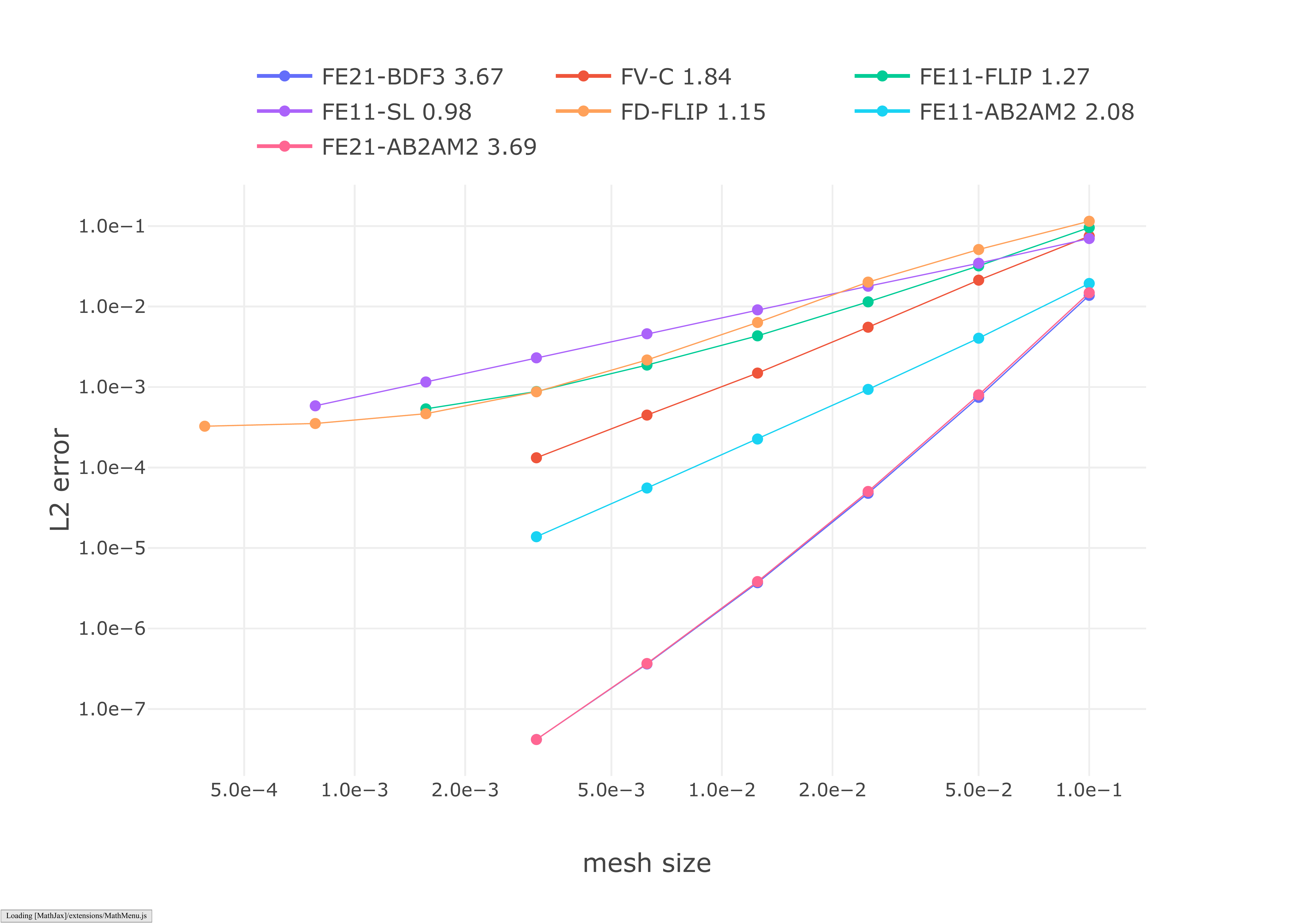}\\
    \parbox{.45\linewidth}{\centering Tri mesh}\hfill
    \parbox{.45\linewidth}{\centering Regular grid}
    \caption{Convergence of five methods on a square domain. In the legend we report the convergence obtained by linear fit in log space.}
    \label{fig:convergence}
\end{figure}

\begin{figure}
    \centering\scriptsize
    \includegraphics[width=.45\linewidth]{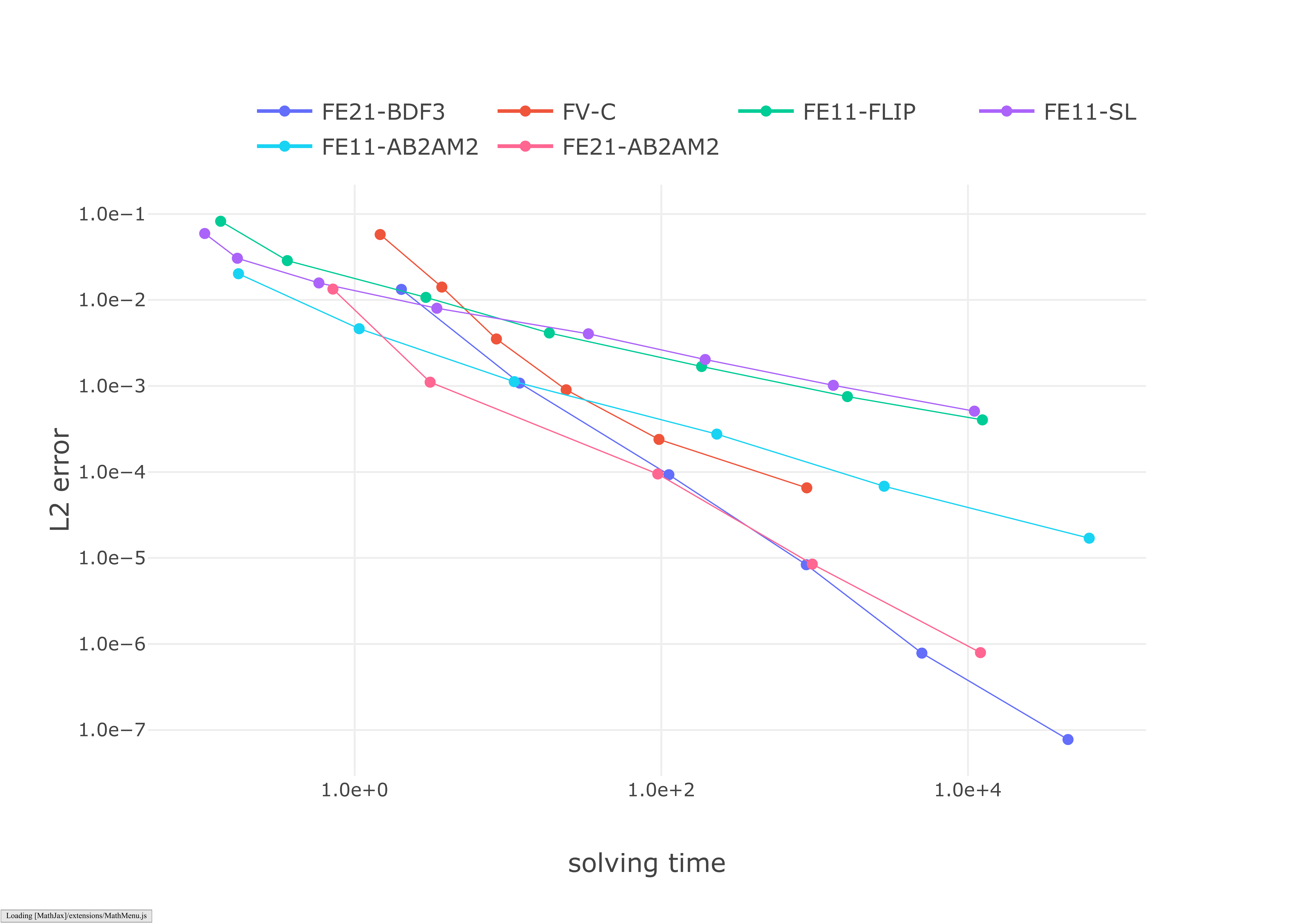}\hfill
    \includegraphics[width=.45\linewidth]{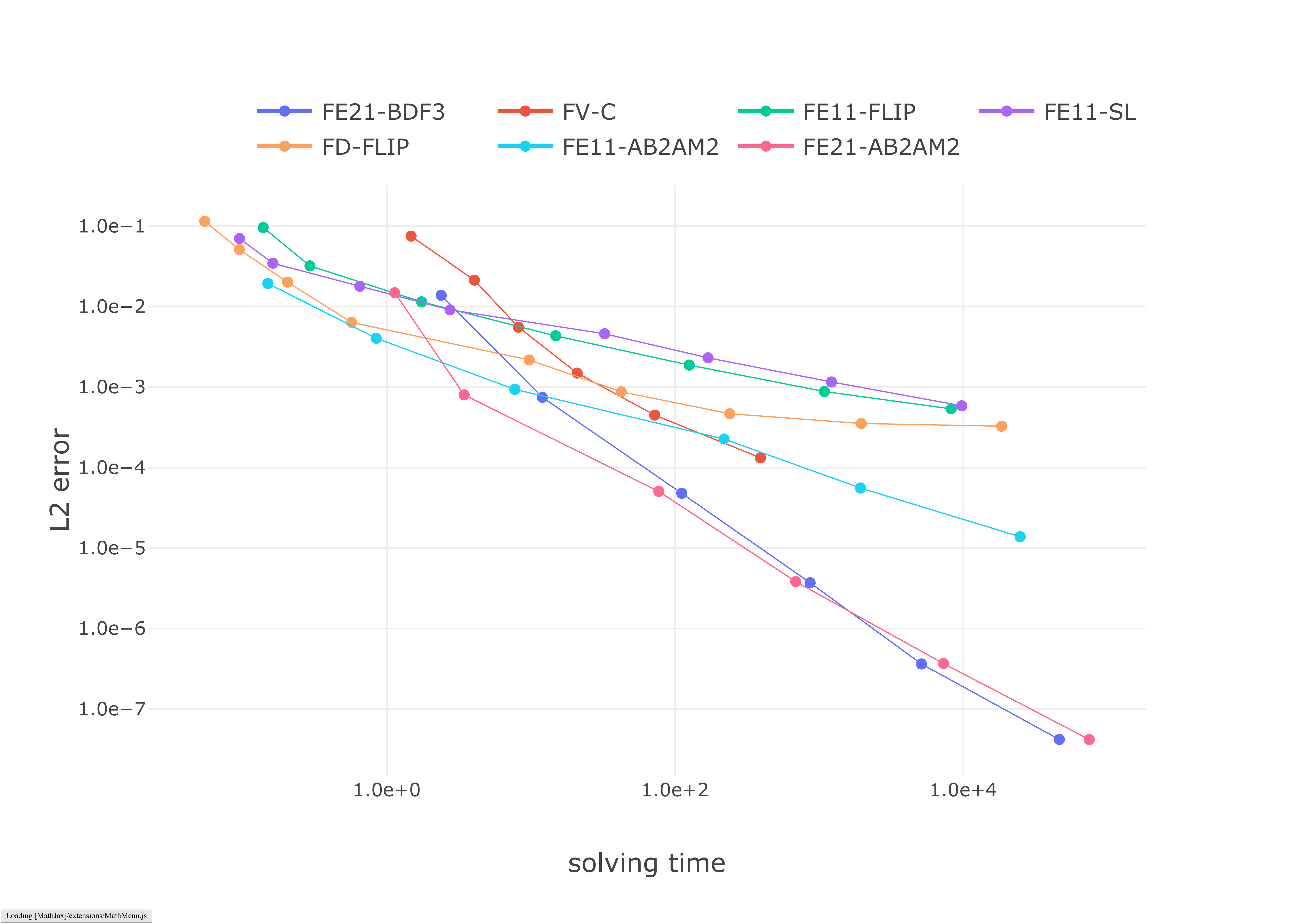}\\
    \parbox{.45\linewidth}{\centering Tri mesh}\hfill
    \parbox{.45\linewidth}{\centering Regular grid}
    \caption{Performance of 5 methods on a square domain.}
    \label{fig:time-vs-error}
\end{figure}

We implemented the mesh-based methods using the PolyFEM library \cite{polyfem}. For FV-C, we rely on the implementation in Ansys \cite{ansys2016ansys}. For FD-FLIP, we use a separate framework specialized for MAC-grids. We report the convergence of the five different methods on a simple square domain (Figure~\ref{fig:convergence}). As expected, FE21-BDF3 and FE21-AB2AM2 converges cubically, FV-C and FE11-AB2AM2 quadratically, and the other three methods linearly. Since the different methods lead to different system sizes and non-linear optimization, we also compare the time required to solve to a given error (Figure~\ref{fig:time-vs-error}). %The plot shows that FE is the best method providing the best accuracy for a given error.

%% file: 05-testproblems.tex
\section{Commonly Used Test Problems}\label{sec:test-probl}

We study five two-dimensional standard test cases to cover most common variations of fluid flow problems and various boundary conditions:
driven-cavity (classical problem with only Dirichlet boundary conditions, Section~\ref{sec:cavity}),
airfoil (complex internal boundary, Section~\ref{sec:airfoil}),
open cavity (different viscosities, Section~\ref{sec:ocavity}),
% Corner flow (Neumann boundary conditions, Section~\ref{sec:L}),
vortex street (complex turbulent behavior, Section~\ref{sec:vortex}),
and a drag force computation (Section~\ref{sec:drag}).
For a complete description of our experimental setup we refer to Appendix \ref{app:setup}.

\paragraph{Evaluation}
%\DZ{We need to say something about not including the here few cases when closed-form solutions are known, Taylor-Green, Couette, there are others; especially given that we use Taylor-Green in the next section}

% For these problems, there are no known analytical solutions. (We did verify convergence of all our implementations using the classical Taylor-Green vortex solution in Appendix \ref{app:simple-ex}).

We compute a numerical reference solution using a denser mesh using the same CFL number (i.e., we use a smaller time step) and a more accurate method: that is, we define the error as
\[
    e^2 = \int_\Omega \|u_h(x, T) - u^\star(x, T)\|^2\, dx,
\]
where $u_h(x, T)$ is the solution of a given method and $u^\star(x, T)$ is the reference solution computed using FE32-BDF4 on a denser mesh. This scheme uses $P_3/Q_3$ basis for velocity and $P_2/Q_2$ basis for pressure, and BDF order 4 for time integration. Note that this scheme uses both a time and space discretization of higher order than all the methods we include in the study, in addition to being applied to a denser mesh. We note that for many applications other error measures are relevant (e.g., error in velocity gradients, determining the fluid forces). However, for many commonly used low-order methods, velocity is discontinuous, and convergence of gradients can be shown only in specific discrete norms, i.e., effectively, in addition to numerical method itself, one needs to choose a discretization of the gradient to define the norm \cite{eymard2000finite}.  This makes fair comparisons between discretizations and methods more difficult as the optimal choice of velocity gradient discretization may be method-dependent.  For this reason, we focus on the solution error. 
We also note that the equation residual provides a stringent test for how well the equation is satisfied, but can be easily applied only to higher-order methods for which both gradients and second derivatives can be computed pointwise; 
we present some experiments in Appendix~\ref{app:residual}, for a particular choice of approximation of the high order derivatives.

We acknowledge that using a dense mesh with FE32-BDF4 as ``ground-truth'' potentially might give an unfair advantage to FE21-BDF3. We selected FE32-BDF4 as it gives the lower errors in the experiments having a ground truth solution, and it is reasonable to expect that in other cases its accuracy remains high on fine meshes. 

\paragraph{Boundary Geometry}
To facilitate the comparison for FD-FLIP (where complex boundaries need to be captured with a  level-set), and to avoid favoring unstructured meshes which are easier to adapt to complex boundaries, we choose most of our domains to have straight boundaries. For all methods except FD-FLIP we run experiments on both a regular quad mesh and an unstructured triangular mesh. We adapt the mesh size for every experiment to ensure that all the methods have similar running times and report the resulting errors. Neumann boundary conditions are imposed differently for different methods as discussed in Appendix \ref{app:neumann}.

\paragraph{Time step}
We use $dt=C_2h^2$ for FE11-AB2AM2, $dt=C_3h^2$ for FE21-AB2AM2, and $dt=C_1h$ for all other methods, where $h$ is the mesh size and the choices of constants $C_i$ are in Table \ref{tab:dt}.

\begin{center}
\begin{table}[!htbp]\label{tab:dt}
\centering
\caption{Choices of $dt$}
\begin{tabular}{c|c|c|c|c|c}
\toprule
Problem & Driven Cavity & Airfoil & Open Cavity & Vortex Street & Drag Force\\
\midrule
$C_1$ & 0.1 & 0.036 & 0.2 & 0.29 & 0.12 \\
$C_2$ & 0.125 & 0.0013 & 0.5 & 0.082 & 0.014 \\
$C_3$ & 0.0625 & 0.0013 & 0.5 & 0.082 & 0.014 \\
\bottomrule
\end{tabular}
\end{table}
\end{center}

%\DZ{General: a lot of discussion here seems to be qualitative, and we do not mention this explicitly, and it is not consistent.   For an applied math audience, qualitative discussion is possible, but needs to be more specific. E.g., when we talk about numerical viscosity, this is a cause or interpretation of the fact that the flow is closer to laminar. A better way to phrase this would be that compared to the accurate solution, the flow is closer to laminar, or has lower vorticity. I think though that unless we can point to a specific qualitative phenomenon distinguishing the reference solution from a particular approximate one,  we are likely to be asked to quantify.}

\subsection{Driven Cavity}\label{sec:cavity}
\begin{figure}
    \includegraphics[width=.45\linewidth]{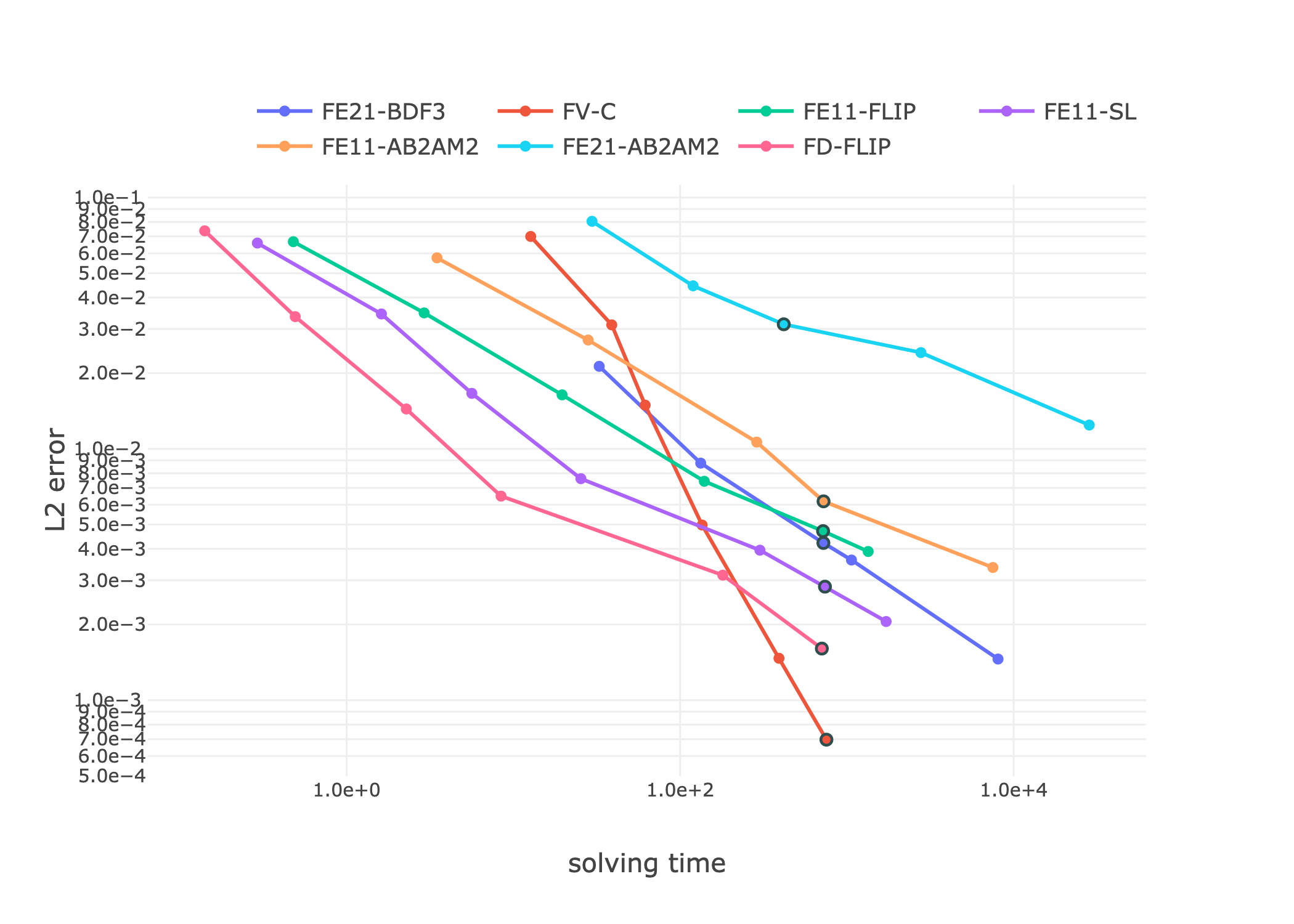}\hfill
    \includegraphics[width=.45\linewidth]{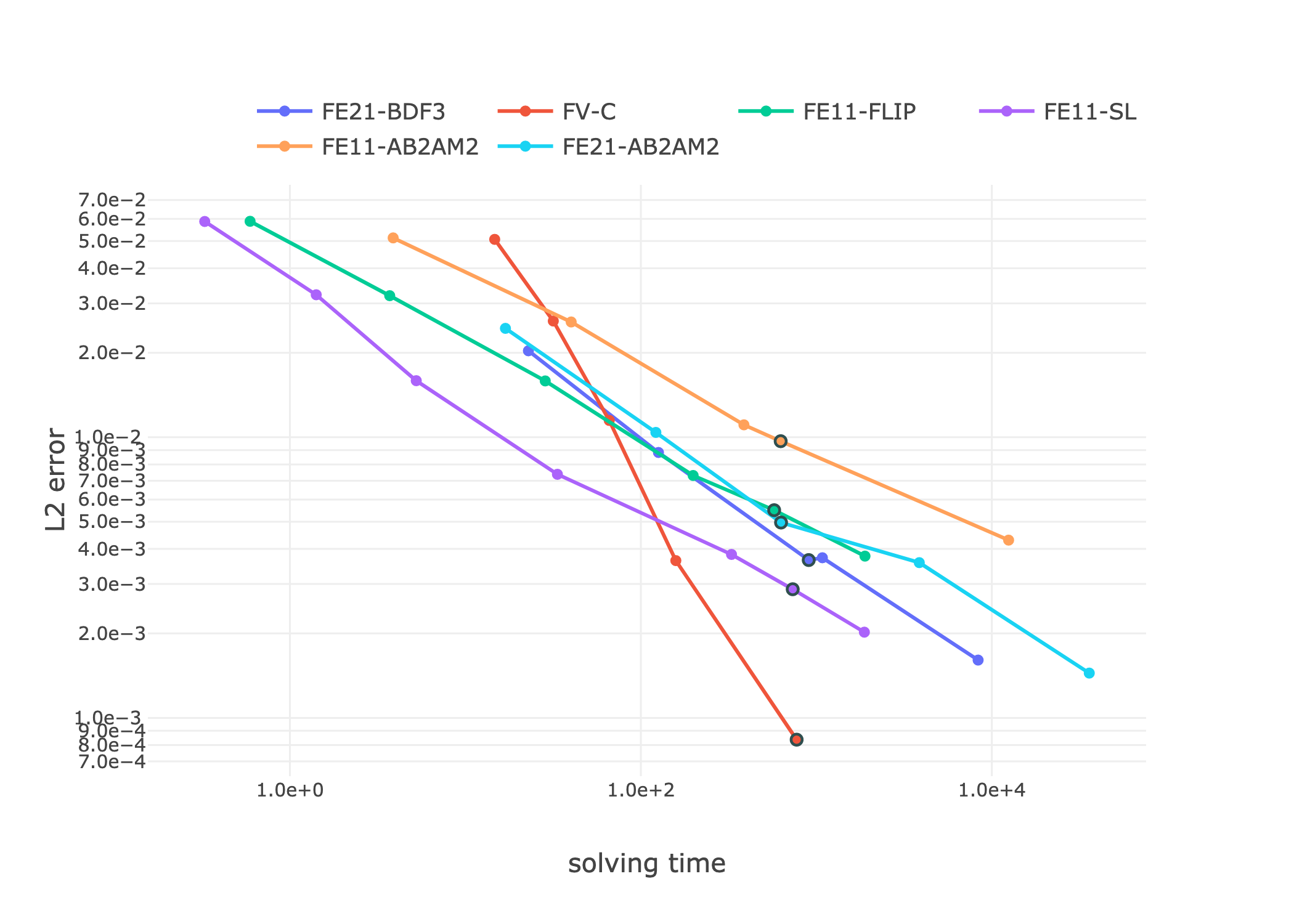}\par
    \parbox{.45\linewidth}{\centering Regular grid}\hfill
    \parbox{.45\linewidth}{\centering Triangle mesh}
    \caption{Error versus solve time for the five different methods for the driven cavity.  Figure~\ref{fig:cavity} shows the solution for the highlighted points.} 
    \label{fig:cavity-time}
\end{figure}

For this test, the domain is a unit square, the initial velocity is zero, zero Dirichlet boundary condition is imposed on the left/bottom/right side, and $(\sin(\frac{\pi}{2}t), 0)$ on the top side.  We run the experiments with $\nu = 0.1$, $T=2$. For this simple setup, all methods produce similar results (Figure~\ref{fig:cavity}). 
%All methods use a similar amount of memory, with FV using up to 1.3Gb \DZ{not sure it makes sense to talk about memory numbers w/o context -- resolution, solver}, and MAC-FLIP using as little as 0.13Gb.
In this experiment, FV-C has a slightly smaller error compared with the other methods. Our experiments show that FV-C is the best method (measured by time versus accuracy) at the coarser resolutions (Figure~\ref{fig:cavity-time}). %However, as we change the boundary conditions to $C^0$ and $C^1$, FE exhibits quadratic and cubic convergence (Section~\ref{app:smothness-conv}).

\input{pics/cavity/plot}

\subsubsection{Boundary Condition Smoothness}\label{app:smothness-conv}

The boundary condition for this test in a standard form is not continuous; consequently, high-order methods like FV-C and FE21-BDF3 do not reach the theoretical convergence rate. To study how the smoothness of the boundary condition affects the convergence, we solve the driven-cavity problem with three different boundary conditions for the top boundary:

$$
    v=(1, 0), \qquad
    v=(4(1-x)x, 0), \qquad \text{and} \qquad
    v=\Big(50\exp\Big(\frac{1}{x(1-x)}\Big), 0\Big).
$$

\begin{figure}
    \centering\scriptsize
    \includegraphics[width=.3\linewidth]{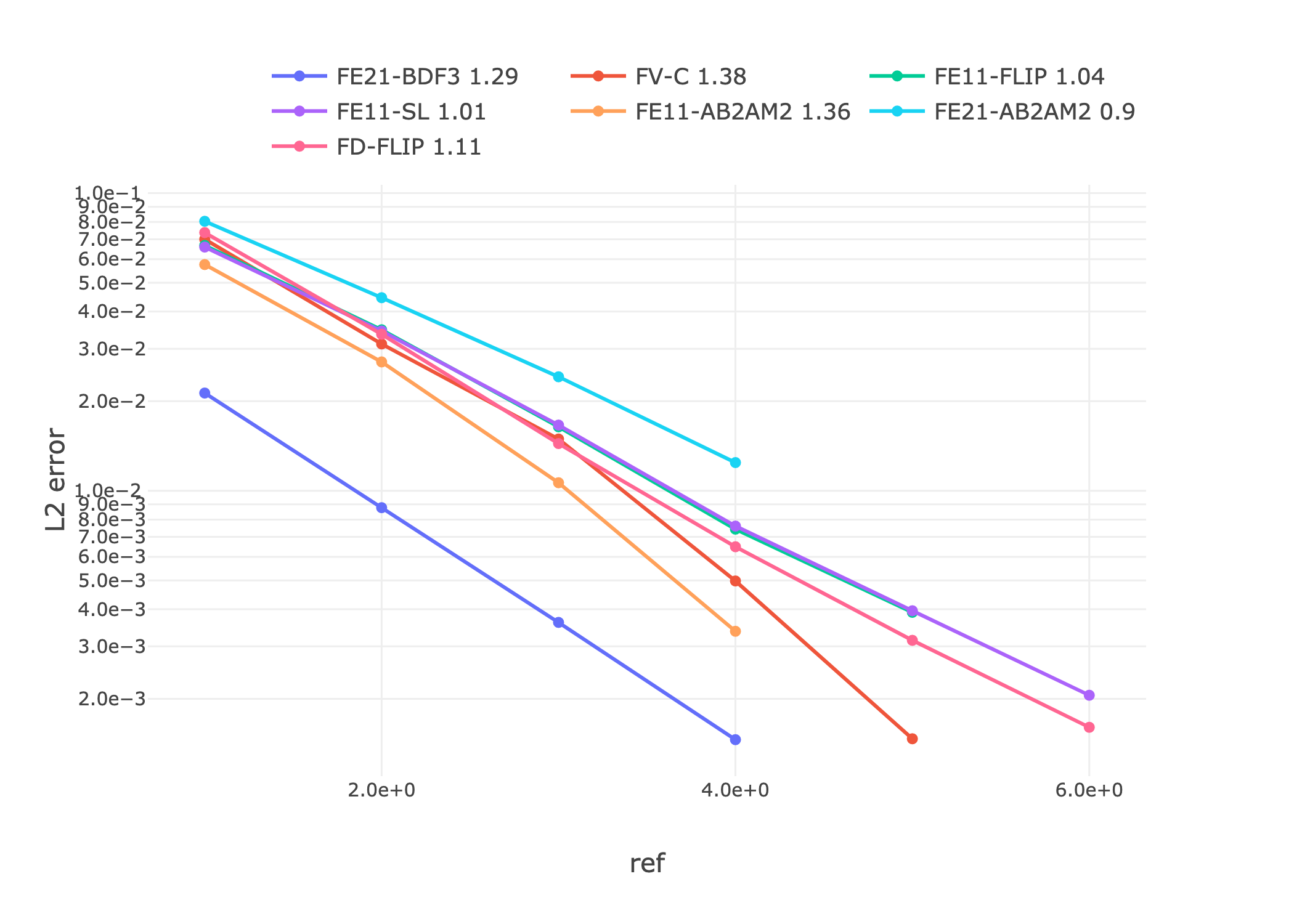}\hfill
    \includegraphics[width=.3\linewidth]{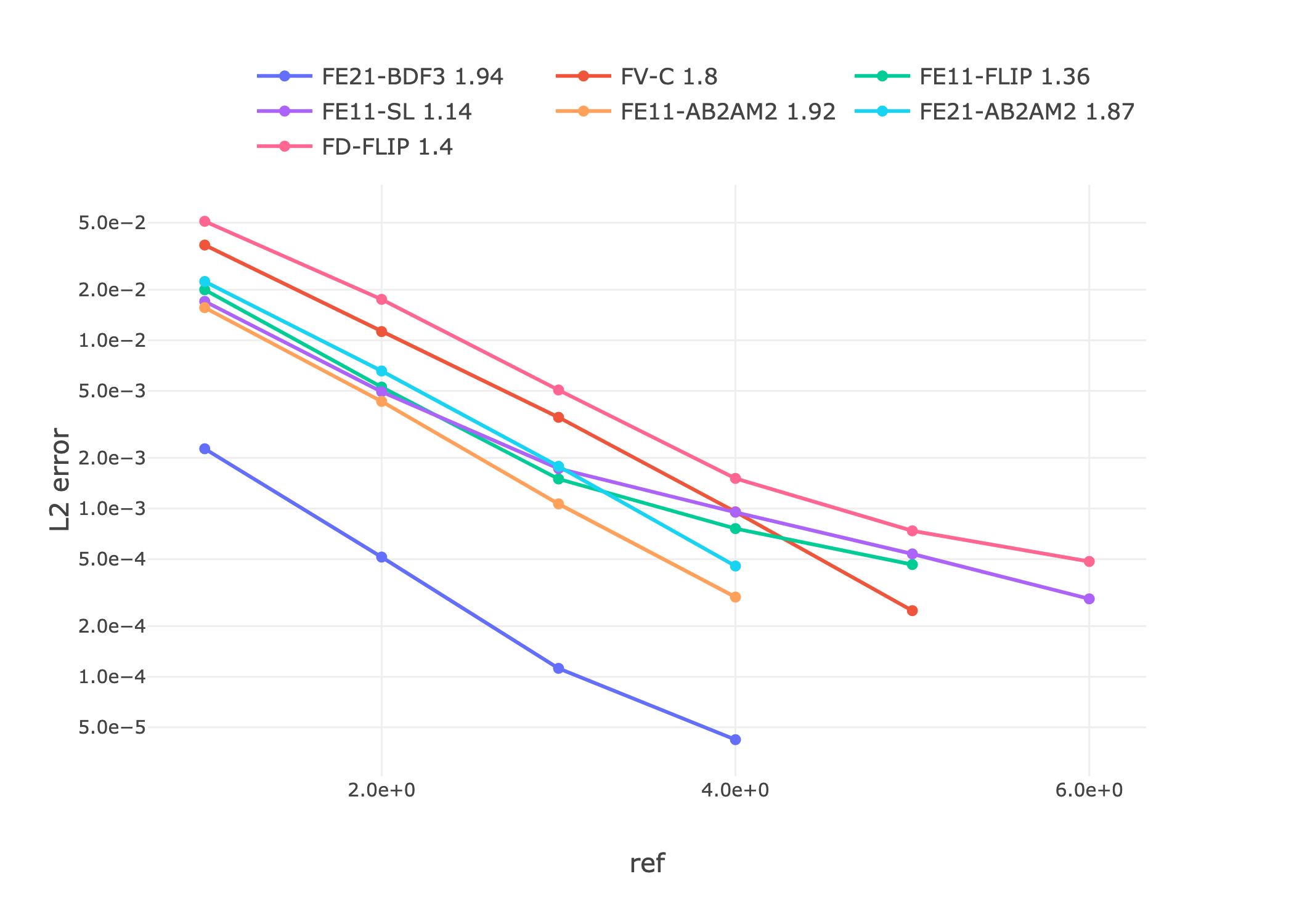}\hfill
    \includegraphics[width=.3\linewidth]{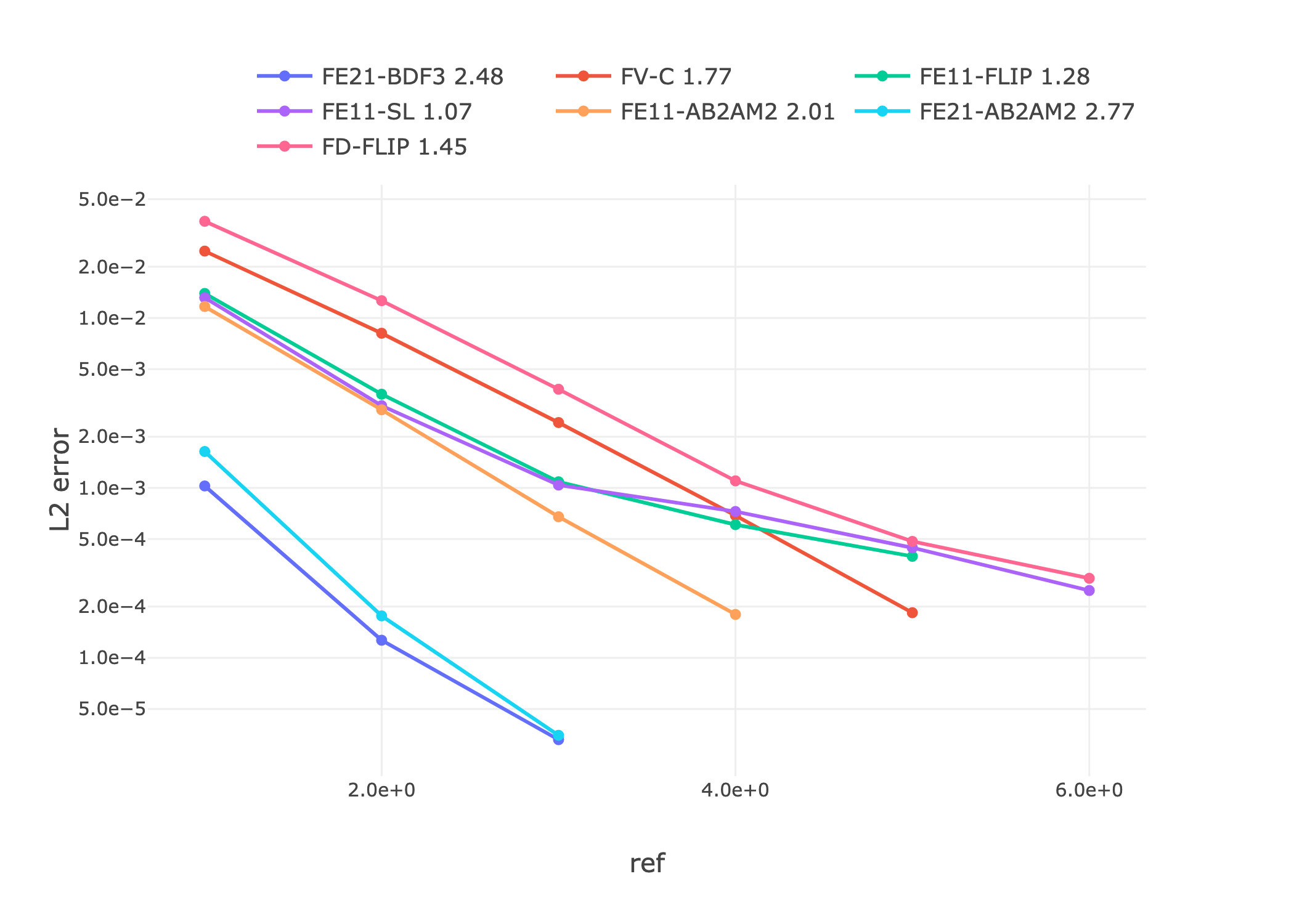}\par
    \parbox{.3\linewidth}{\centering $v=1$}\hfill
    \parbox{.3\linewidth}{\centering $v=4(1-x)x$}\hfill
    \parbox{.3\linewidth}{\centering $v=50\exp\left(\frac{1}{x(1-x)}\right)$}
    \caption{Error vs. mesh size of all methods in the driven cavity problem with different boundary conditions.}
    \label{fig:cavity-convergence-h}
\end{figure}

\begin{figure}
    \centering\scriptsize
    \includegraphics[width=.3\linewidth]{pics/cavity/cavity_discontinuous_quad-time-error}\hfill
    \includegraphics[width=.3\linewidth]{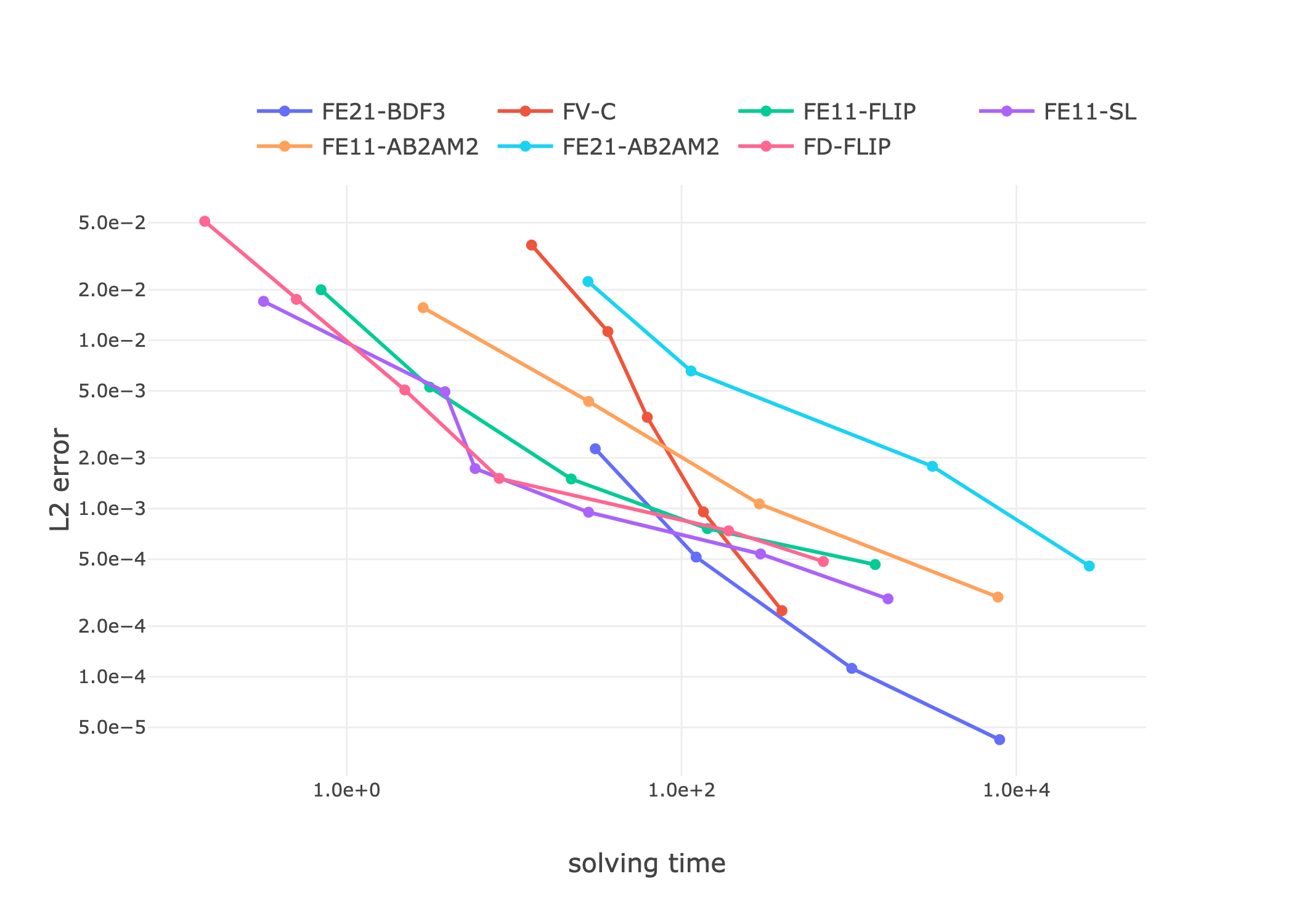}\hfill
    \includegraphics[width=.3\linewidth]{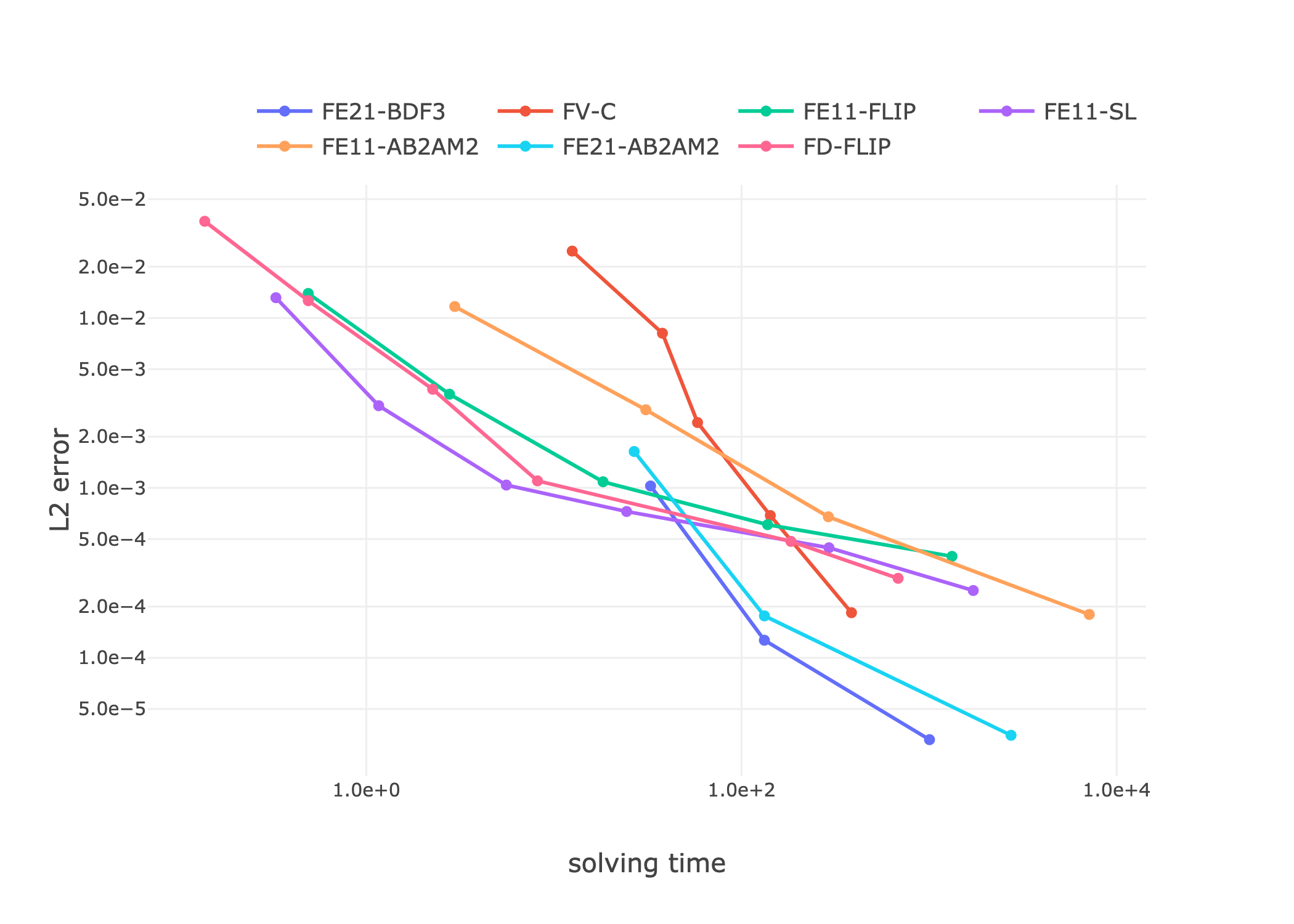}\par
    \parbox{.3\linewidth}{\centering $v=1$}\hfill
    \parbox{.3\linewidth}{\centering $v=4(1-x)x$}\hfill
    \parbox{.3\linewidth}{\centering $v=50\exp(\frac{1}{x(1-x)})$}
    \caption{Error vs. running time of all methods in the driven cavity problem with different boundary conditions.}
    \label{fig:cavity-convergence-time}
\end{figure}

Figure~\ref{fig:cavity-convergence-h} shows that for discontinuous Dirichlet boundary condition, no method can achieve convergence rate higher than one. As we increase the continuity of the boundary conditions to $C^0$ FE21-BDF3, FV-C, FE11-AB2AM2, and FE21-AB2AM2 show quadratic convergence, and finally, when we use a smooth function, FE21-BDF3 and FE21-AB2AM2 have cubic convergence. Since the solve time of the different methods is mostly independent from the boundary conditions, the reduction in the convergence rate affects the error corresponding to a given time budget (Figure~\ref{fig:cavity-convergence-time}): for $v=1$ all methods are similar, and as smoothness of the boundary condition increases FE21-BDF3 and FE21-AB2AM2 gets more efficient.

\subsection{Airfoil}\label{sec:airfoil}

In this experiment (Figure~\ref{fig:airfoil}), we embed the NACA 5012 airfoil in a background mesh and specify left-to-right inflow velocity of $v=1-e^{-5t}$ and zero initial velocity. We apply  zero Dirichlet boundary condition on the airfoil surface. We run our simulations with $\nu=0.0002$ and $T=4$. For this more complex geometry, the methods produce different results (Figure~\ref{fig:airfoil}, closeups),  FE11-FLIP and FE11-SL exhibit additional numerical viscosity, compared to FD-FLIP. We note that our error measure in this case is not able to capture some of the major artefacts in the solutions computed with higher-order splitting methods. When using AB2AM2, the necessary number of time steps is very high, thus to match the wall-clock time for different methods, we have to use coarse meshes leading to large errors and artefacts.
%Since only FE and FV can solve for steady-state directly (i.e., solve for equilibrium by dropping the time-dependent derivative), we confirm that the equilibrium solution matches their results for $T=4$. 
% For this experiment, we excluded the quad mesh as the shape of the airfoil is highly irregular. \DZ{unclear - why is this a problem for quad meshes? Just state this was done on triangle meshes only without any arguments}

\input{pics/airfoil/plot}

\subsection{Open Cavity}\label{sec:ocavity}
For this test, we compare different methods for several viscosities (Figure~\ref{fig:ocavity}). The fluid flows from left to right;  the velocity is  $v=(1-e^{-5t}, 0)$ on the left-most vertical boundary,  the right side has zero Neumann boundary condition, and the remaining boundaries have zero velocity.  For low Reynold number, all methods exhibit similar behavior. As the Reynolds number increases, FE11-SL and FE11-FLIP have higher numerical viscosity, and the qualitative stream behavior is different, with FE11-FLIP showing non-physical turbulence-like behavior in the middle. These differences have a small impact on the error, which stays within one order of magnitude independently from $\nu$.

\input{pics/ocavity/plot}

\subsection{Vortex Street}\label{sec:vortex}

\input{pics/vortex/plot-tri}

\begin{figure}
    \centering\scriptsize
    \includegraphics[width=.4\linewidth]{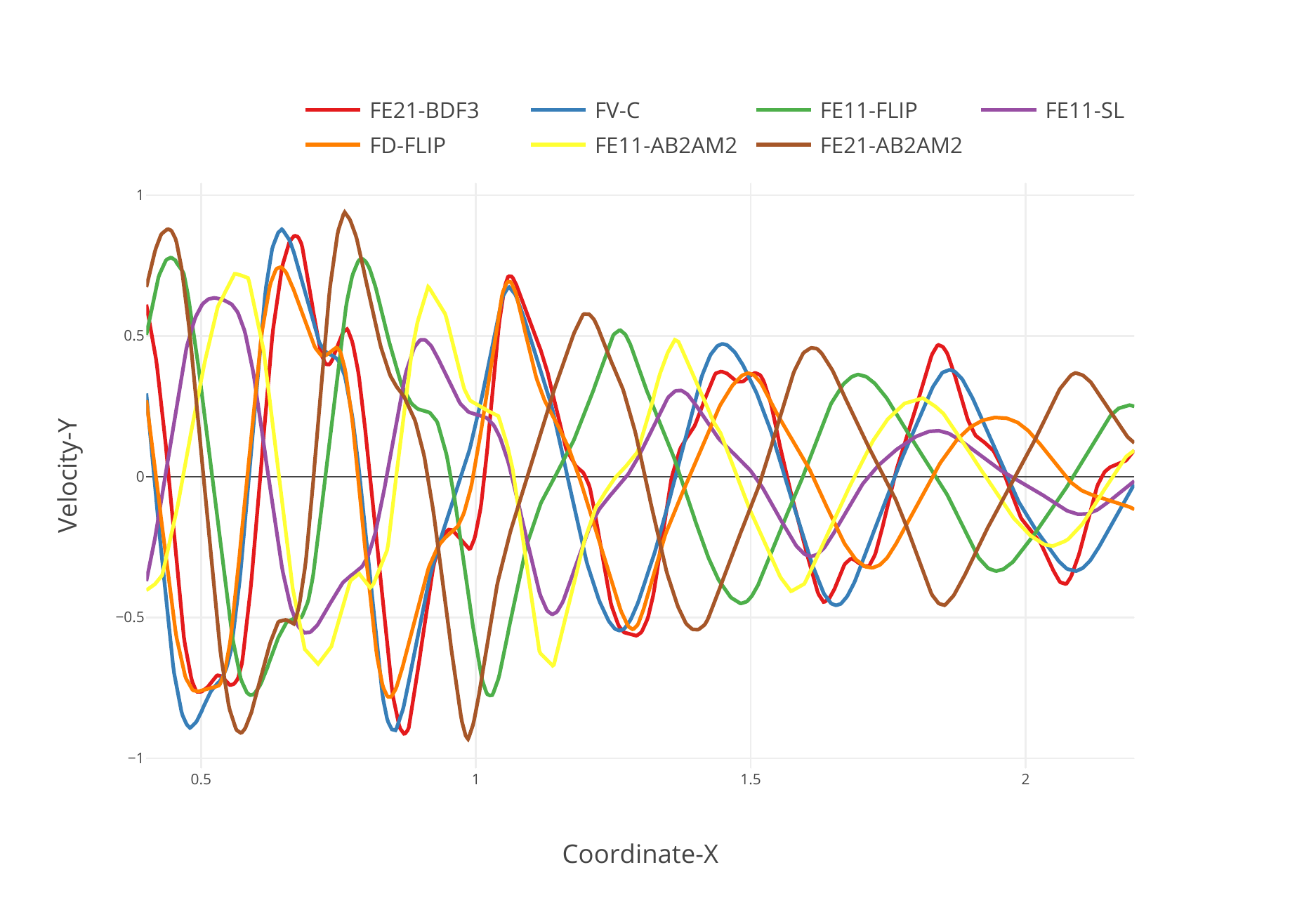}
    \caption{$y$-component of the velocity at $t=6s$ along an horizontal line passing trough the middle of the domain for the results in Figure~\ref{fig:vortex-tri}.}
    \label{fig:cross-sec}
\end{figure}

We run five methods on a pipe with an obstacle located off-axis, to generate a vortex. The fluid has viscosity $0.001$, the top, bottom, and circle have zero Dirichlet boundary conditions, velocity on the left side is $u(0,y) = (6(1-e^{-5t})(0.41-y)y/0.1681, 0)$, and  zero Neumann on the right side \footnote{We used the setup from \url{http://www.featflow.de/en/benchmarks/cfdbenchmarking/flow/dfg_benchmark2_re100.html}}. Figure~\ref{fig:vortex-tri} shows the results at the end of the simulation, when the vortex is fully formed. For a qualitative comparison, we plot the $y$-component of the velocity along a horizontal line passing through the middle of the domain (Figure~\ref{fig:cross-sec}); all methods manifest similar behavior, and the dissipation is similar. It is interesting to note that some methods have a significant ``phase shift''  with respect to others.

\begin{figure}
    \centering\scriptsize
    \includegraphics[width=.25\linewidth]{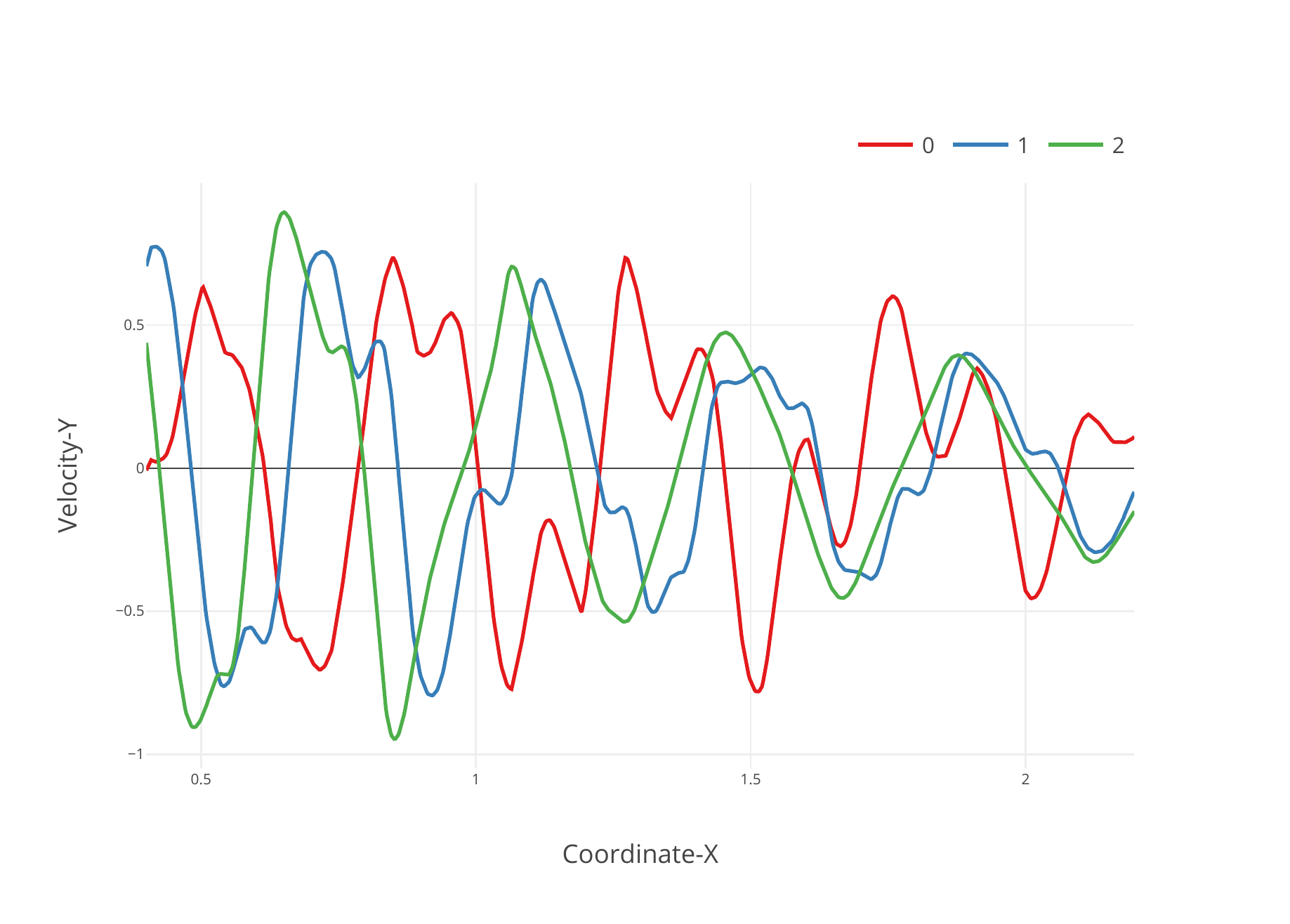}\hfill
    \includegraphics[width=.25\linewidth]{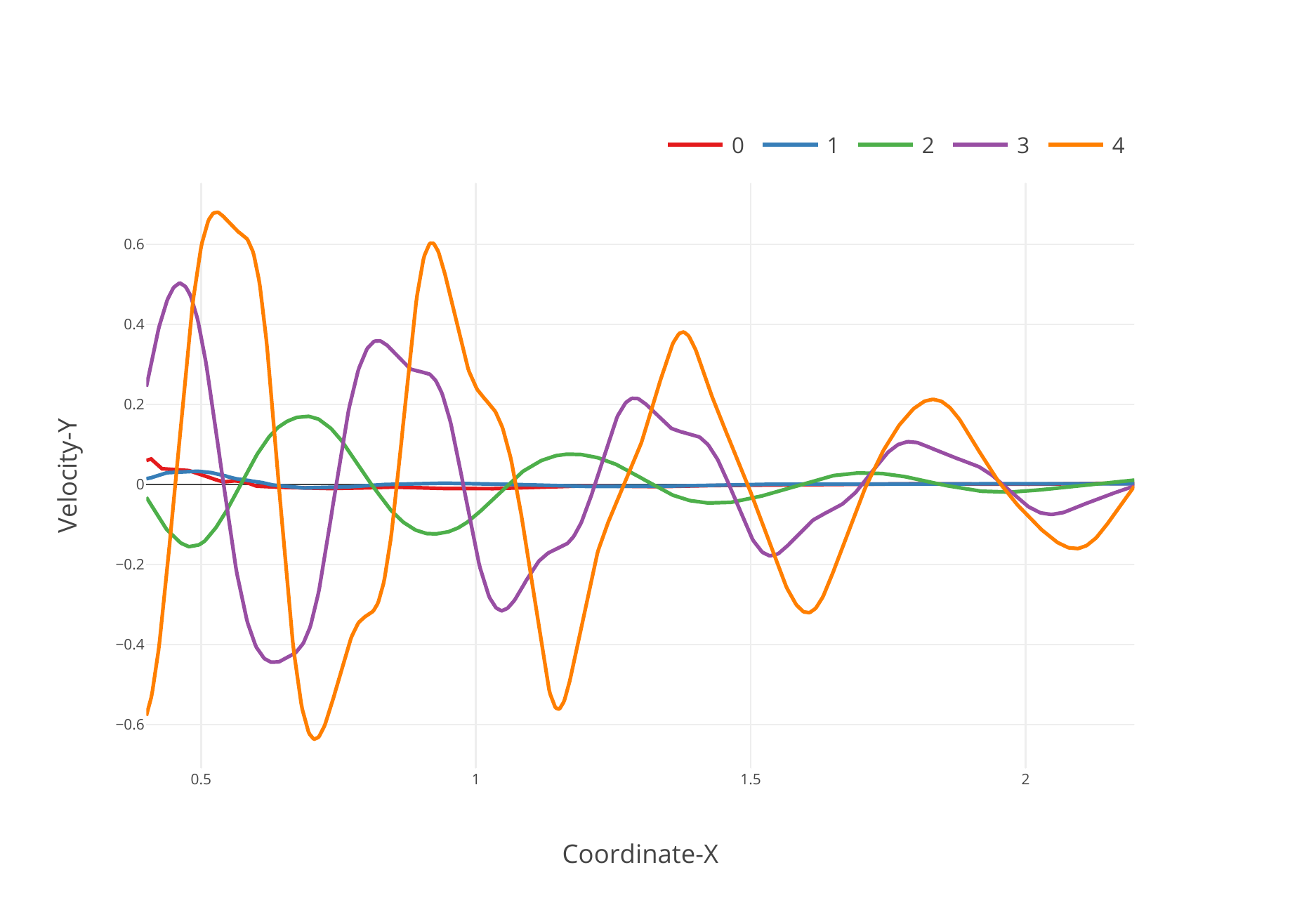}\hfill
    \includegraphics[width=.25\linewidth]{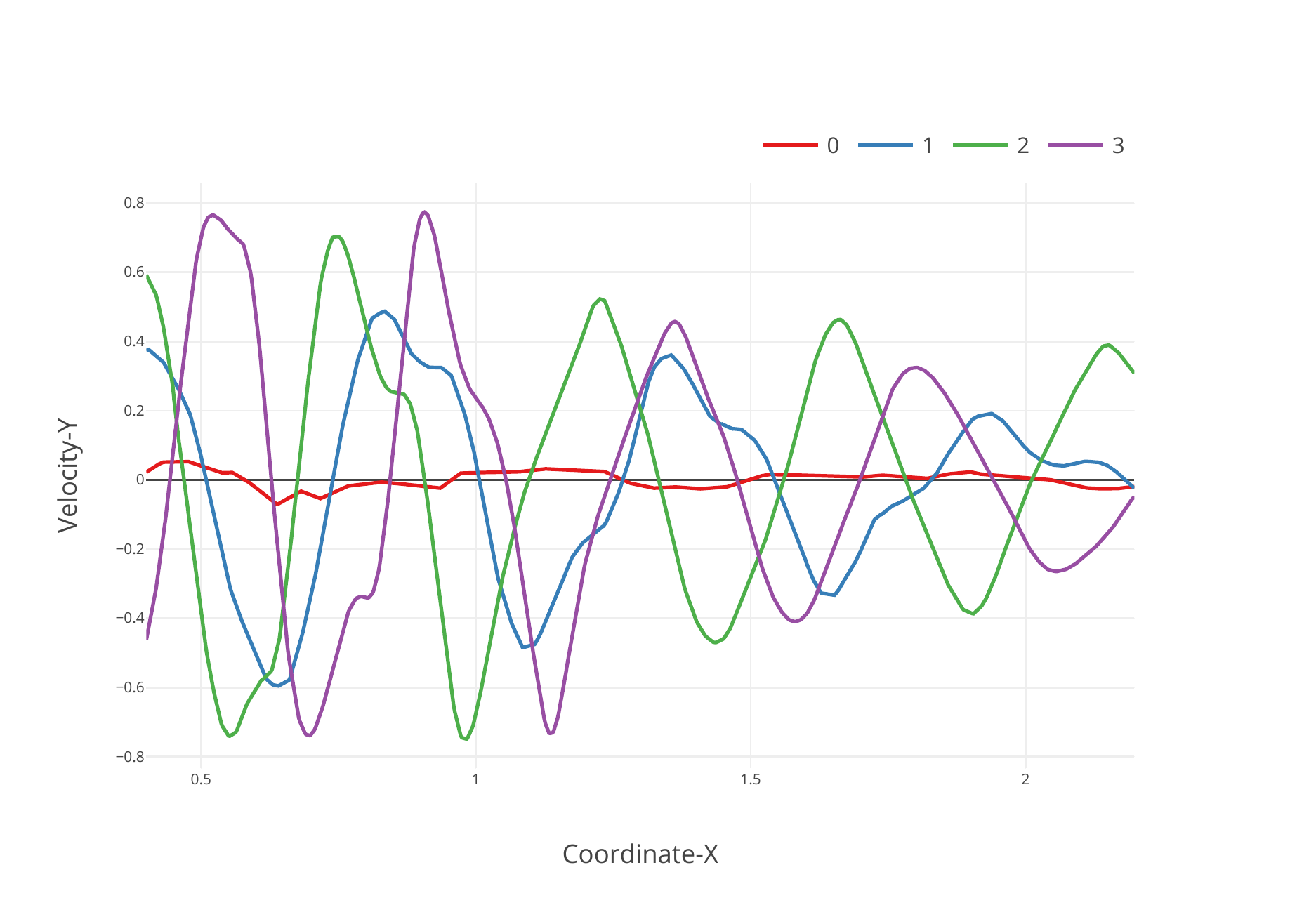}\\
    \parbox{.25\linewidth}{\centering FE21-BDF3}\hfill
    \parbox{.25\linewidth}{\centering FE11-SL}\hfill
    \parbox{.25\linewidth}{\centering FE11-FLIP}\\
    \includegraphics[width=.25\linewidth]{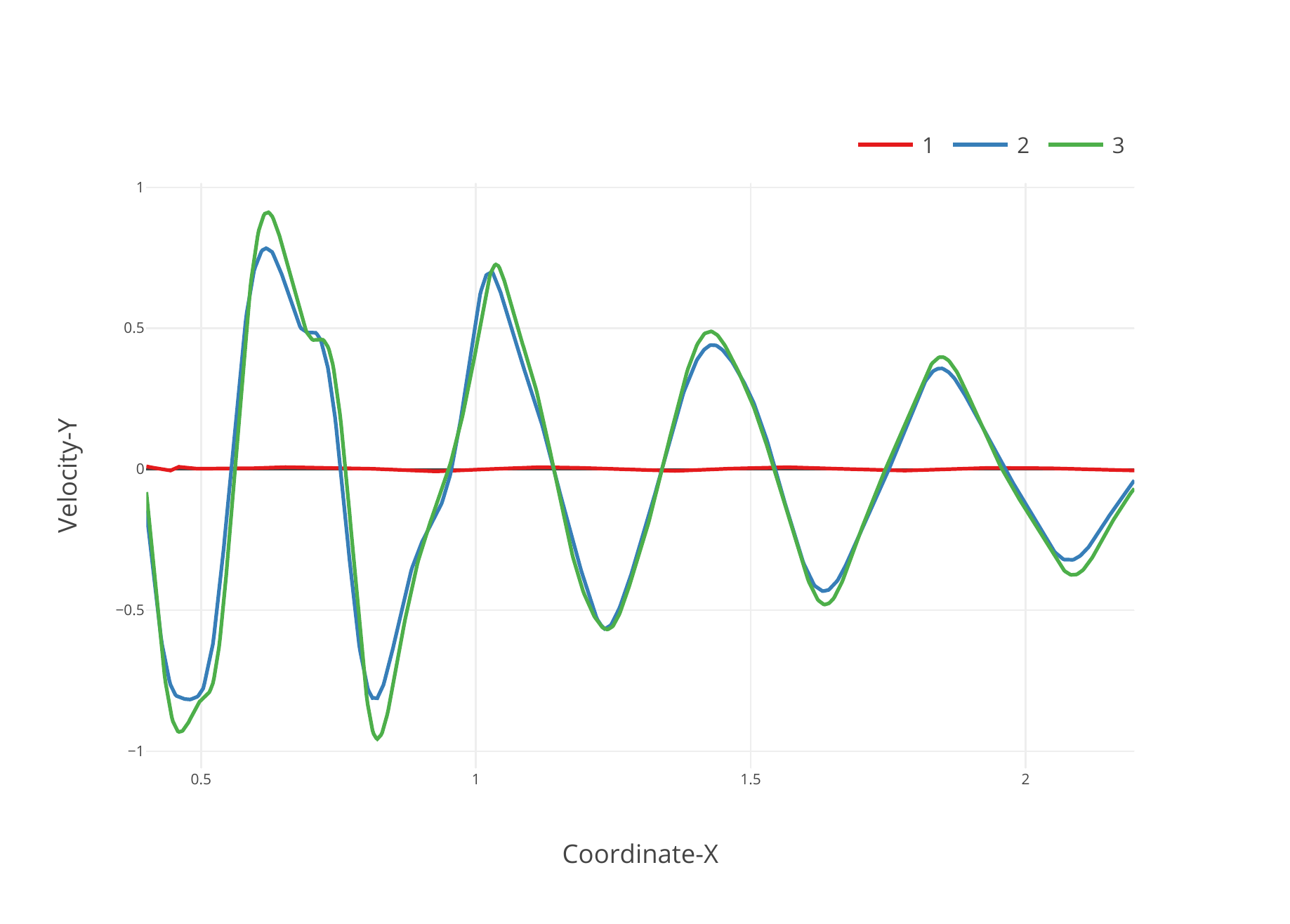}\hfill
    \includegraphics[width=.25\linewidth]{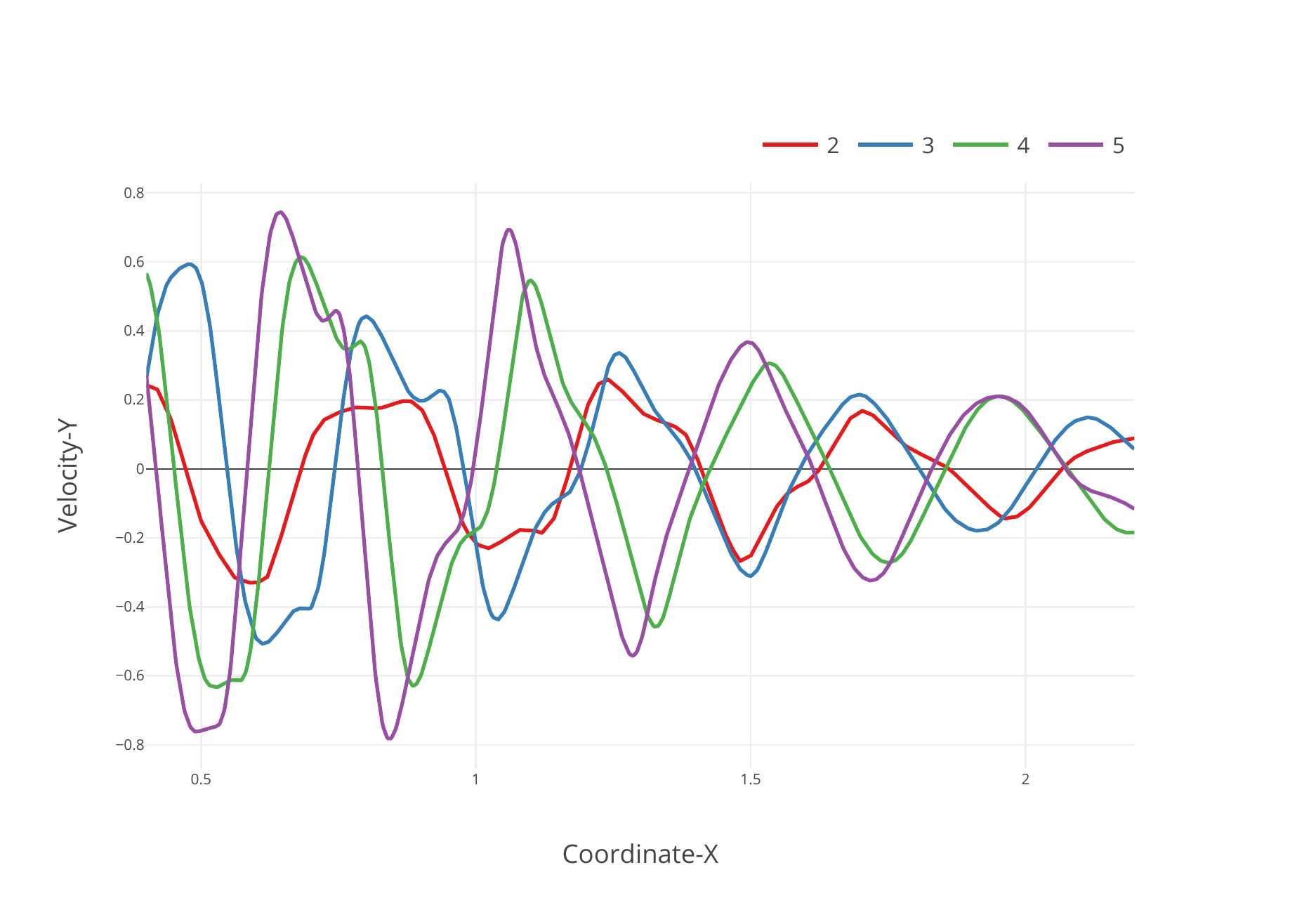}\hfill
    \includegraphics[width=.25\linewidth]{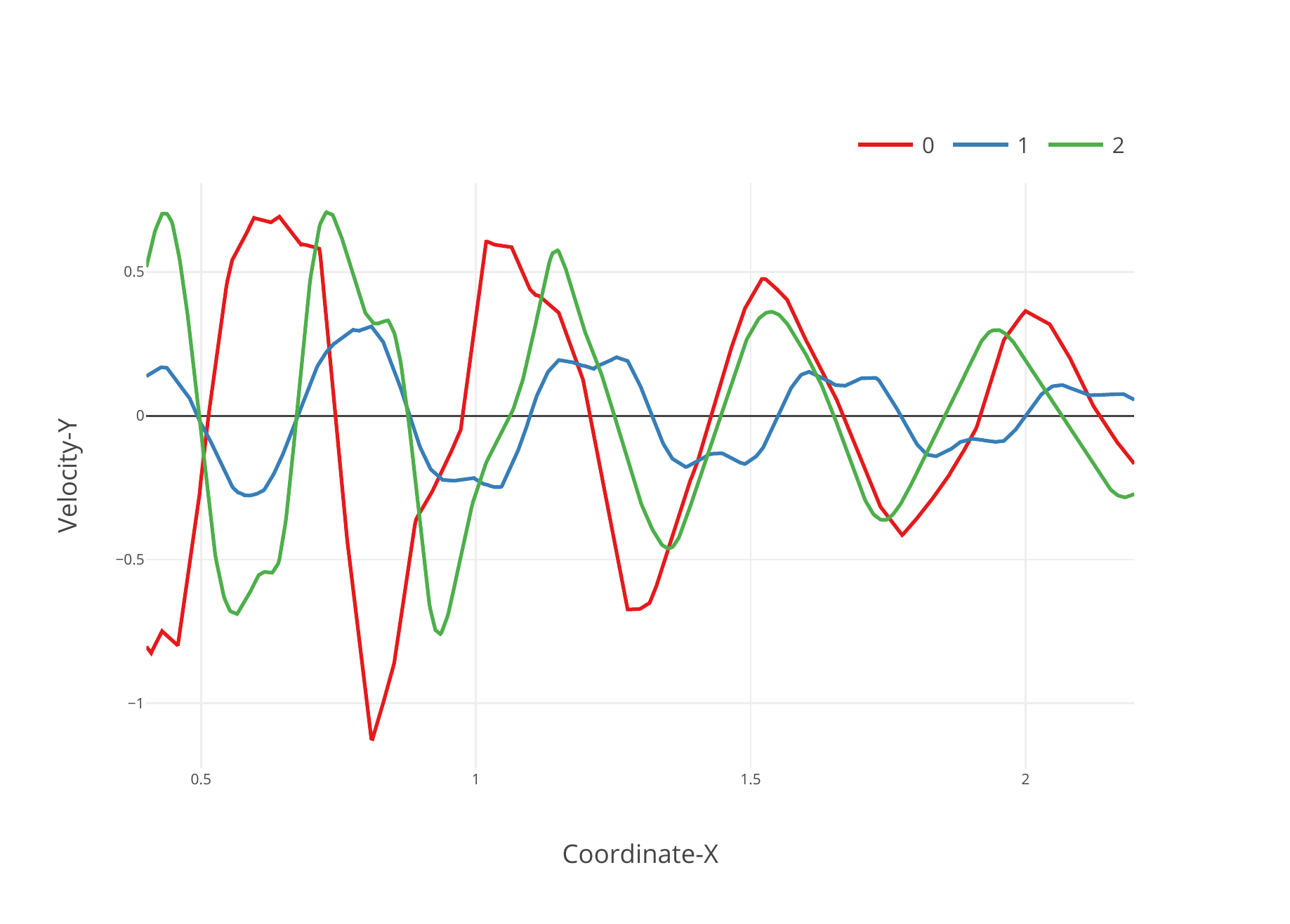}\hfill
    \includegraphics[width=.25\linewidth]{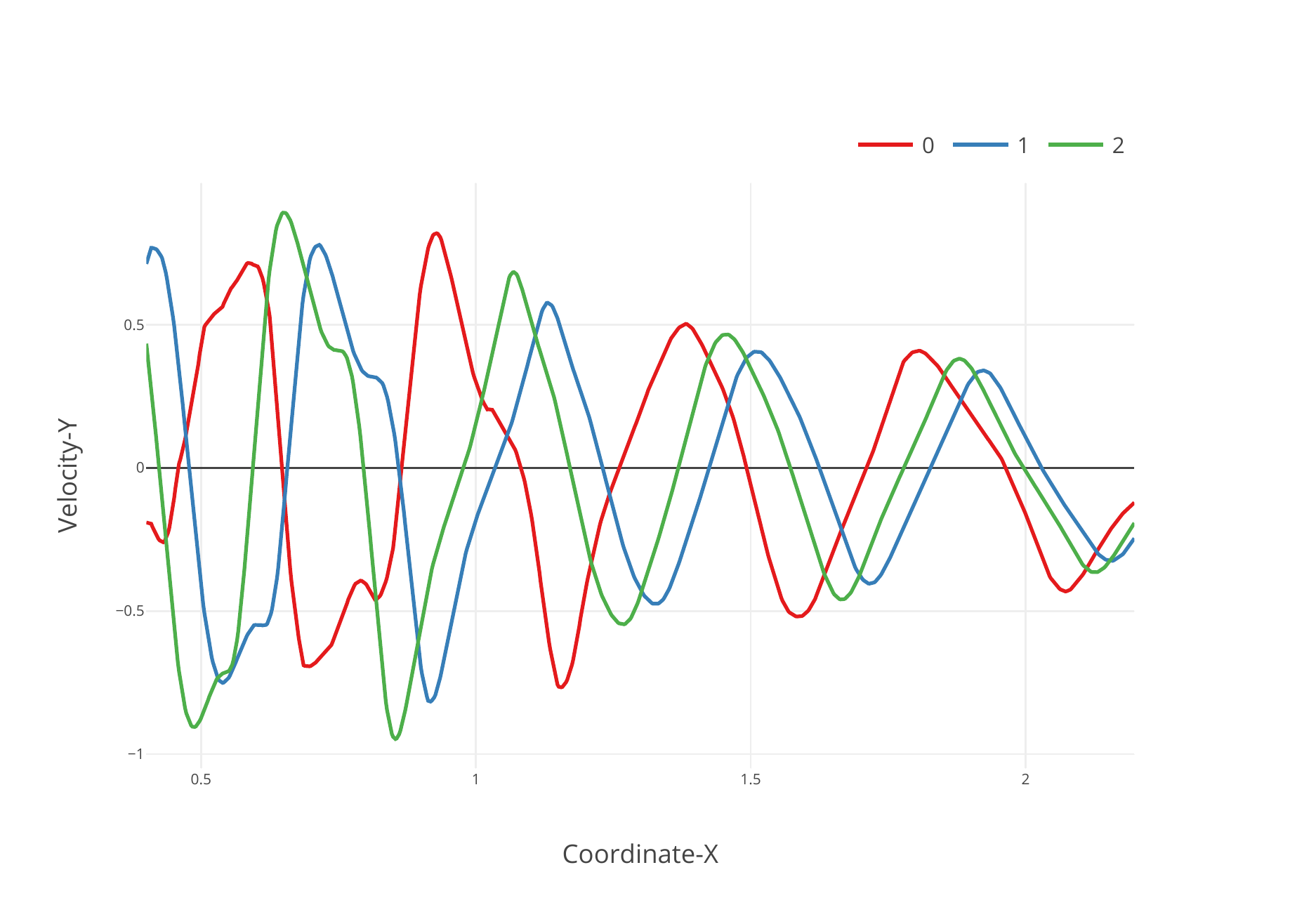}\\
    \parbox{.25\linewidth}{\centering FV-C}\hfill
    \parbox{.25\linewidth}{\centering FD-FLIP}\hfill
    \parbox{.25\linewidth}{\centering FE11-AB2AM2}\hfill
    \parbox{.25\linewidth}{\centering FE21-AB2AM2}
    \caption{$y$-component of the velocity at $t=6s$ for different level of refinement.}
    \label{fig:dissipation}
\end{figure}

The behavior of the different methods varies depending on the mesh resolution (Figure~\ref{fig:dissipation}): only FE21-BDF3, FV-C, and FE21-AB2AM2 exhibit similar dissipation on coarse meshes. In contrast, the other methods have high numerical dissipation at coarse resolution.

For this specific example, we also consider a quad boundary layer mesh: the resolution of the quad mesh is higher around the obstacle and gets coarser on the left-hand side of the domain. As the boundary-adapted structure of the quad mesh was hand-crafted, we can only increase its resolution by uniform refinement, and thus  it is impossible to tune the resolution to match running time across different methods.
Additionally, FD-FLIP cannot take advantage of the adaptivity as it is designed to work on regular grids. Figure~\ref{fig:vortex-quad} shows that different methods produce similar results (i.e., the errors are within an order of magnitude) on the adaptive quad mesh; however, FE11-FLIP, FV-C, and FD-FLIP are significantly faster.

\input{pics/vortex/plot-quad}

\subsection{Drag Force}\label{sec:drag}

\begin{figure}\centering\scriptsize
    \includegraphics[width=.32\linewidth]{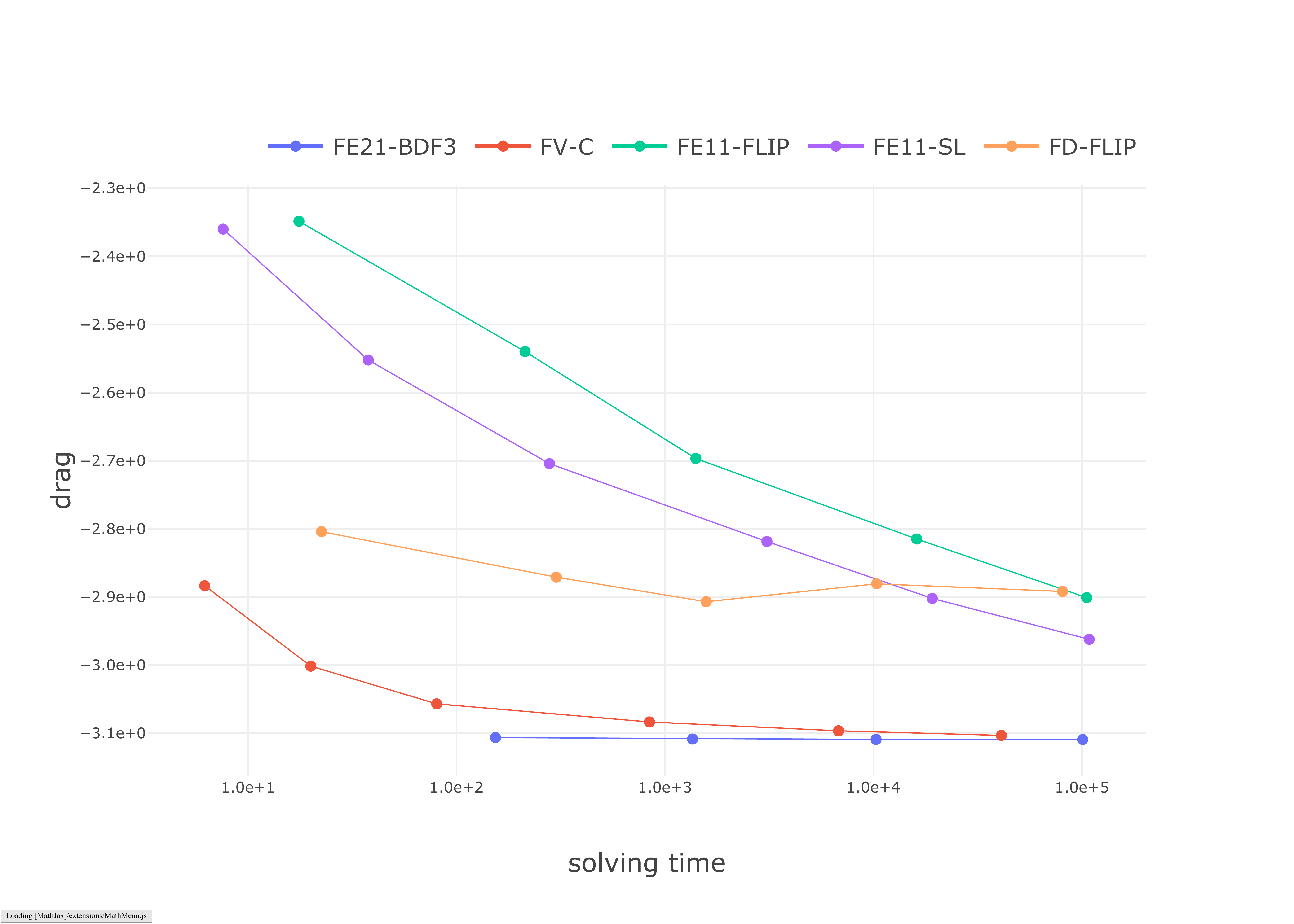}\hfill
    \includegraphics[width=.32\linewidth]{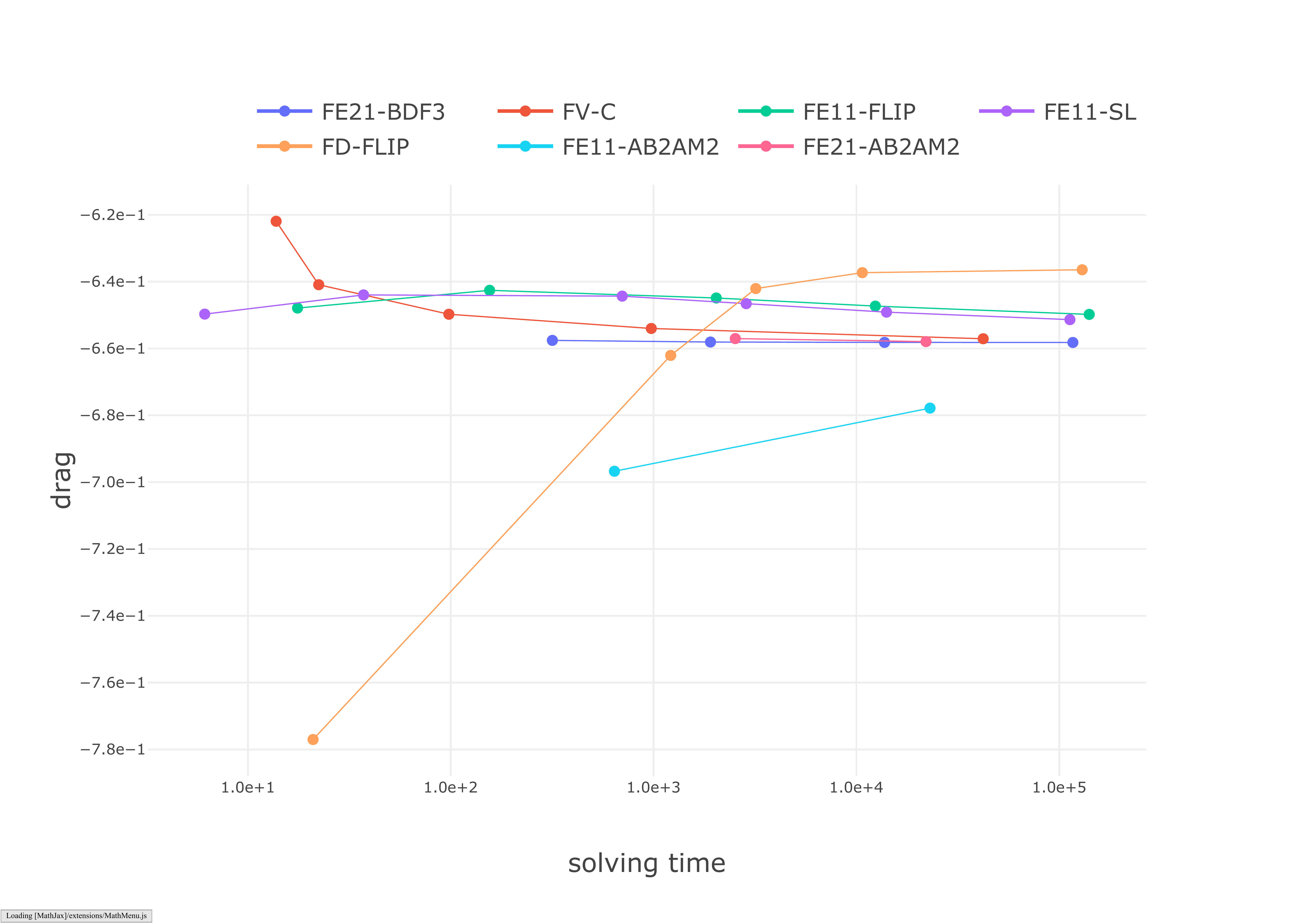}\hfill
    \includegraphics[width=.32\linewidth]{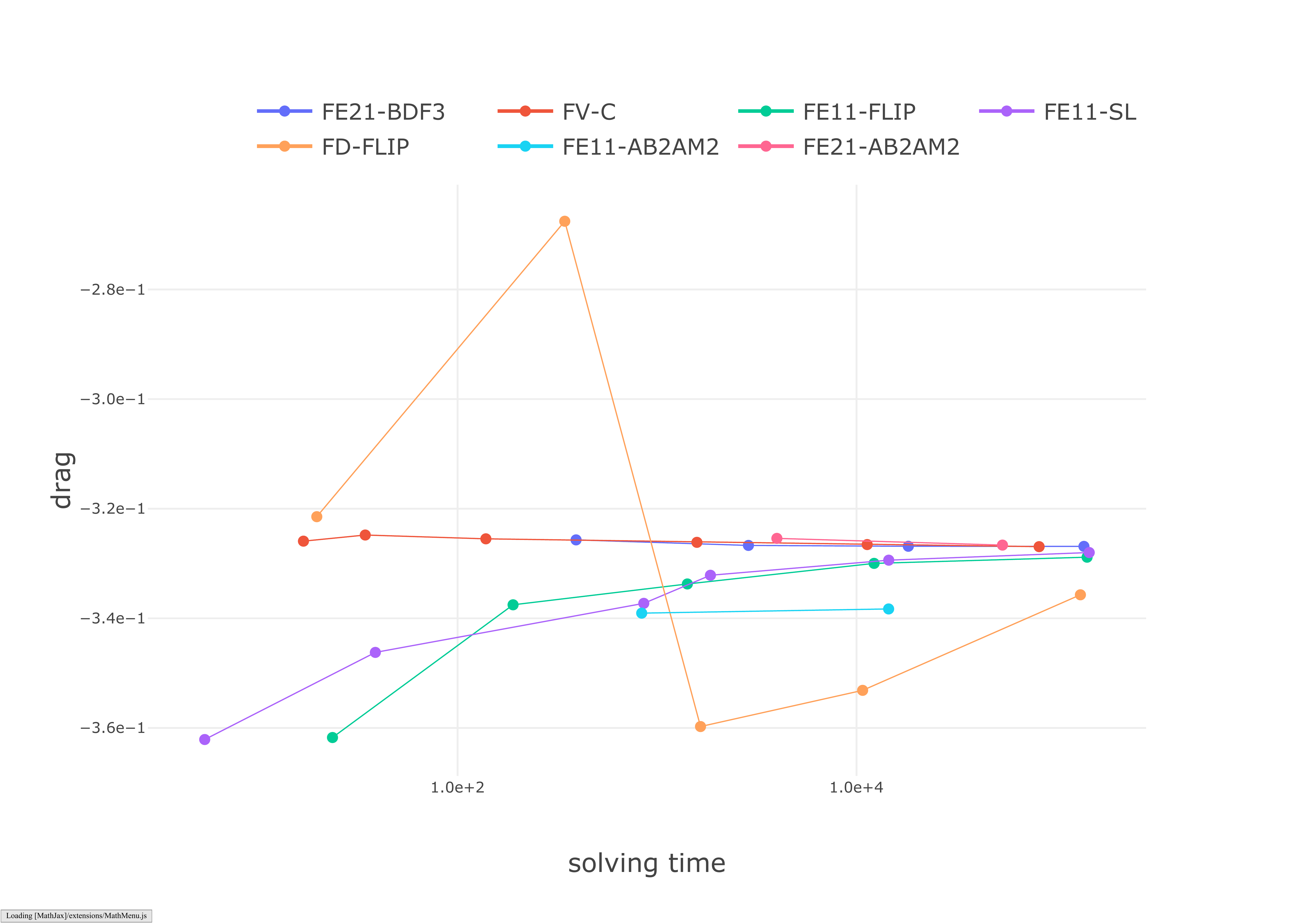}\par
    % \parbox{.32\linewidth}{\centering$\nu=0.1$}\hfill
    % \parbox{.32\linewidth}{\centering$\nu=0.01$}\hfill
    % \parbox{.32\linewidth}{\centering$\nu=0.001$}\par
%     \caption{Drag force introduced by the different method over solve time.}
%     \label{fig:drag-value}
% \end{figure}

% \begin{figure}\centering\scriptsize
    \includegraphics[width=.32\linewidth]{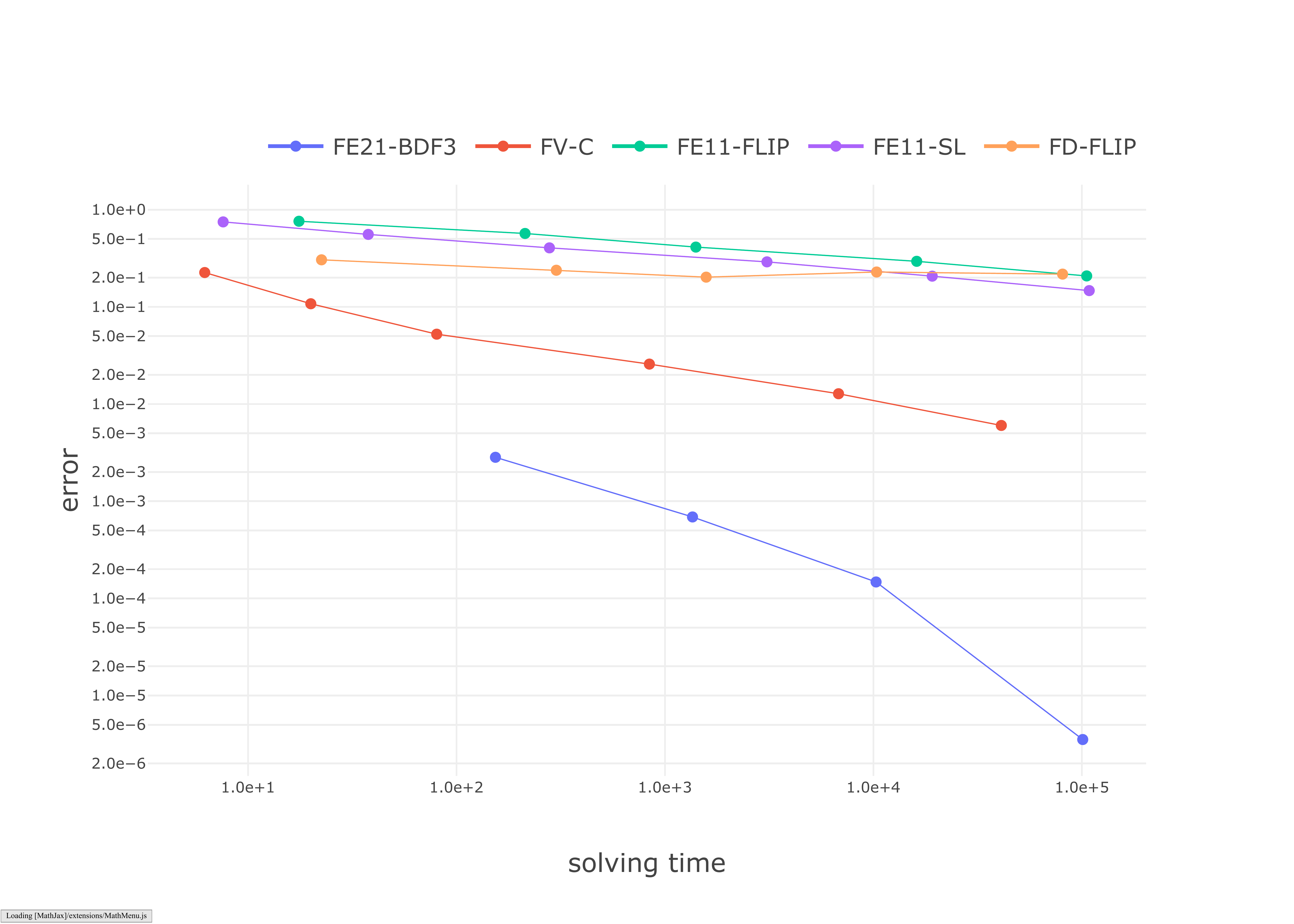}\hfill
    \includegraphics[width=.32\linewidth]{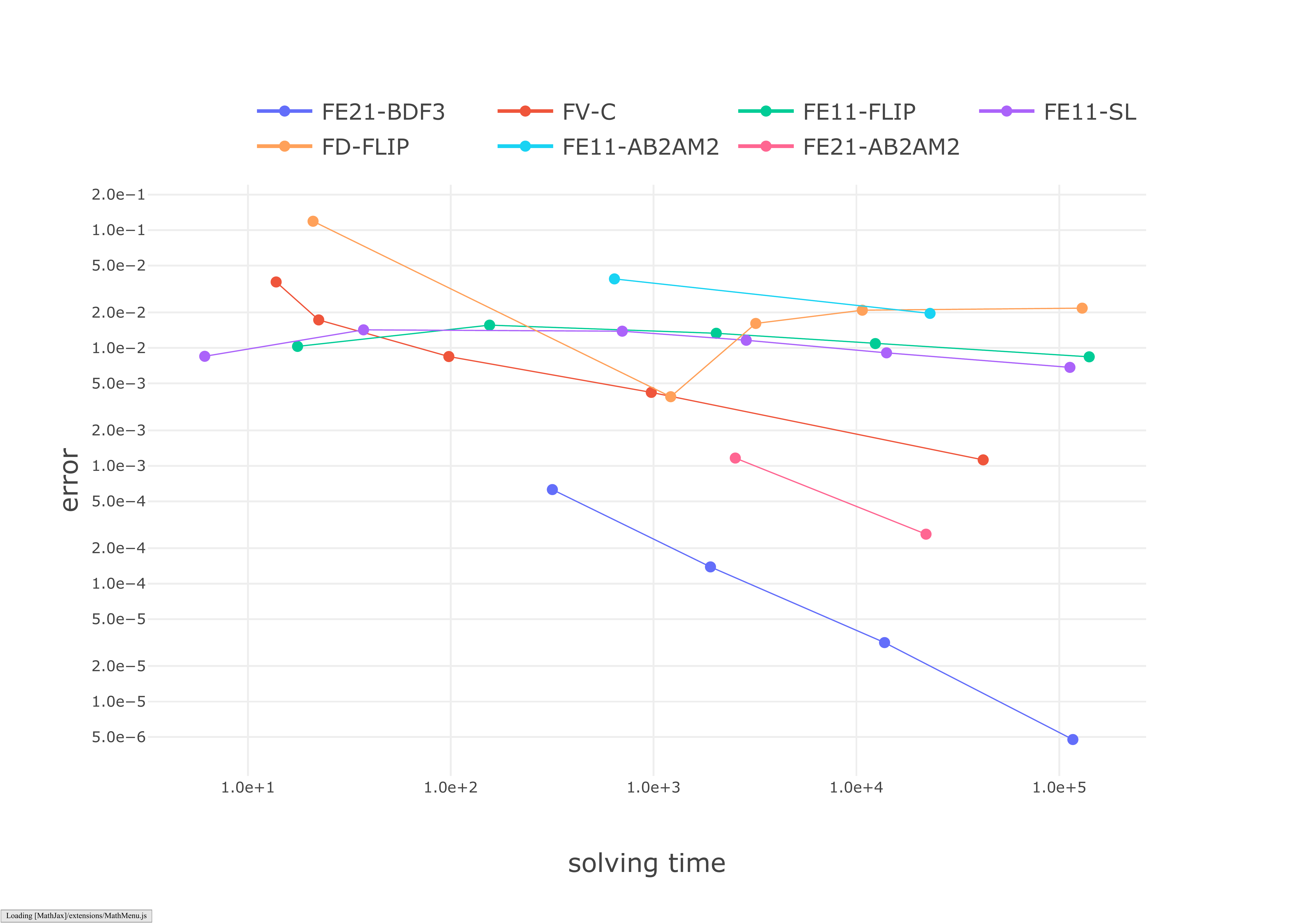}\hfill
    \includegraphics[width=.32\linewidth]{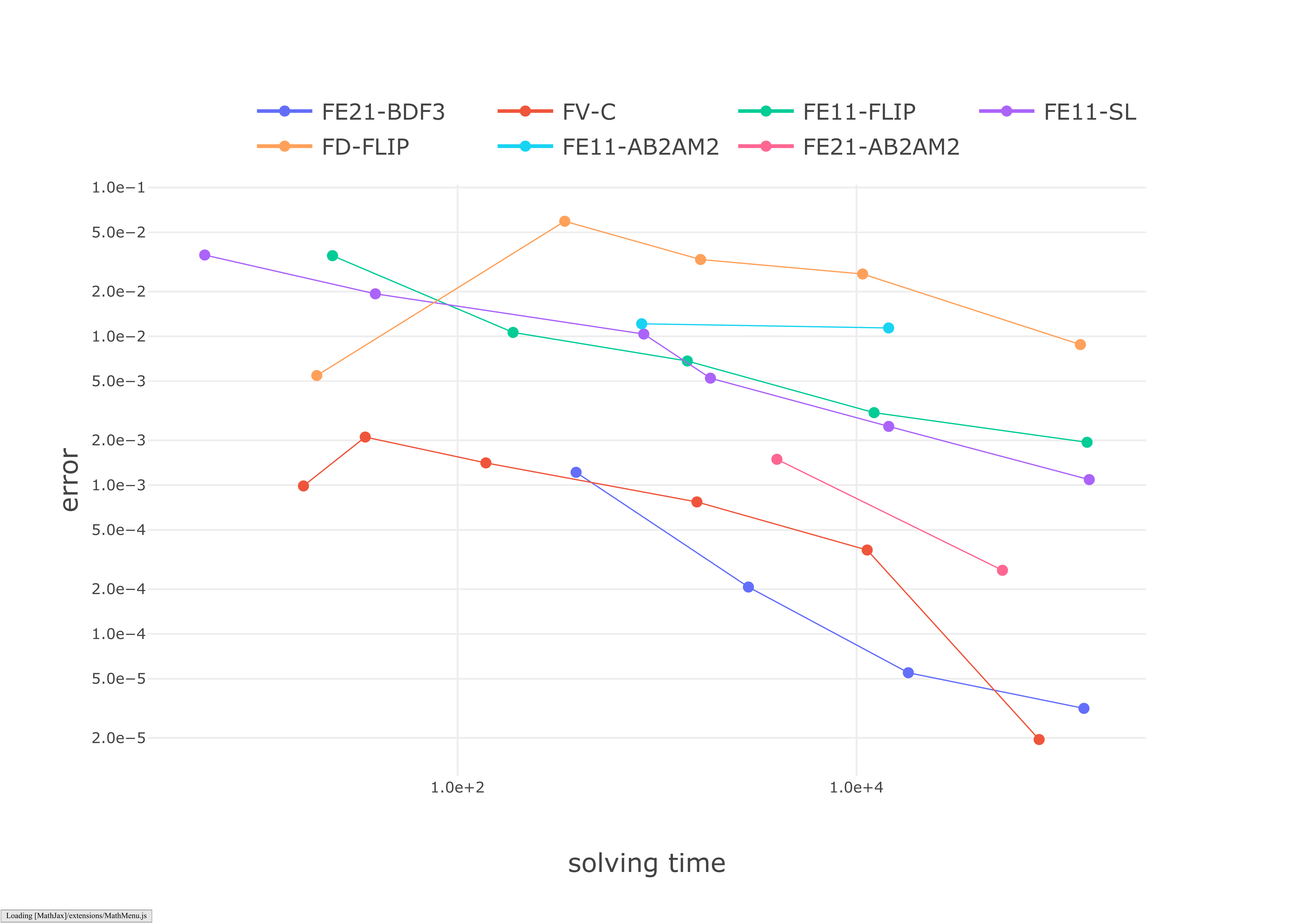}\par
    \parbox{.32\linewidth}{\centering$\nu=0.1$}\hfill
    \parbox{.32\linewidth}{\centering$\nu=0.01$}\hfill
    \parbox{.32\linewidth}{\centering$\nu=0.001$}
    \caption{Drag force (top) and error (bottom) introduced by the different method over solve time.}
    \label{fig:drag}
\end{figure}

For quantitative comparisons, we embed a circle of radius $r=0.05$ into a large square background mesh (the size of the square is $1$) to avoid boundary effects near the obstacle.  The flow moves with velocity $v=1$ from left to right (imposed as a  Dirichlet condition on the boundary of the square) and measure drag force on the circle (zero Dirichlet boundary condition on the circle boundary). 
To compute the drag force more accurately, we generate an adaptive mesh with high resolution close  to the circle. 
%We use FE32-BDF4 solution on a fine mesh with average edge length $0.0025816$ as a reference.
Figure~\ref{fig:drag} shows a comparison of the drag force for three viscosity values. We note that FD-FLIP is designed to work with regular non-adaptive grids, which explains the high error at coarse resolutions.

For high viscosity, only FE21-BDF3 has low (below 1\%) error, while for $\nu=0.001$, both FE21-BDF3 and FV-C have similar errors. For the same running time, FE21-BDF3 has lower error than other methods. We note that we could not compute certain solutions for  FE21-AB2AM2 as it did not terminate after our cutoff wall-clock time of 48 hours.

%% file: pics/cavity/plot.tex
    \begin{figure}
        \centering\scriptsize
        Regular grid\\
        \includegraphics[width=.25\linewidth]{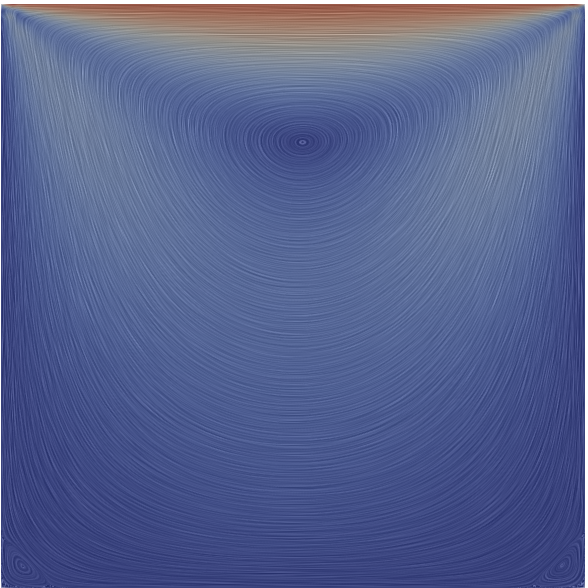}\hfill
        \includegraphics[width=.25\linewidth]{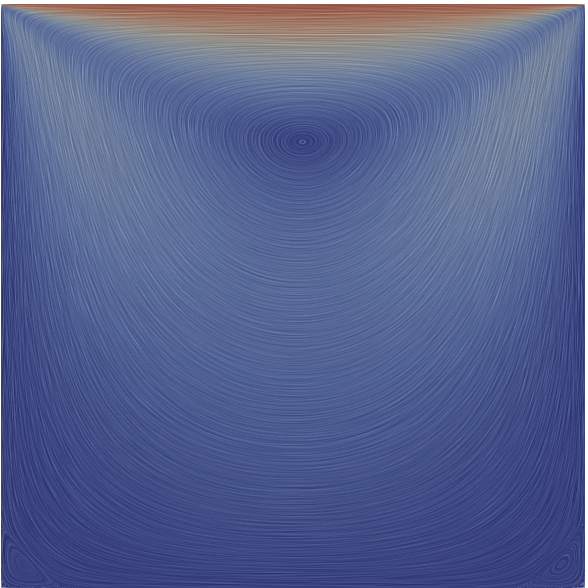}\hfill
        \includegraphics[width=.25\linewidth]{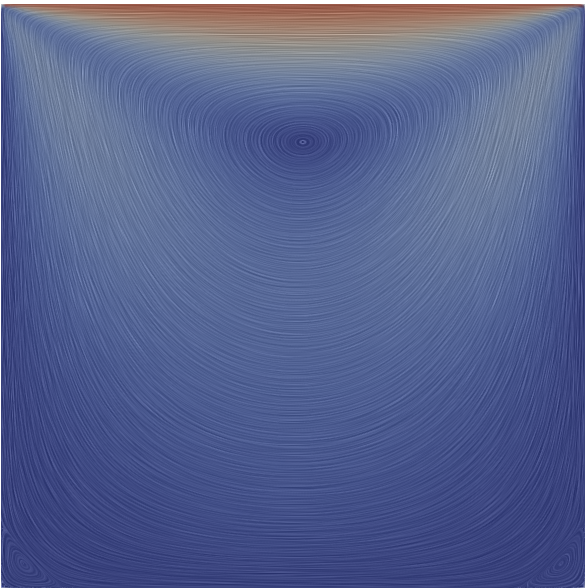}\\
        \parbox{.25\linewidth}{\centering FE21-BDF3 12m, $e=4.22 \times 10^{-3}$}\hfill
        \parbox{.25\linewidth}{\centering FE11-SL 12m, $e=2.83 \times 10^{-3}$}\hfill
        \parbox{.25\linewidth}{\centering FE11-FLIP 12m, $e=4.71 \times 10^{-3}$}\\
        \includegraphics[width=.25\linewidth]{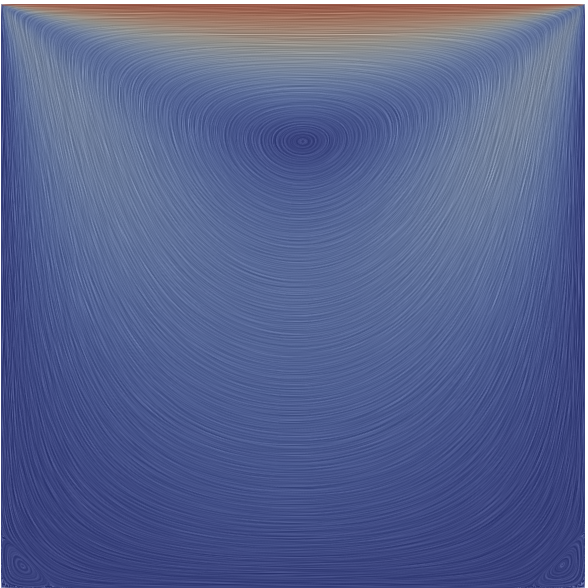}\hfill
        \includegraphics[width=.25\linewidth]{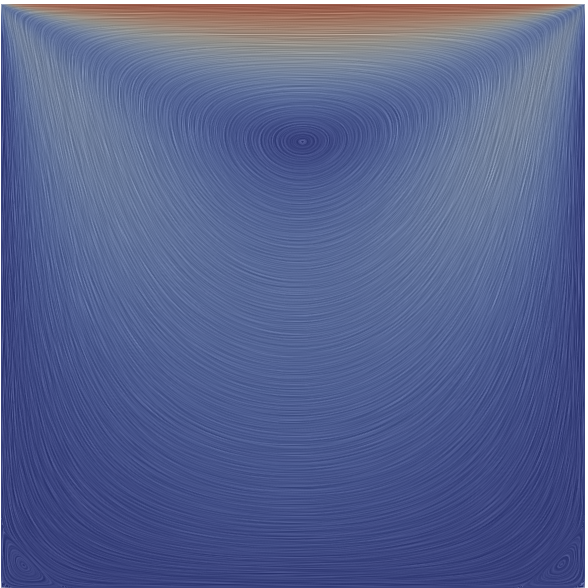}\hfill
        \includegraphics[width=.25\linewidth]{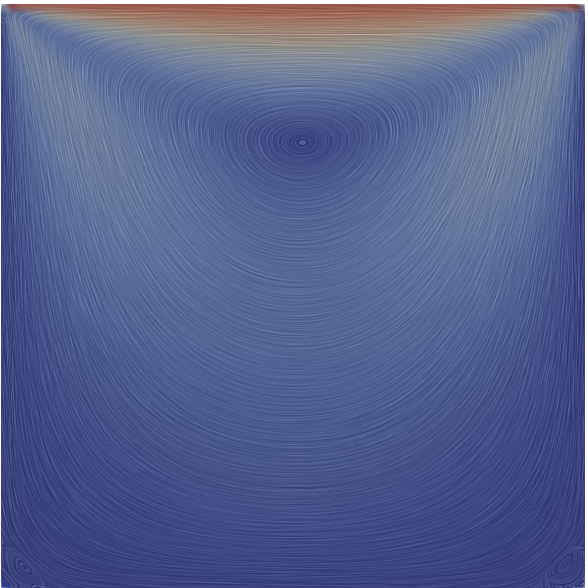}\hfill
        \includegraphics[width=.25\linewidth]{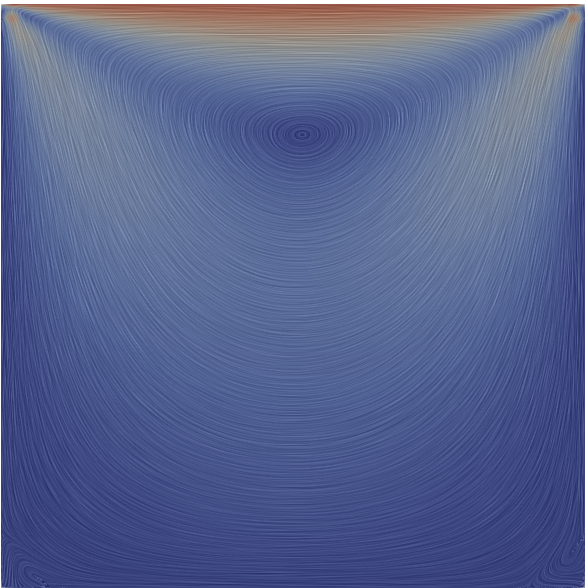}\\
        \parbox{.25\linewidth}{\centering FV-C 13m, $e=6.97 \times 10^{-4}$}\hfill
        \parbox{.25\linewidth}{\centering FD-FLIP 12m, $e=1.60 \times 10^{-3}$}\hfill
        \parbox{.25\linewidth}{\centering FE11-AB2AM2 12m, $e=6.18 \times 10^{-3}$}\hfill
        \parbox{.25\linewidth}{\centering FE21-AB2AM2 7m, $e=3.13 \times 10^{-2}$}\\[2em]

        Triangle mesh\\
        \parbox{.25\linewidth}{\includegraphics[width=\linewidth]{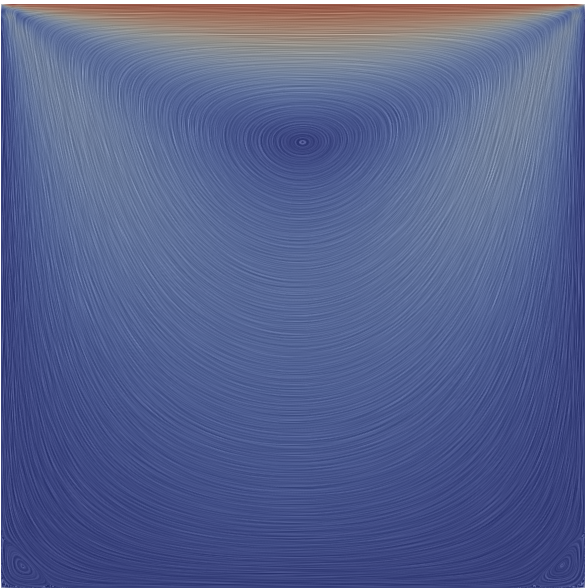}}\hfill
        \parbox{.25\linewidth}{\includegraphics[width=\linewidth]{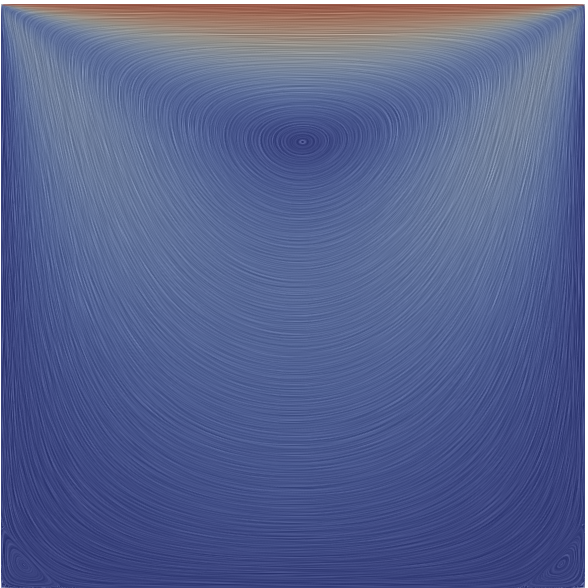}}\hfill
        \parbox{.25\linewidth}{\includegraphics[width=\linewidth]{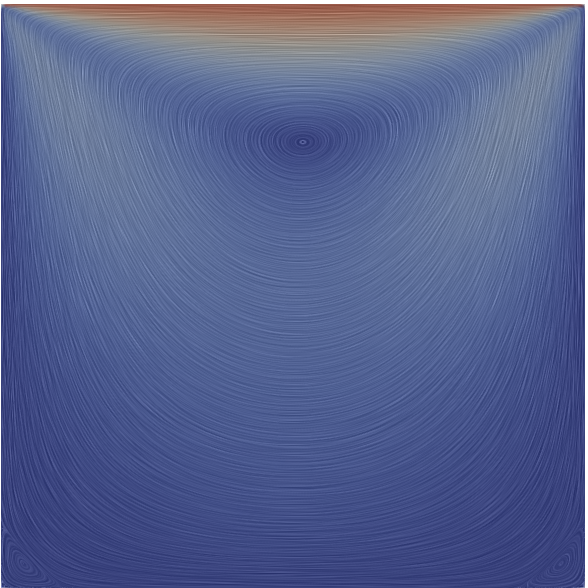}}\\
        \parbox{.25\linewidth}{\centering FE21-BDF3 15m, $e=3.65 \times 10^{-3}$}\hfill
        \parbox{.25\linewidth}{\centering FE11-SL 12m, $e=2.87 \times 10^{-3}$}\hfill
        \parbox{.25\linewidth}{\centering FE11-FLIP 10m, $e=5.50 \times 10^{-3}$}\\
        \parbox{.25\linewidth}{\includegraphics[width=\linewidth]{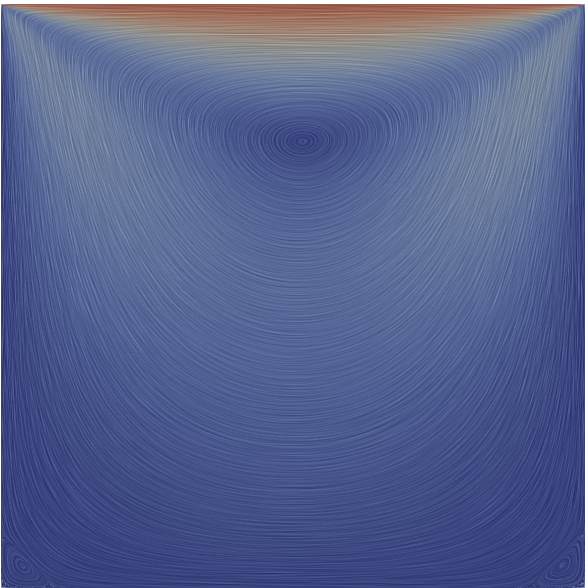}}\hfill
        \parbox{.25\linewidth}{\includegraphics[width=\linewidth]{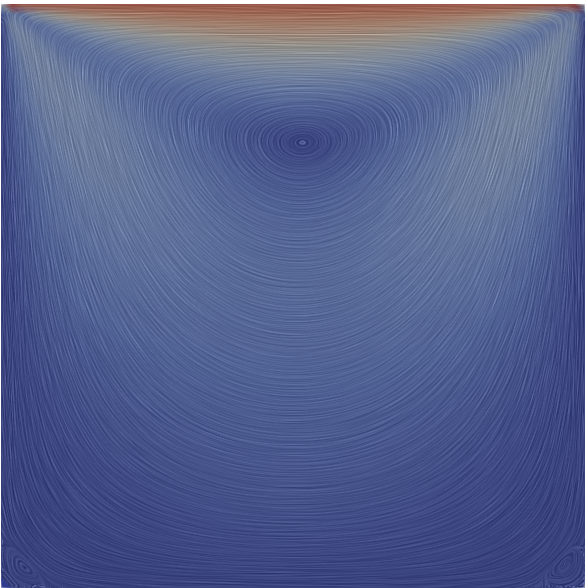}}\hfill
        \parbox{.25\linewidth}{\includegraphics[width=\linewidth]{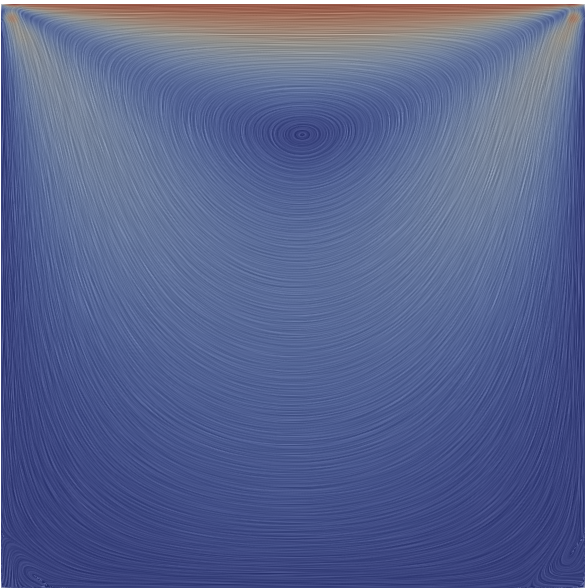}}\\
        \parbox{.25\linewidth}{\centering FV-C 13m, $e=8.37 \times 10^{-4}$}\hfill
        \parbox{.25\linewidth}{\centering FE11-AB2AM2 10m, $e=9.68 \times 10^{-3}$}\hfill
        \parbox{.25\linewidth}{\centering FE21-AB2AM2 10m, $e=4.97 \times 10^{-3}$}\\
        \caption{Result of the driven cavity at time $t=T=2$.}
        \label{fig:cavity}
    \end{figure}

%% file: pics/airfoil/plot.tex
    \begin{figure}
        \centering\scriptsize
        \includegraphics[width=.24\linewidth]{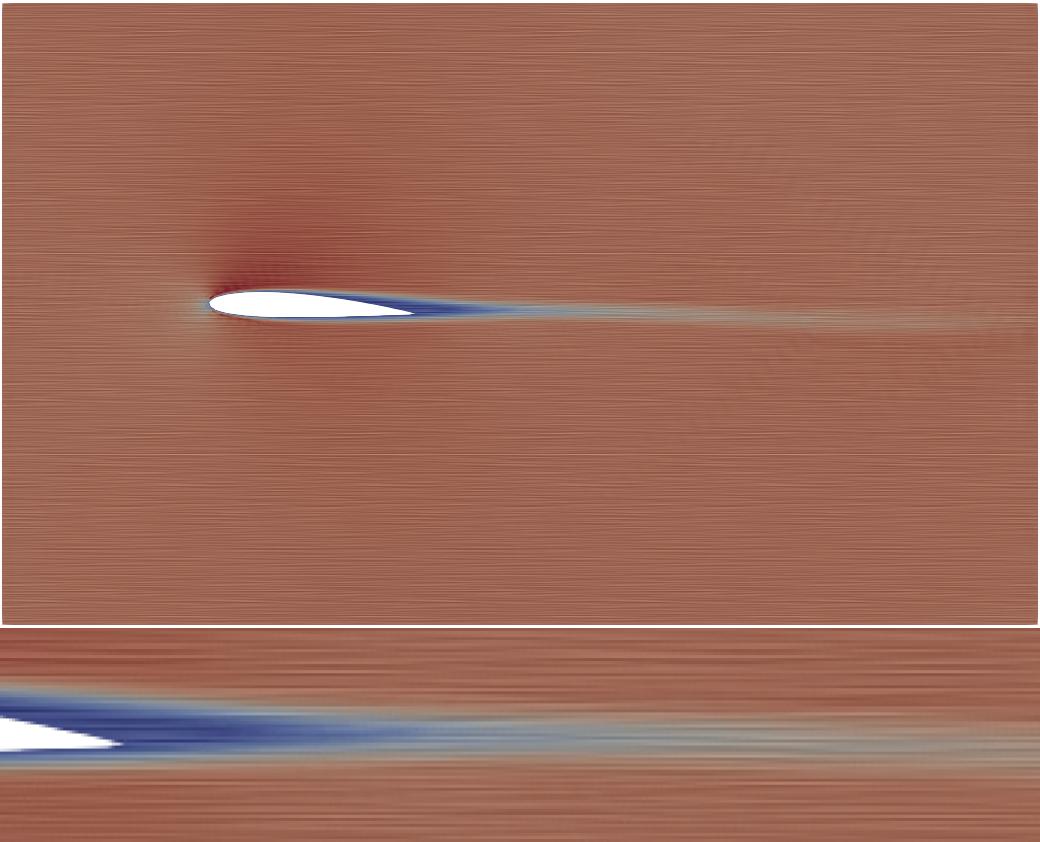}\hfill
        \includegraphics[width=.24\linewidth]{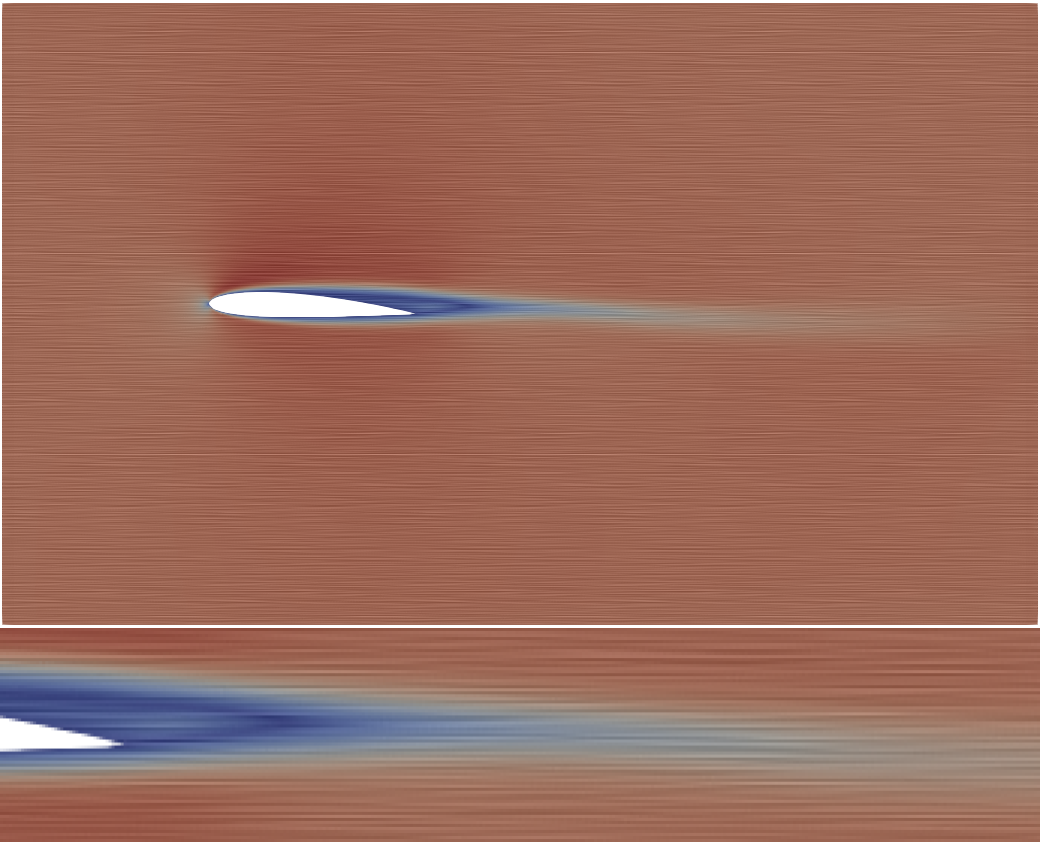}\hfill
        \includegraphics[width=.24\linewidth]{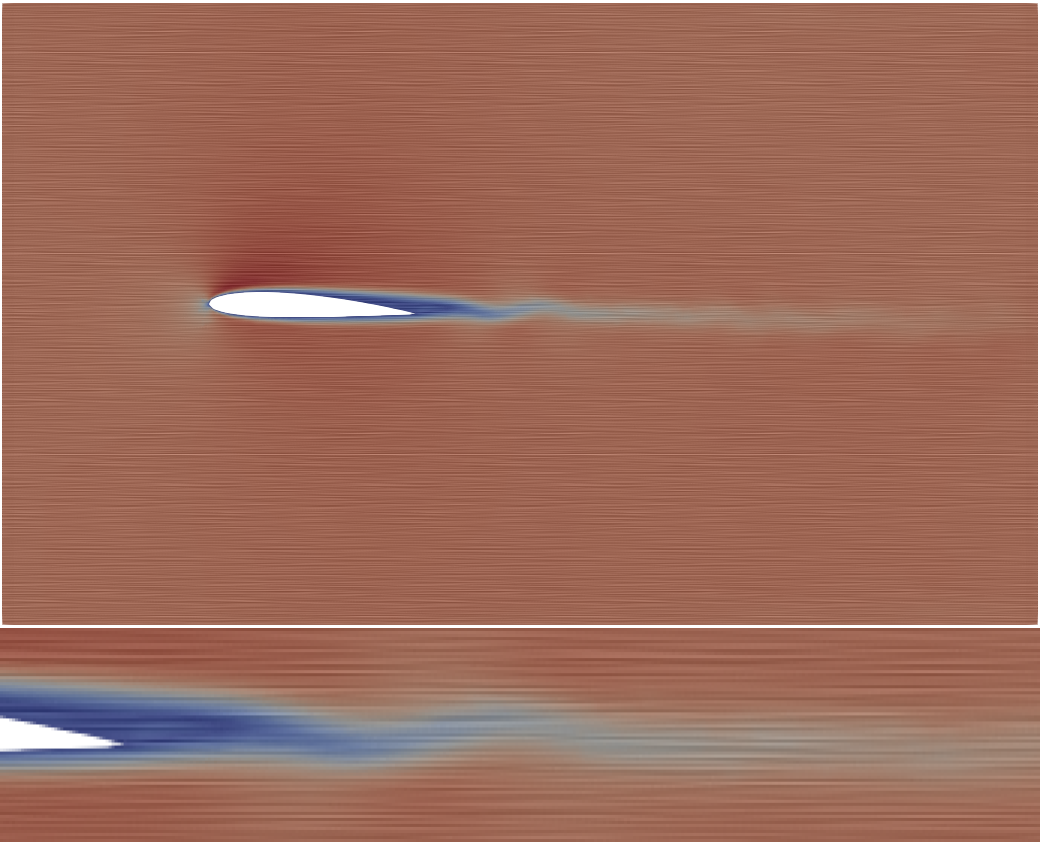}\\
        \parbox{.24\linewidth}{\centering FE21-BDF3 31m, $e=3.13 \times 10^{-2}$}\hfill
        \parbox{.24\linewidth}{\centering FE11-SL 26m, $e=1.04 \times 10^{-1}$}\hfill
        \parbox{.24\linewidth}{\centering FE11-FLIP 25m, $e=3.73 \times 10^{-2}$}\\
        \includegraphics[width=.24\linewidth]{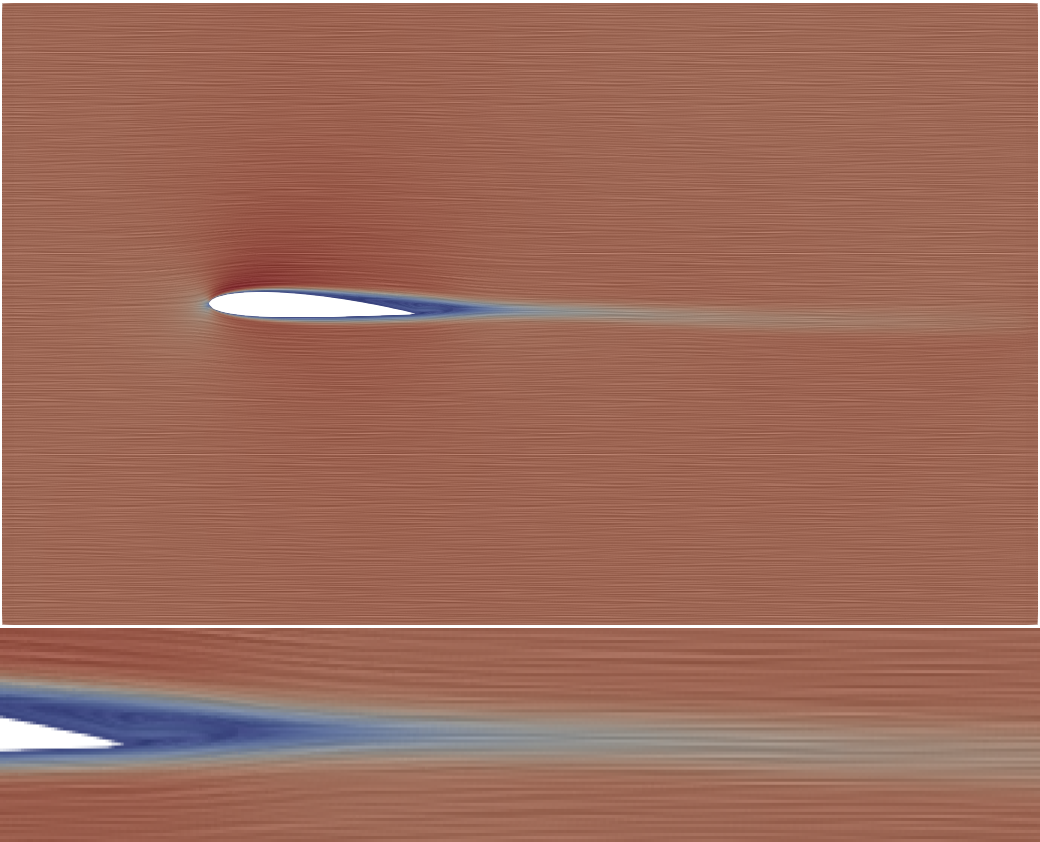}\hfill
        \includegraphics[width=.24\linewidth]{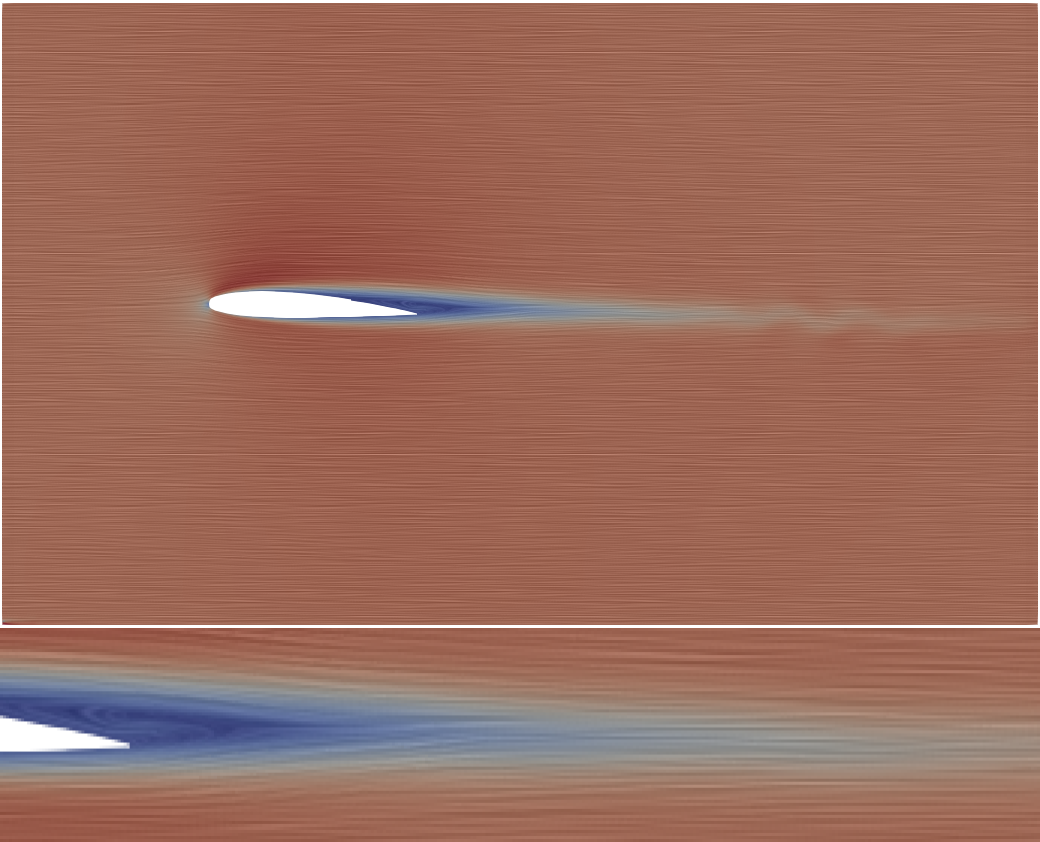}\hfill
        \includegraphics[width=.24\linewidth]{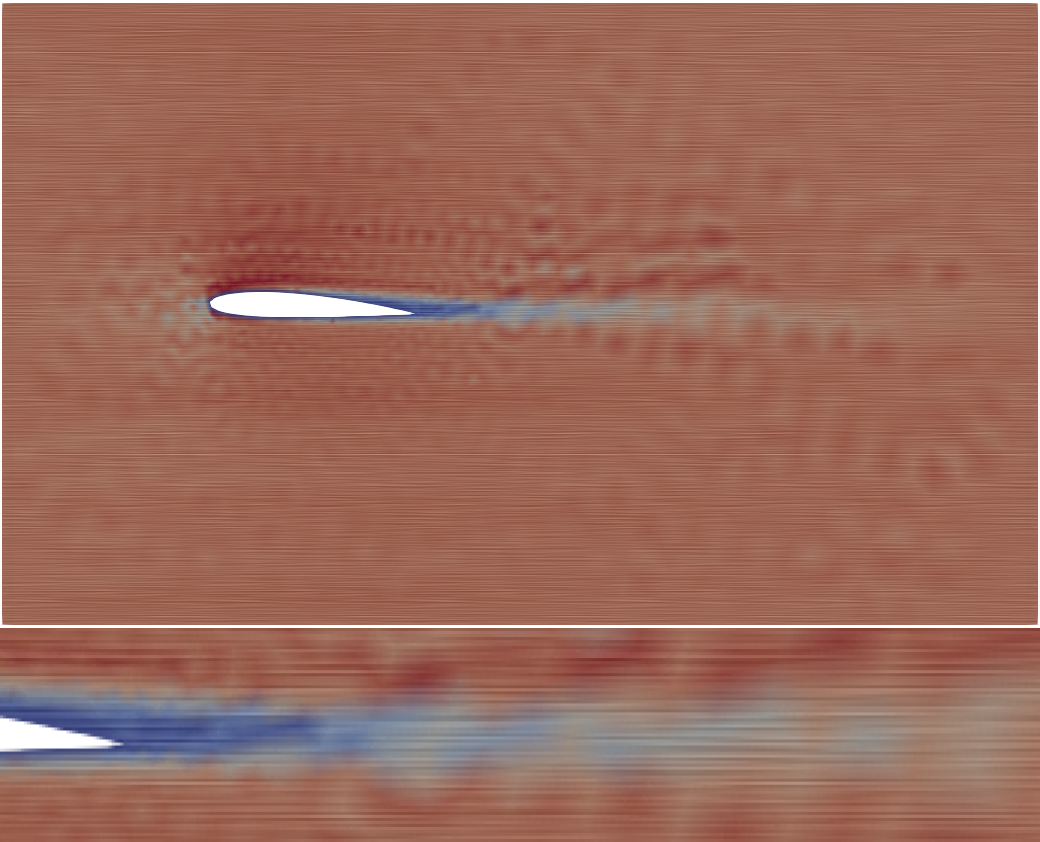}\hfill
        \includegraphics[width=.24\linewidth]{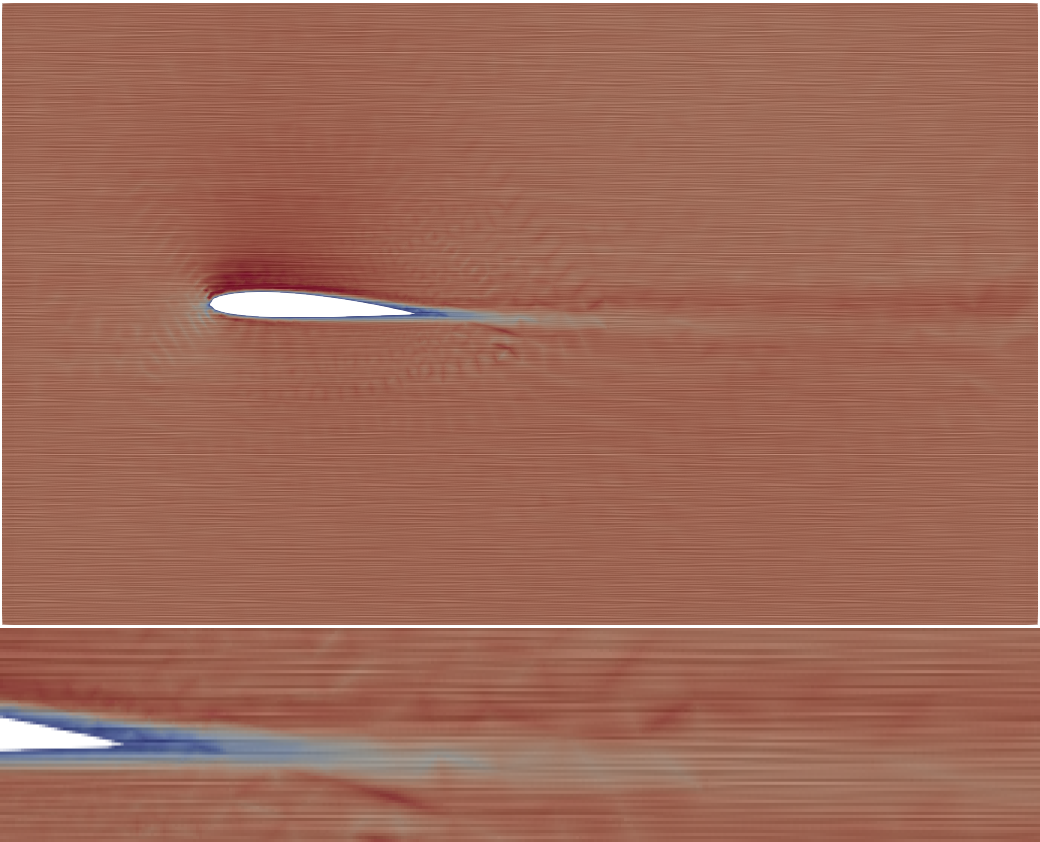}\\
        \parbox{.24\linewidth}{\centering FV-C 25m, $e=9.14 \times 10^{-3}$}\hfill
        \parbox{.24\linewidth}{\centering FD-FLIP 29m, $e=7.66 \times 10^{-2}$}\hfill
        \parbox{.24\linewidth}{\centering FE11-AB2AM2 27m, $e=5.91 \times 10^{-2}$}\hfill
        \parbox{.24\linewidth}{\centering FE21-AB2AM2 15m, $e=1.55 \times 10^{-1}$}
        \caption{Simulation results for $T=4$ for the airfoil mesh. Bottom shows a closeup of the back of the wing. The error is computed only in a subset of the domain containing the airfoil ($[-0.2,1.2]\times[-0.2,0.25]$).}
        \label{fig:airfoil}
    \end{figure}

%% file: pics/ocavity/plot.tex
    \begin{figure}
    {\centering\scriptsize
            \parbox{0.04\linewidth}{\centering\rotatebox{90}{$\nu=0.0005$}}\hfill
            \parbox{0.95\linewidth}{\centering
            Regular grid\\
            \includegraphics[width=.13\linewidth]{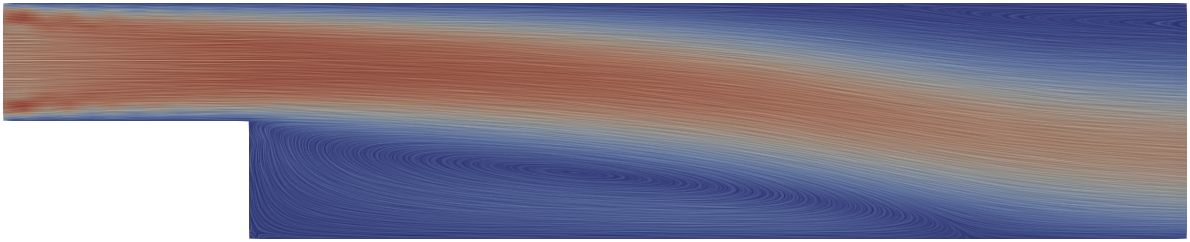}\hfill
            \includegraphics[width=.13\linewidth]{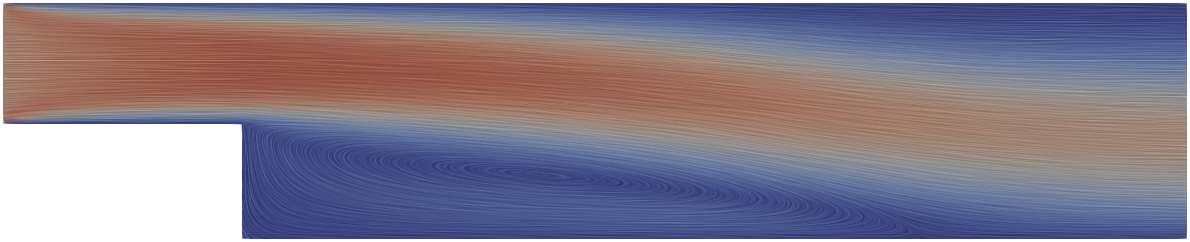}\hfill
            \includegraphics[width=.13\linewidth]{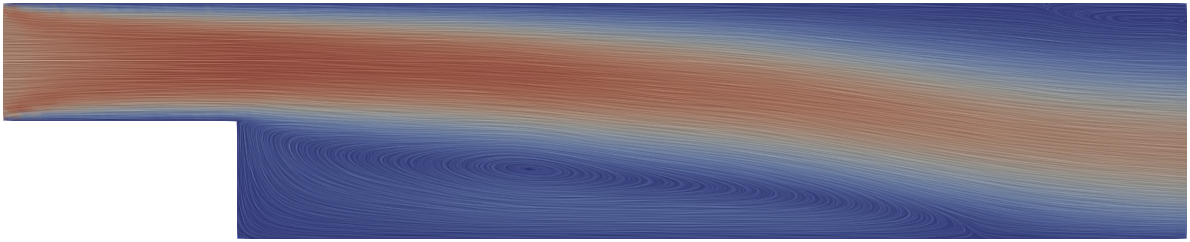}\hfill
            \includegraphics[width=.13\linewidth]{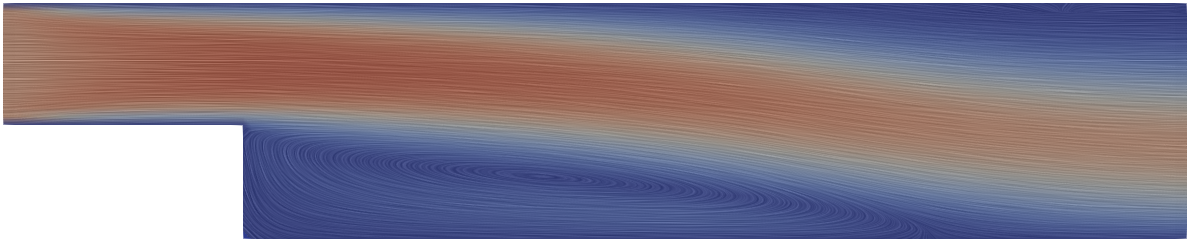}\hfill
            \includegraphics[width=.13\linewidth]{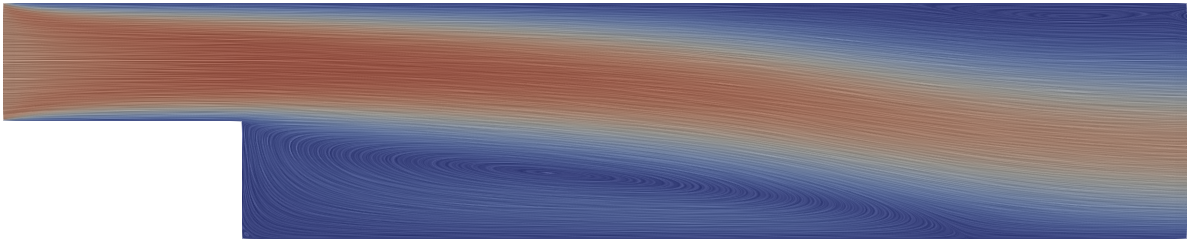}\hfill
            \includegraphics[width=.13\linewidth]{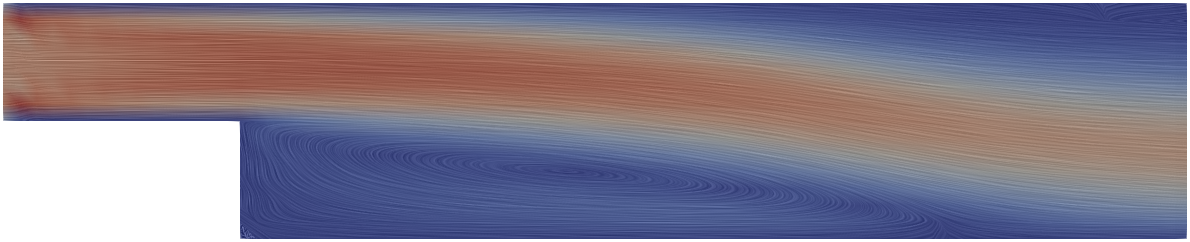}\hfill
            \includegraphics[width=.13\linewidth]{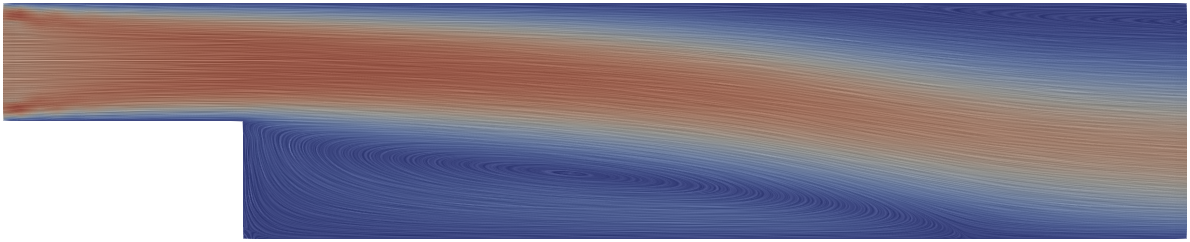}\\
            \parbox{.13\linewidth}{\centering\tiny FE21-BDF3 8m, $e=9.51 \times 10^{-3}$}\hfill
            \parbox{.13\linewidth}{\centering\tiny FE11-SL 8m, $e=1.95 \times 10^{-2}$}\hfill
            \parbox{.13\linewidth}{\centering\tiny FE11-FLIP 10m, $e=1.41 \times 10^{-2}$}\hfill
            \parbox{.13\linewidth}{\centering\tiny FV-C 9m, $e=2.18 \times 10^{-2}$}\hfill
            \parbox{.13\linewidth}{\centering\tiny FD-FLIP 9m, $e=1.21 \times 10^{-2}$}\hfill
            \parbox{.13\linewidth}{\centering\tiny FE11-AB2AM2 8m, $e=1.80 \times 10^{-2}$}\hfill
            \parbox{.13\linewidth}{\centering\tiny FE21-AB2AM2 8m, $e=6.46 \times 10^{-3}$}\\[1em]
            Triangle mesh\\
            \parbox{.13\linewidth}{\includegraphics[width=\linewidth]{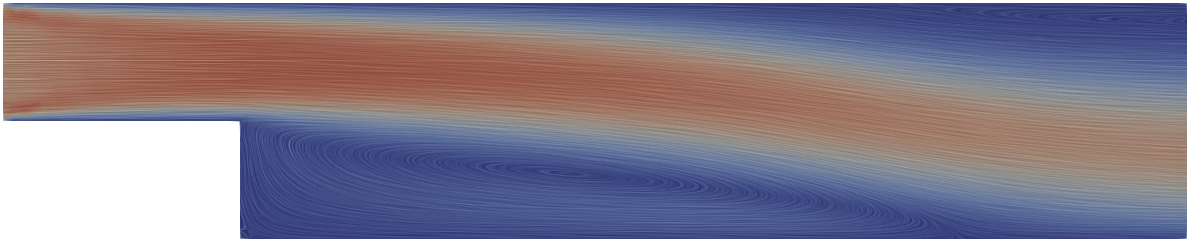}}\hfill
            \parbox{.13\linewidth}{\includegraphics[width=\linewidth]{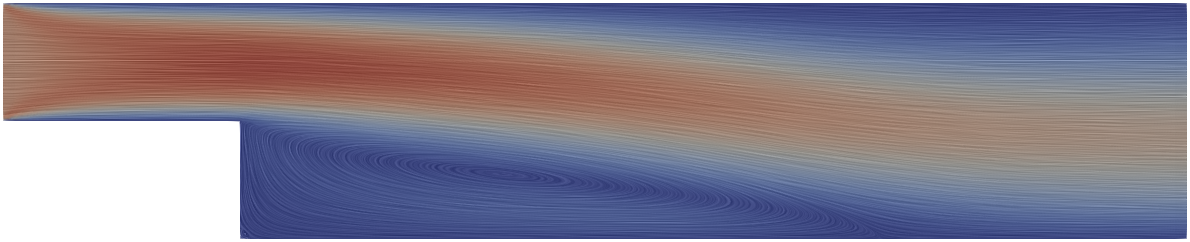}}\hfill
            \parbox{.13\linewidth}{\includegraphics[width=\linewidth]{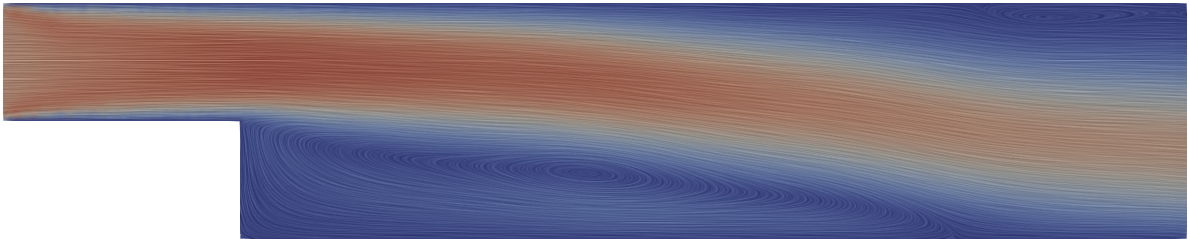}}\hfill
            \parbox{.13\linewidth}{\includegraphics[width=\linewidth]{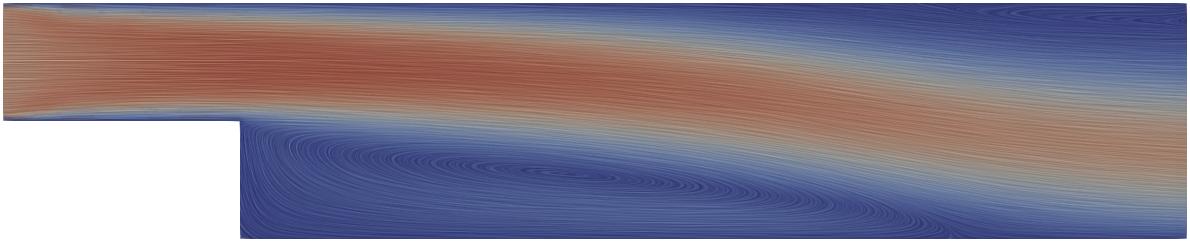}}\hfill
            \parbox{.13\linewidth}{\includegraphics[width=\linewidth]{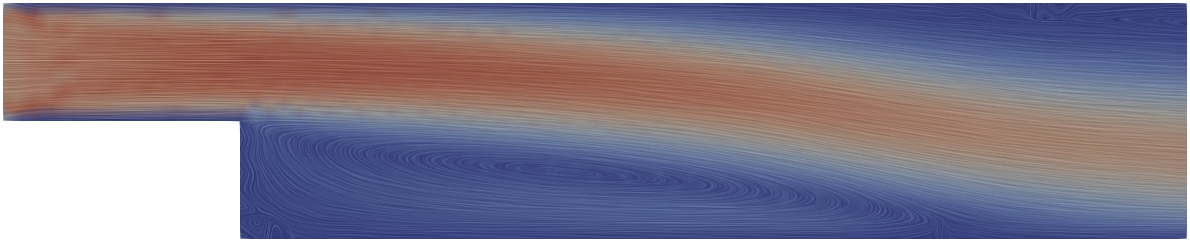}}\hfill
            \parbox{.13\linewidth}{\includegraphics[width=\linewidth]{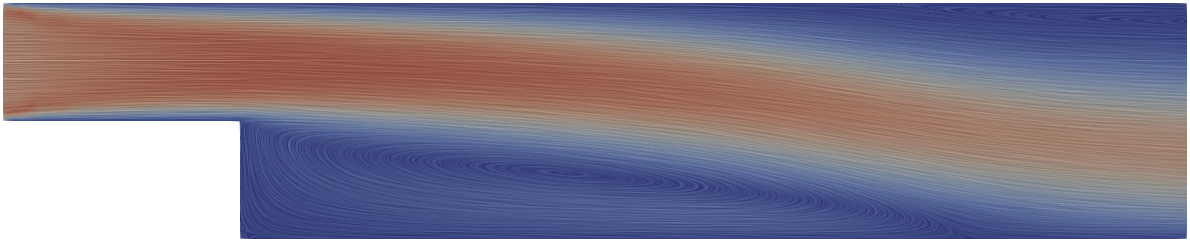}}\hfill
            \parbox{.13\linewidth}{~}\\
            \parbox{.13\linewidth}{\centering\tiny FE21-BDF3 9m, $e=7.49 \times 10^{-3}$}\hfill
            \parbox{.13\linewidth}{\centering\tiny FE11-SL 9m, $e=4.11 \times 10^{-2}$}\hfill
            \parbox{.13\linewidth}{\centering\tiny FE11-FLIP 8m, $e=1.09 \times 10^{-2}$}\hfill
            \parbox{.13\linewidth}{\centering\tiny FV-C 8m, $e=1.37 \times 10^{-2}$}\hfill
            \parbox{.13\linewidth}{\centering\tiny FE11-AB2AM2 10m, $e=1.72 \times 10^{-2}$}\hfill
            \parbox{.13\linewidth}{\centering\tiny FE21-AB2AM2 8m, $e=5.31 \times 10^{-3}$}\hfill
            \parbox{.13\linewidth}{~}
        }\\[2em]
            %%%%%%%%%%%%%%%%
            \parbox{0.04\linewidth}{\centering\rotatebox{90}{$\nu=0.001$}}\hfill
            \parbox{0.95\linewidth}{\centering
            Regular grid\\
            \includegraphics[width=.13\linewidth]{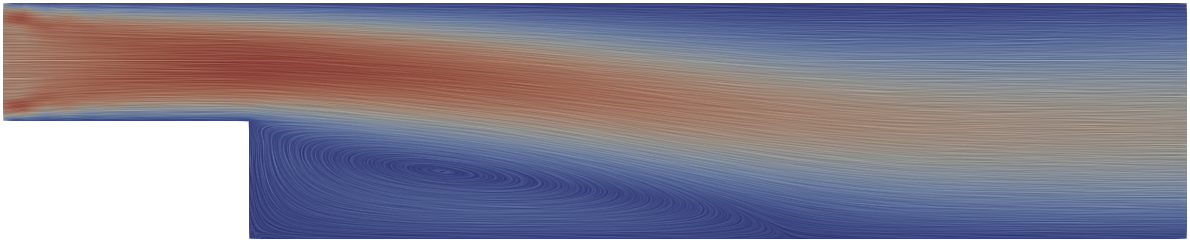}\hfill
            \includegraphics[width=.13\linewidth]{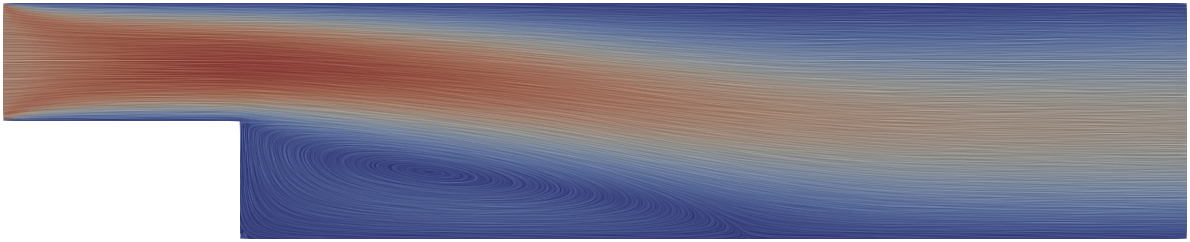}\hfill
            \includegraphics[width=.13\linewidth]{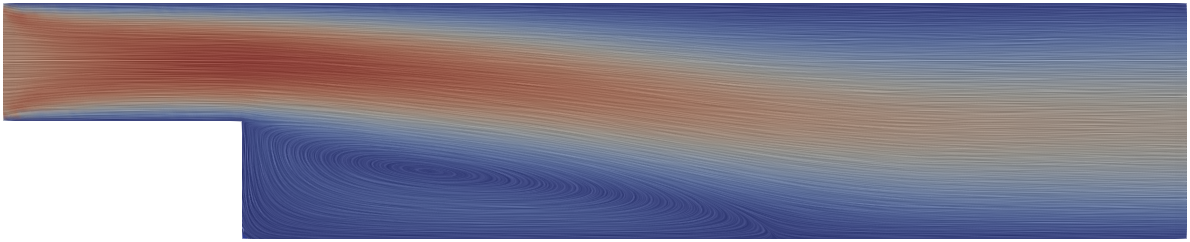}\hfill
            \includegraphics[width=.13\linewidth]{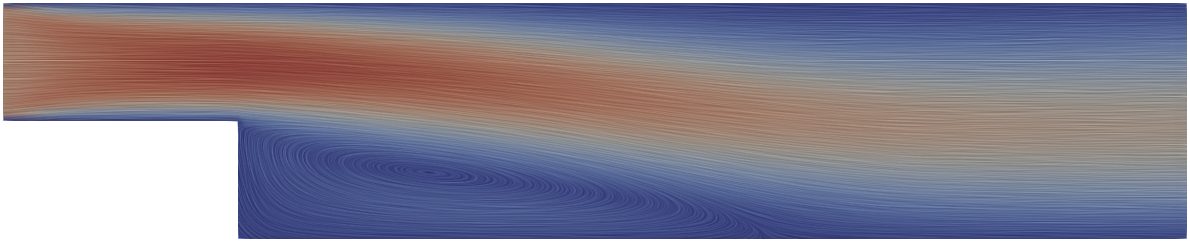}\hfill
            \includegraphics[width=.13\linewidth]{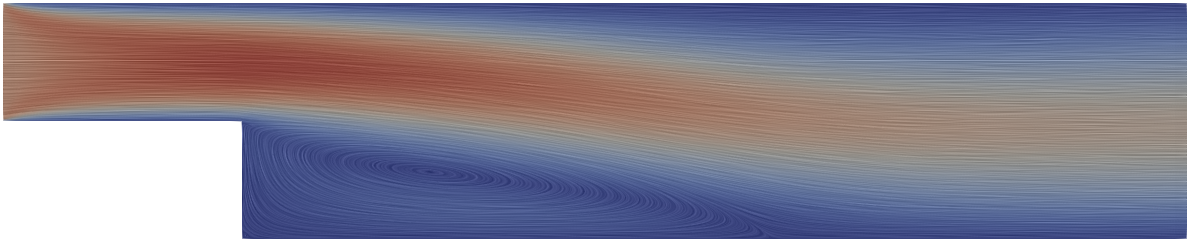}\hfill
            \includegraphics[width=.13\linewidth]{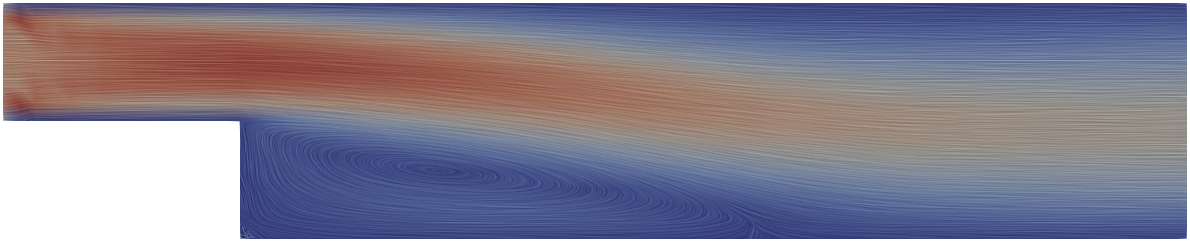}\hfill
            \includegraphics[width=.13\linewidth]{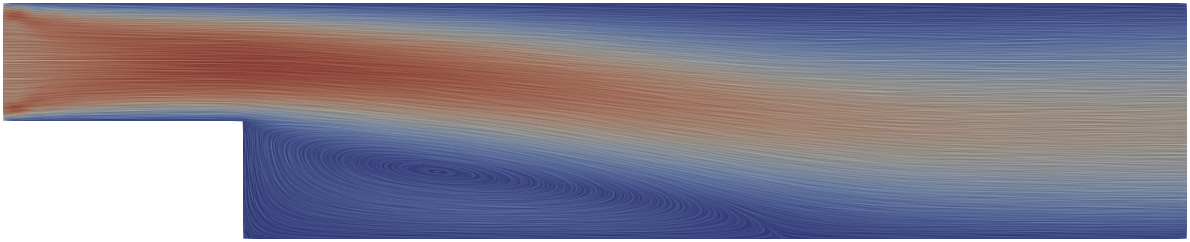}\\
            \parbox{.13\linewidth}{\centering\tiny FE21-BDF3 8m, $e=8.07 \times 10^{-3}$}\hfill
            \parbox{.13\linewidth}{\centering\tiny FE11-SL 8m, $e=1.39 \times 10^{-2}$}\hfill
            \parbox{.13\linewidth}{\centering\tiny FE11-FLIP 7m, $e=9.32 \times 10^{-3}$}\hfill
            \parbox{.13\linewidth}{\centering\tiny FV-C 9m, $e=8.29 \times 10^{-3}$}\hfill
            \parbox{.13\linewidth}{\centering\tiny FD-FLIP 13m, $e=5.09 \times 10^{-3}$}\hfill
            \parbox{.13\linewidth}{\centering\tiny FE11-AB2AM2 8m, $e=1.62 \times 10^{-2}$}\hfill
            \parbox{.13\linewidth}{\centering\tiny FE21-AB2AM2 9m, $e=5.67 \times 10^{-3}$}\\[1em]
            Triangle mesh\\
            \parbox{.13\linewidth}{\includegraphics[width=\linewidth]{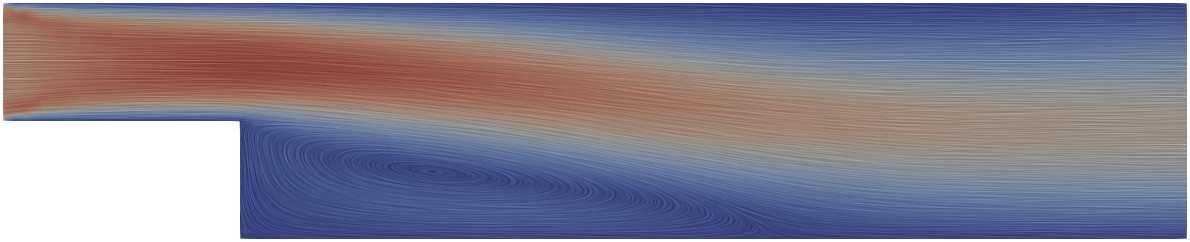}}\hfill
            \parbox{.13\linewidth}{\includegraphics[width=\linewidth]{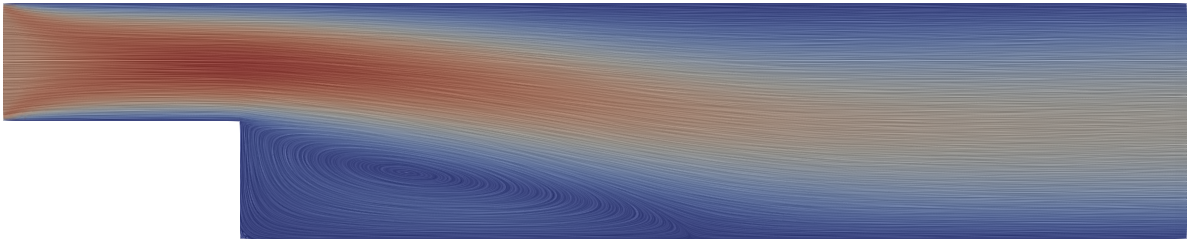}}\hfill
            \parbox{.13\linewidth}{\includegraphics[width=\linewidth]{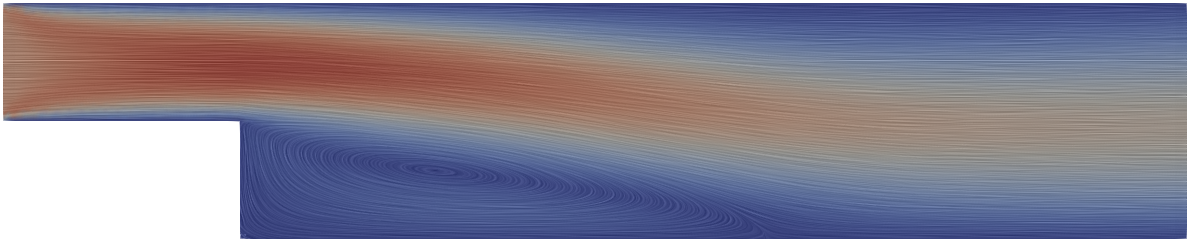}}\hfill
            \parbox{.13\linewidth}{\includegraphics[width=\linewidth]{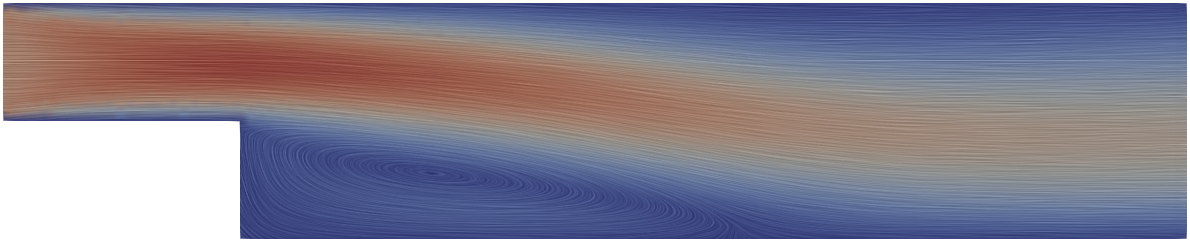}}\hfill
            \parbox{.13\linewidth}{\includegraphics[width=\linewidth]{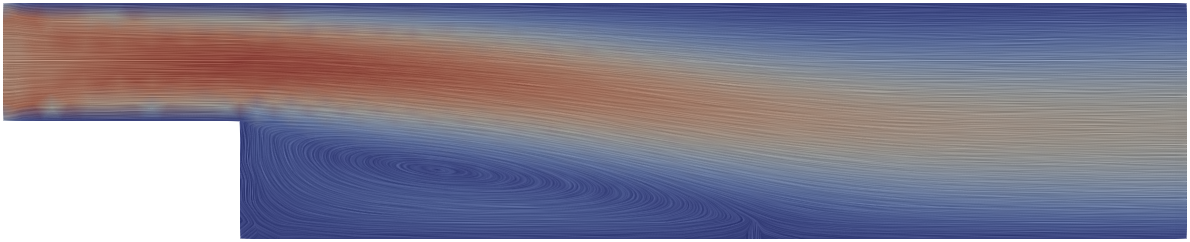}}\hfill
            \parbox{.13\linewidth}{\includegraphics[width=\linewidth]{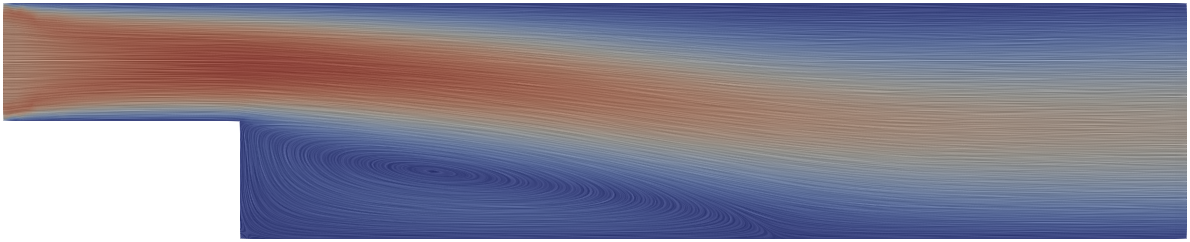}}\hfill
            \parbox{.13\linewidth}{~}\\
            \parbox{.13\linewidth}{\centering\tiny FE21-BDF3 9m, $e=6.94 \times 10^{-3}$}\hfill
            \parbox{.13\linewidth}{\centering\tiny FE11-SL 8m, $e=2.98 \times 10^{-2}$}\hfill
            \parbox{.13\linewidth}{\centering\tiny FE11-FLIP 8m, $e=7.10 \times 10^{-3}$}\hfill
            \parbox{.13\linewidth}{\centering\tiny FV-C 6m, $e=1.11 \times 10^{-2}$}\hfill
            \parbox{.13\linewidth}{\centering\tiny FE11-AB2AM2 8m, $e=1.78 \times 10^{-2}$}\hfill
            \parbox{.13\linewidth}{\centering\tiny FE21-AB2AM2 9m, $e=5.13 \times 10^{-3}$}\hfill
            \parbox{.13\linewidth}{~}
        }\\[2em]
            %%%%%%%%%%%%%%%%
            \parbox{0.04\linewidth}{\centering\rotatebox{90}{$\nu=0.002$}}\hfill
            \parbox{0.95\linewidth}{\centering
            Regular grid\\
            \includegraphics[width=.13\linewidth]{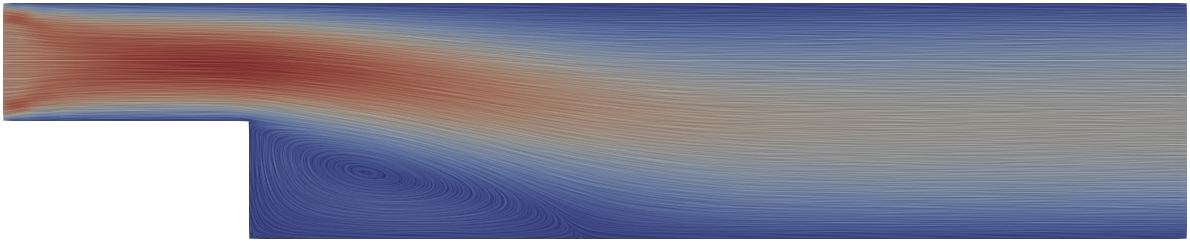}\hfill
            \includegraphics[width=.13\linewidth]{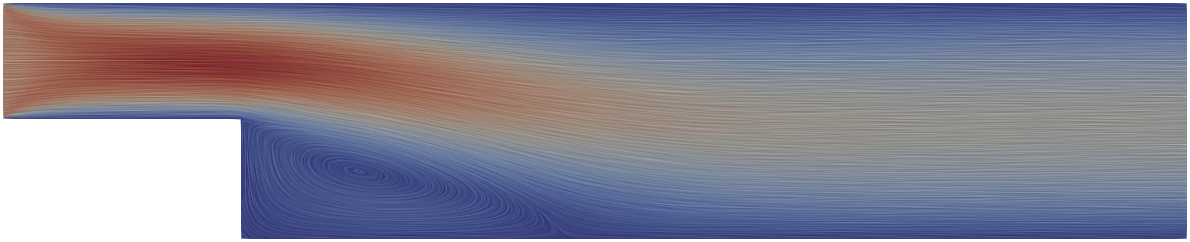}\hfill
            \includegraphics[width=.13\linewidth]{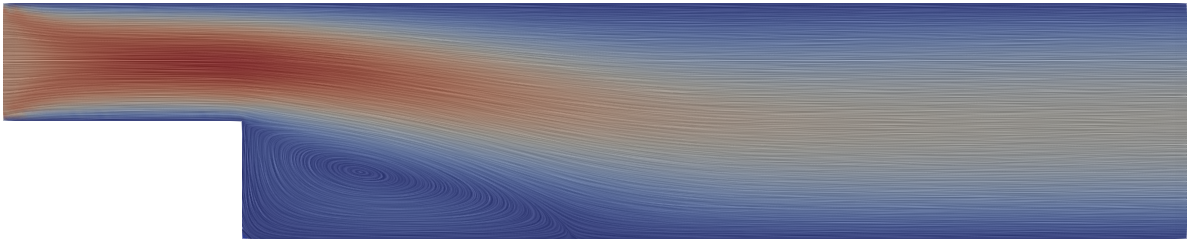}\hfill
            \includegraphics[width=.13\linewidth]{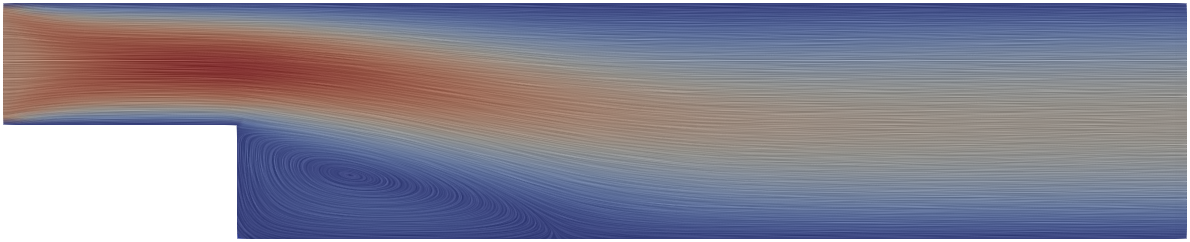}\hfill
            \includegraphics[width=.13\linewidth]{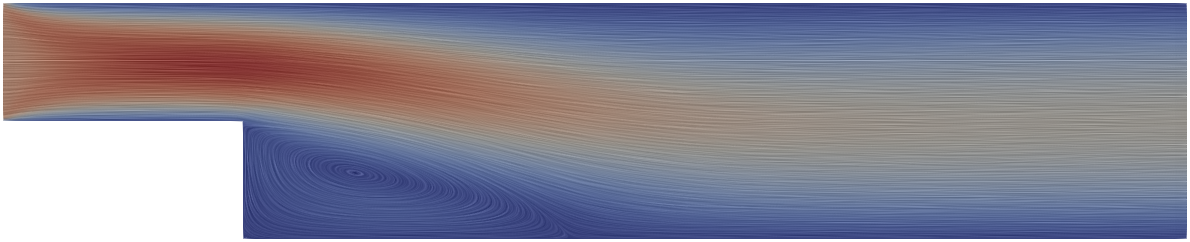}\hfill
            \includegraphics[width=.13\linewidth]{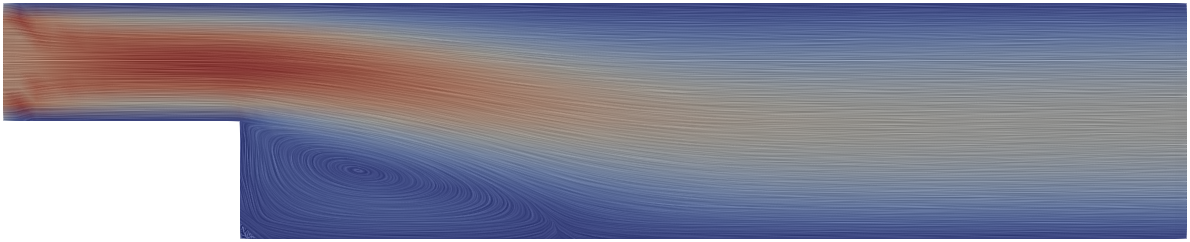}\hfill
            \includegraphics[width=.13\linewidth]{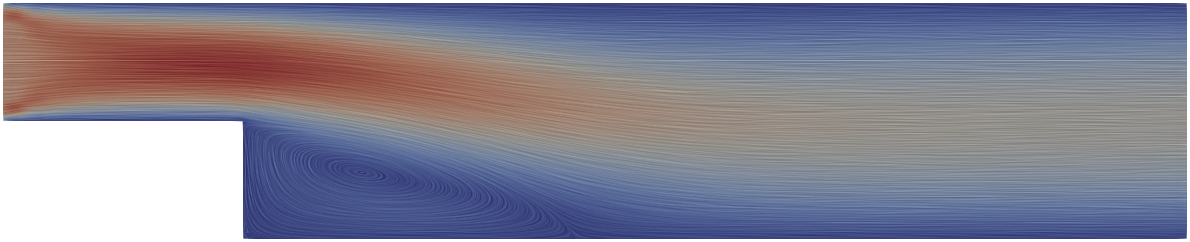}\\
            \parbox{.13\linewidth}{\centering\tiny FE21-BDF3 8m, $e=7.37 \times 10^{-3}$}\hfill
            \parbox{.13\linewidth}{\centering\tiny FE11-SL 8m, $e=9.66 \times 10^{-3}$}\hfill
            \parbox{.13\linewidth}{\centering\tiny FE11-FLIP 7m, $e=6.40 \times 10^{-3}$}\hfill
            \parbox{.13\linewidth}{\centering\tiny FV-C 9m, $e=1.55 \times 10^{-2}$}\hfill
            \parbox{.13\linewidth}{\centering\tiny FD-FLIP 9m, $e=6.49 \times 10^{-3}$}\hfill
            \parbox{.13\linewidth}{\centering\tiny FE11-AB2AM2 9m, $e=1.50 \times 10^{-2}$}\hfill
            \parbox{.13\linewidth}{\centering\tiny FE21-AB2AM2 9m, $e=5.02 \times 10^{-3}$}\\[1em]
            Triangle mesh\\
            \parbox{.13\linewidth}{\includegraphics[width=\linewidth]{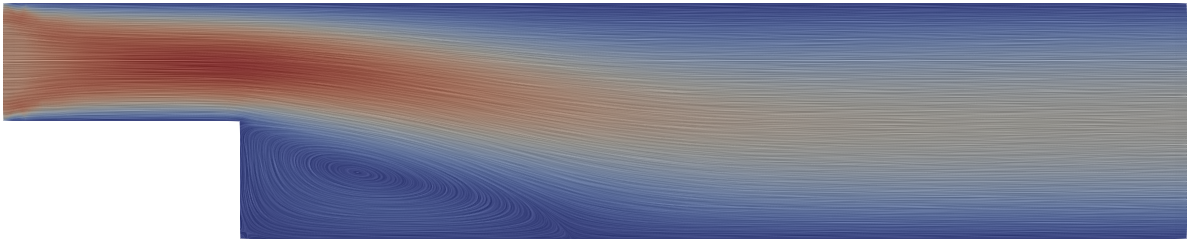}}\hfill
            \parbox{.13\linewidth}{\includegraphics[width=\linewidth]{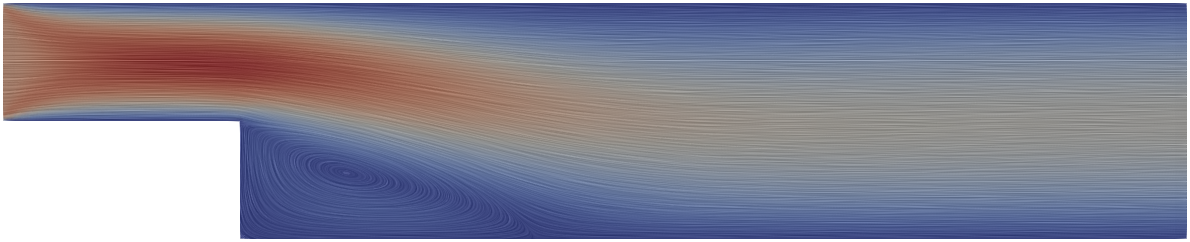}}\hfill
            \parbox{.13\linewidth}{\includegraphics[width=\linewidth]{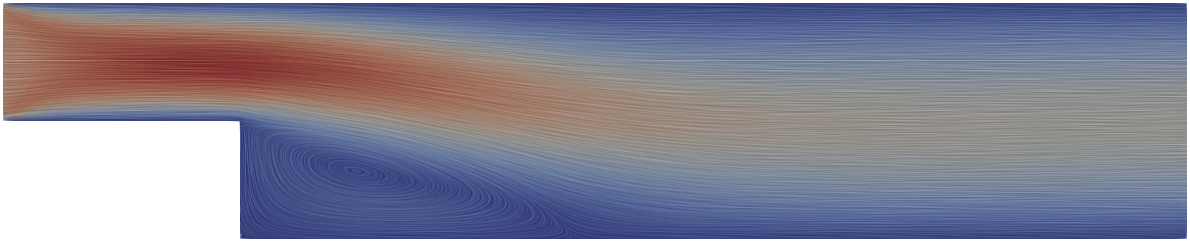}}\hfill
            \parbox{.13\linewidth}{\includegraphics[width=\linewidth]{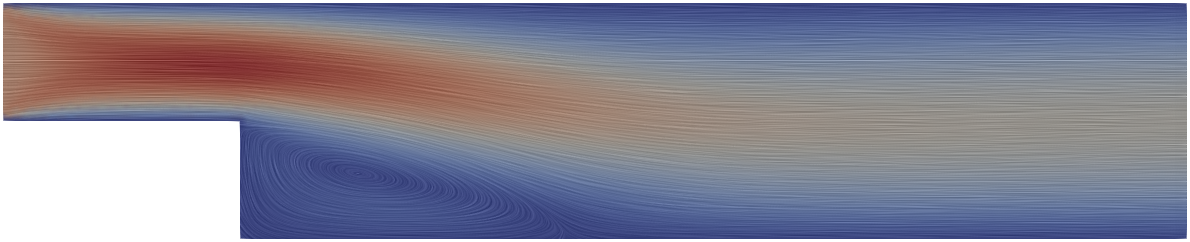}}\hfill
            \parbox{.13\linewidth}{\includegraphics[width=\linewidth]{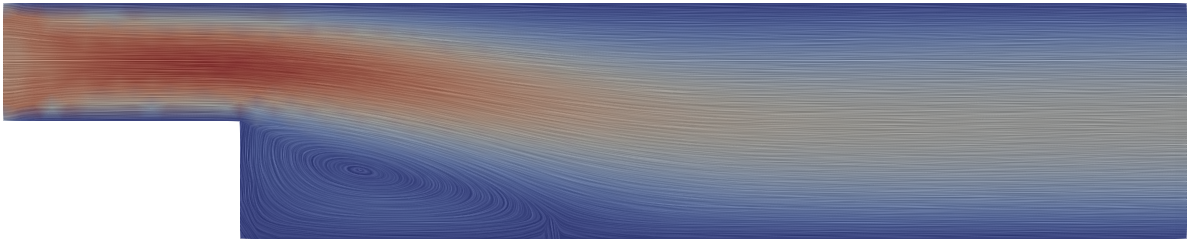}}\hfill
            \parbox{.13\linewidth}{\includegraphics[width=\linewidth]{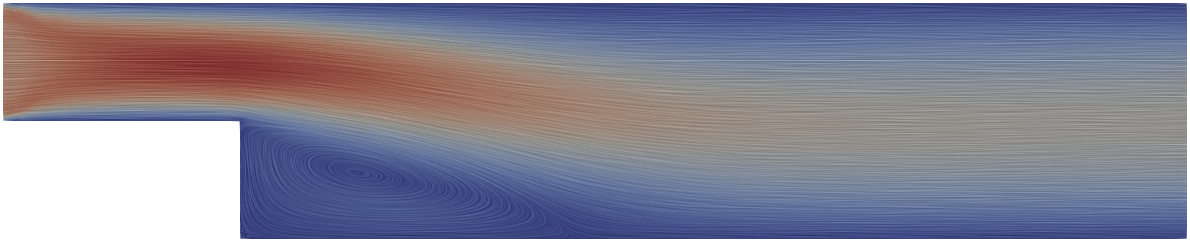}}\hfill
            \parbox{.13\linewidth}{~}\\
            \parbox{.13\linewidth}{\centering\tiny FE21-BDF3 9m, $e=6.82 \times 10^{-3}$}\hfill
            \parbox{.13\linewidth}{\centering\tiny FE11-SL 8m, $e=1.41 \times 10^{-2}$}\hfill
            \parbox{.13\linewidth}{\centering\tiny FE11-FLIP 8m, $e=5.60 \times 10^{-3}$}\hfill
            \parbox{.13\linewidth}{\centering\tiny FV-C 8m, $e=6.50 \times 10^{-3}$}\hfill
            \parbox{.13\linewidth}{\centering\tiny FE11-AB2AM2 8m, $e=1.63 \times 10^{-2}$}\hfill
            \parbox{.13\linewidth}{\centering\tiny FE21-AB2AM2 9m, $e=5.27 \times 10^{-3}$}\hfill
            \parbox{.13\linewidth}{~}
        }\\[2em]
            %%%%%%%%%%%%%%%%
            \parbox{0.04\linewidth}{\centering\rotatebox{90}{$\nu=0.005$}}\hfill
            \parbox{0.95\linewidth}{\centering
            Regular grid\\
            \includegraphics[width=.13\linewidth]{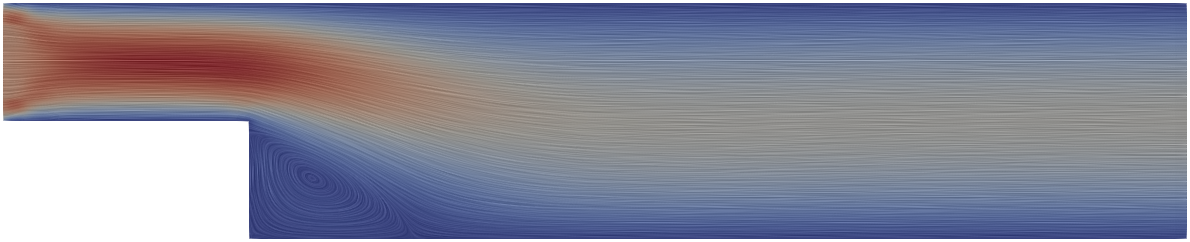}\hfill
            \includegraphics[width=.13\linewidth]{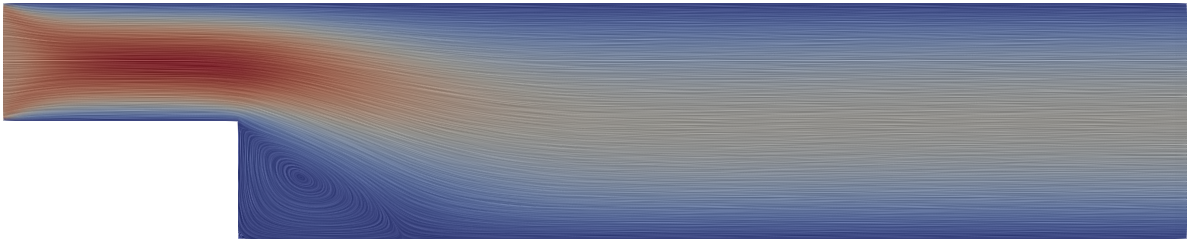}\hfill
            \includegraphics[width=.13\linewidth]{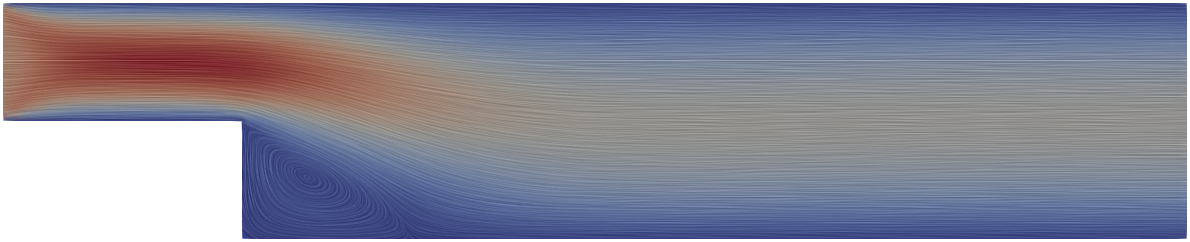}\hfill
            \includegraphics[width=.13\linewidth]{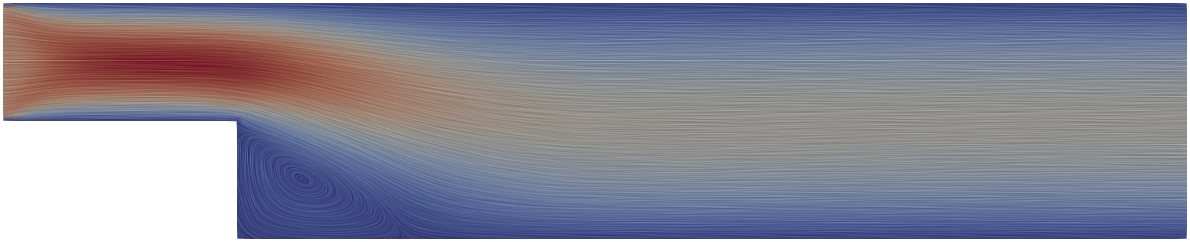}\hfill
            \includegraphics[width=.13\linewidth]{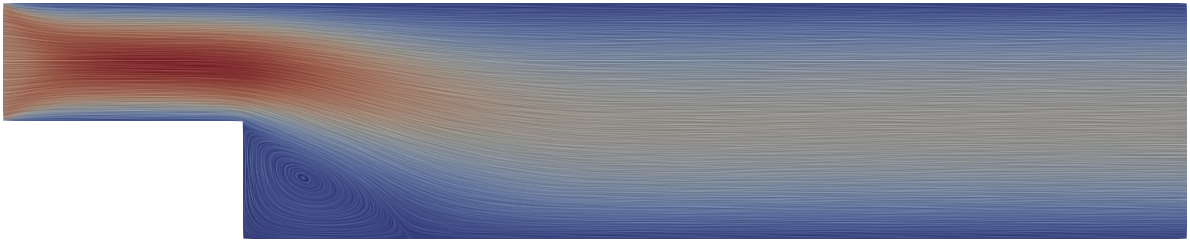}\hfill
            \includegraphics[width=.13\linewidth]{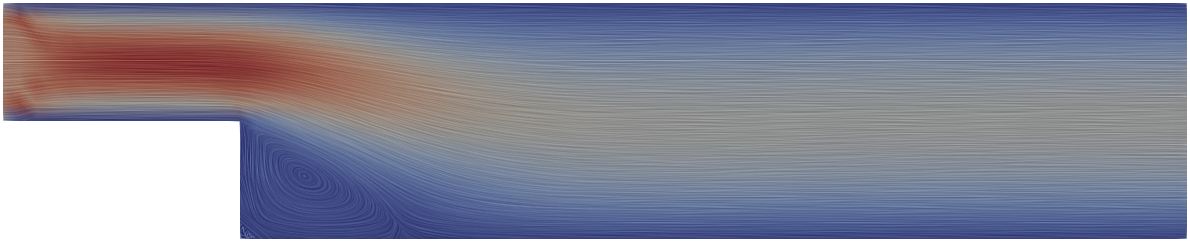}\hfill
            \includegraphics[width=.13\linewidth]{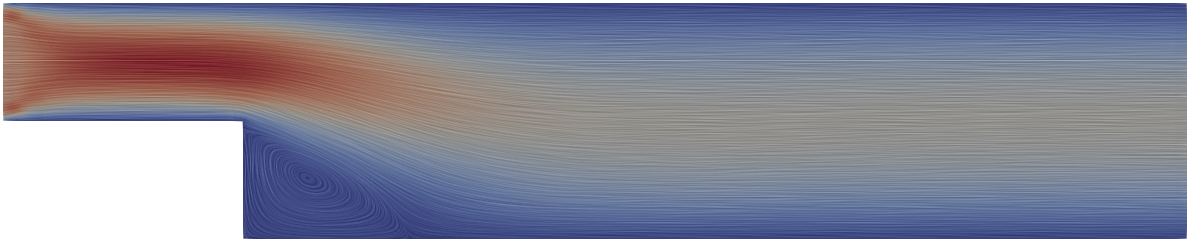}\\
            \parbox{.13\linewidth}{\centering\tiny FE21-BDF3 8m, $e=7.20 \times 10^{-3}$}\hfill
            \parbox{.13\linewidth}{\centering\tiny FE11-SL 8m, $e=7.50 \times 10^{-3}$}\hfill
            \parbox{.13\linewidth}{\centering\tiny FE11-FLIP 7m, $e=5.47 \times 10^{-3}$}\hfill
            \parbox{.13\linewidth}{\centering\tiny FV-C 9m, $e=5.22 \times 10^{-3}$}\hfill
            \parbox{.13\linewidth}{\centering\tiny FD-FLIP 8m, $e=6.18 \times 10^{-3}$}\hfill
            \parbox{.13\linewidth}{\centering\tiny FE11-AB2AM2 9m, $e=1.37 \times 10^{-2}$}\hfill
            \parbox{.13\linewidth}{\centering\tiny FE21-AB2AM2 9m, $e=4.21 \times 10^{-3}$}\\[1em]
            Triangle mesh\\
            \parbox{.13\linewidth}{\includegraphics[width=\linewidth]{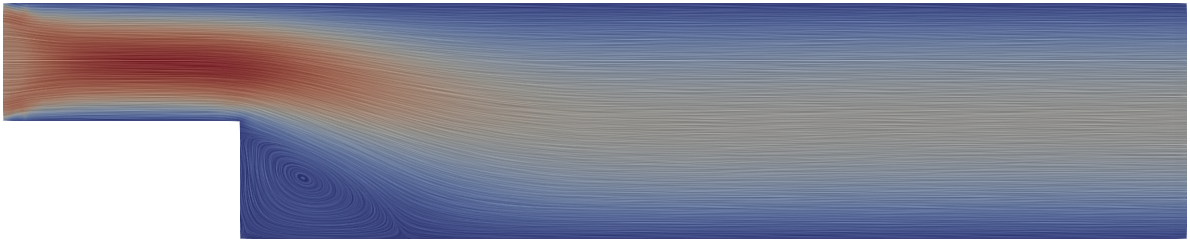}}\hfill
            \parbox{.13\linewidth}{\includegraphics[width=\linewidth]{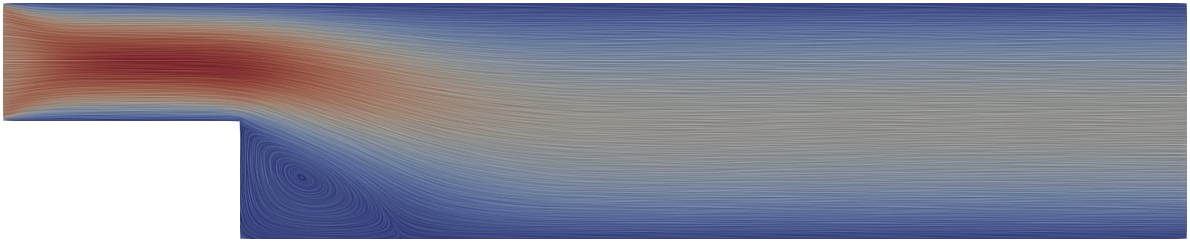}}\hfill
            \parbox{.13\linewidth}{\includegraphics[width=\linewidth]{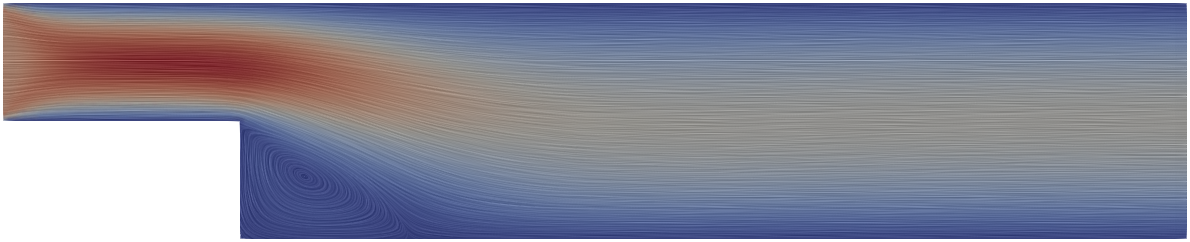}}\hfill
            \parbox{.13\linewidth}{\includegraphics[width=\linewidth]{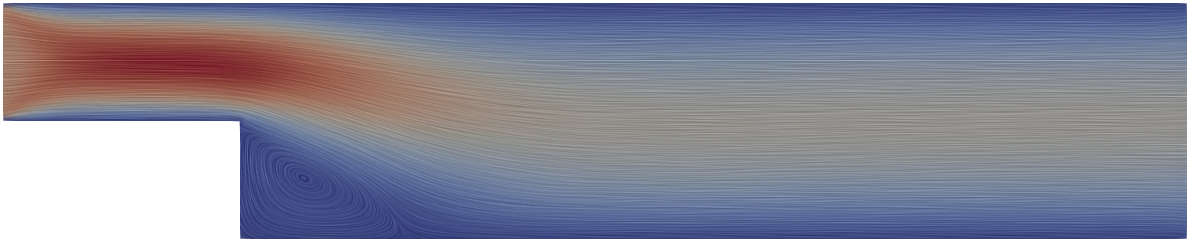}}\hfill
            \parbox{.13\linewidth}{\includegraphics[width=\linewidth]{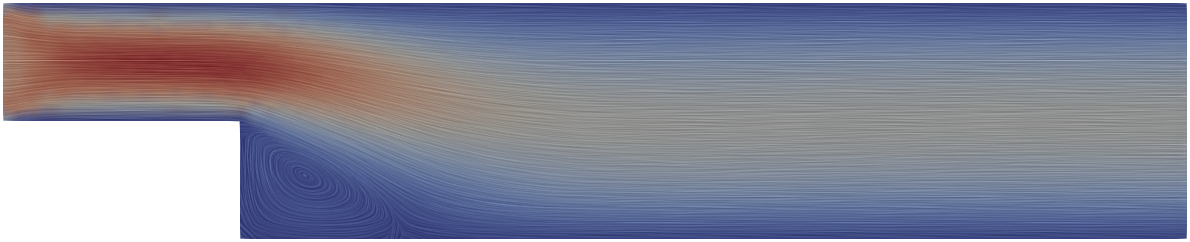}}\hfill
            \parbox{.13\linewidth}{\includegraphics[width=\linewidth]{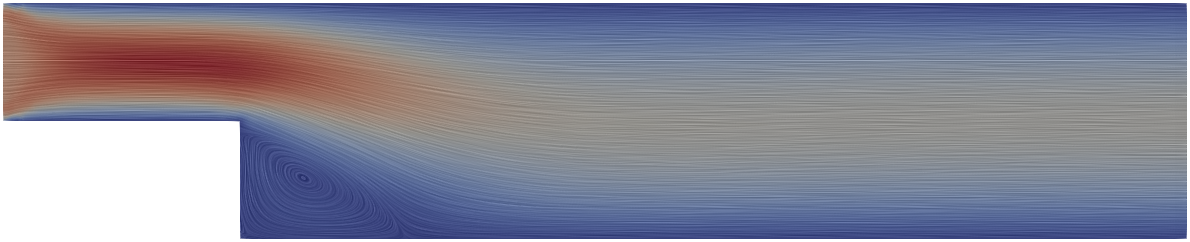}}\hfill
            \parbox{.13\linewidth}{~}\\
            \parbox{.13\linewidth}{\centering\tiny FE21-BDF3 9m, $e=6.57 \times 10^{-3}$}\hfill
            \parbox{.13\linewidth}{\centering\tiny FE11-SL 9m, $e=4.40 \times 10^{-3}$}\hfill
            \parbox{.13\linewidth}{\centering\tiny FE11-FLIP 8m, $e=5.12 \times 10^{-3}$}\hfill
            \parbox{.13\linewidth}{\centering\tiny FV-C 8m, $e=5.26 \times 10^{-3}$}\hfill
            \parbox{.13\linewidth}{\centering\tiny FE11-AB2AM2 10m, $e=1.29 \times 10^{-2}$}\hfill
            \parbox{.13\linewidth}{\centering\tiny FE21-AB2AM2 10m, $e=5.12 \times 10^{-3}$}\hfill
            \parbox{.13\linewidth}{~}
        }\\[2em]
            %%%%%%%%%%%%%%%%
            \parbox{0.04\linewidth}{\centering\rotatebox{90}{$\nu=0.01$}}\hfill
            \parbox{0.95\linewidth}{\centering
            Regular grid\\
            \includegraphics[width=.13\linewidth]{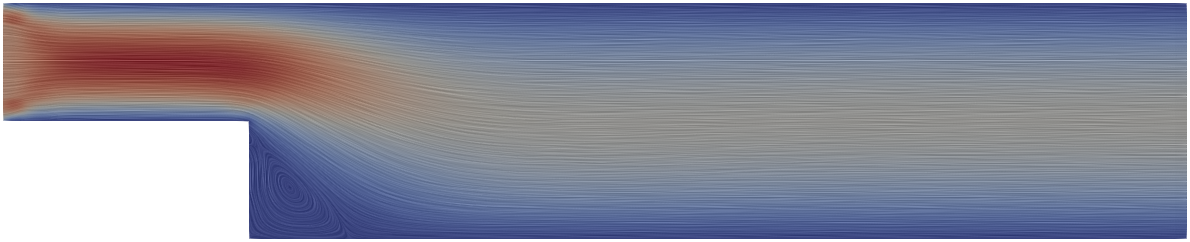}\hfill
            \includegraphics[width=.13\linewidth]{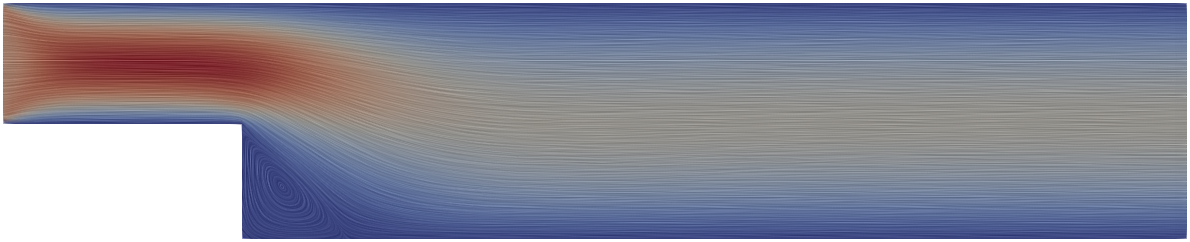}\hfill
            \includegraphics[width=.13\linewidth]{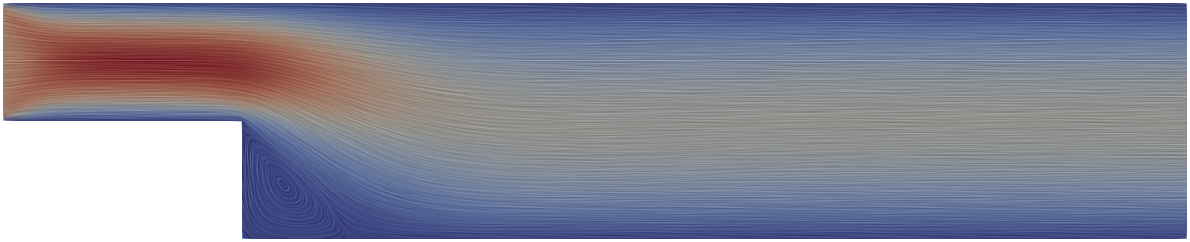}\hfill
            \includegraphics[width=.13\linewidth]{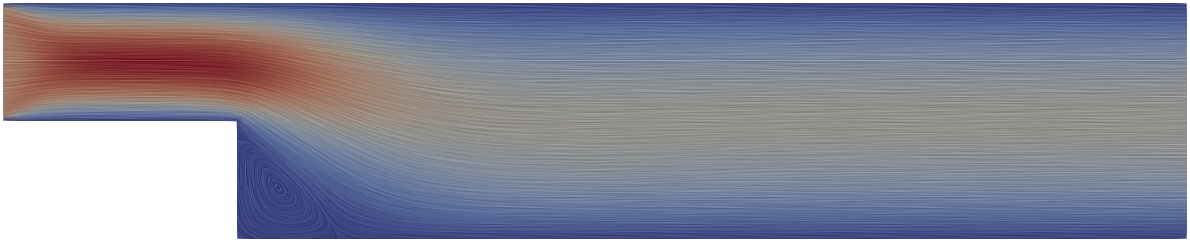}\hfill
            \includegraphics[width=.13\linewidth]{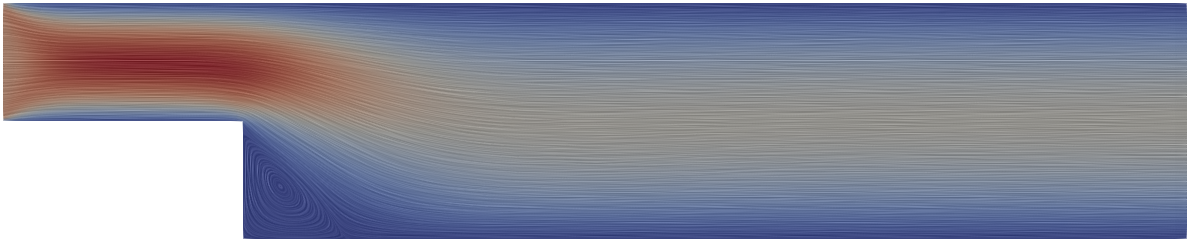}\hfill
            \includegraphics[width=.13\linewidth]{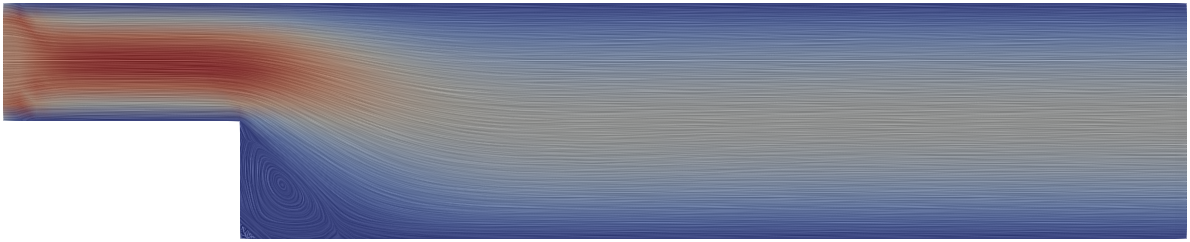}\hfill
            \includegraphics[width=.13\linewidth]{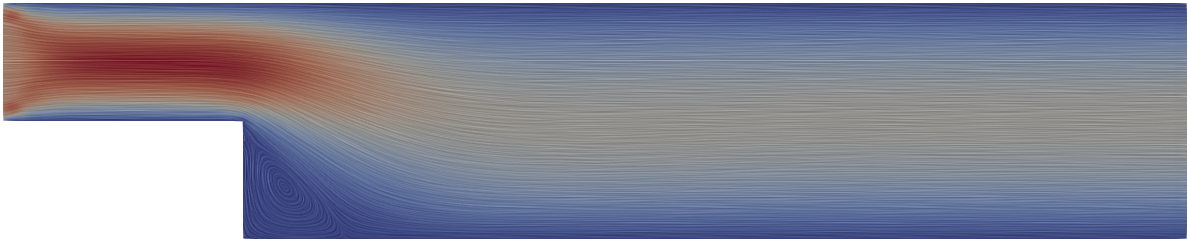}\\
            \parbox{.13\linewidth}{\centering\tiny FE21-BDF3 9m, $e=7.47 \times 10^{-3}$}\hfill
            \parbox{.13\linewidth}{\centering\tiny FE11-SL 7m, $e=7.34 \times 10^{-3}$}\hfill
            \parbox{.13\linewidth}{\centering\tiny FE11-FLIP 7m, $e=5.46 \times 10^{-3}$}\hfill
            \parbox{.13\linewidth}{\centering\tiny FV-C 8m, $e=5.26 \times 10^{-3}$}\hfill
            \parbox{.13\linewidth}{\centering\tiny FD-FLIP 8m, $e=6.47 \times 10^{-3}$}\hfill
            \parbox{.13\linewidth}{\centering\tiny FE11-AB2AM2 8m, $e=1.32 \times 10^{-2}$}\hfill
            \parbox{.13\linewidth}{\centering\tiny FE21-AB2AM2 10m, $e=3.83 \times 10^{-3}$}\\[1em]
            Triangle mesh\\
            \parbox{.13\linewidth}{\includegraphics[width=\linewidth]{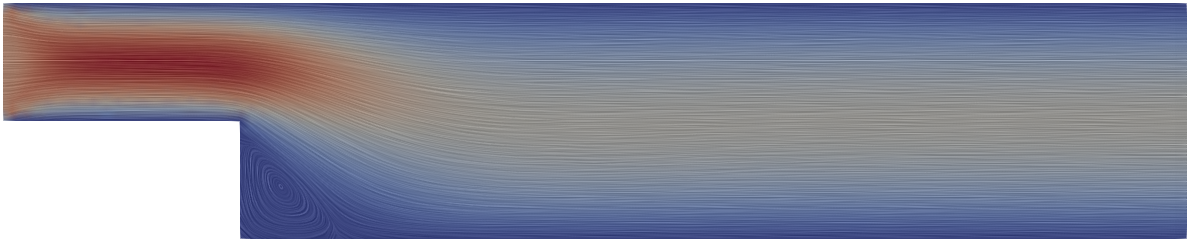}}\hfill
            \parbox{.13\linewidth}{\includegraphics[width=\linewidth]{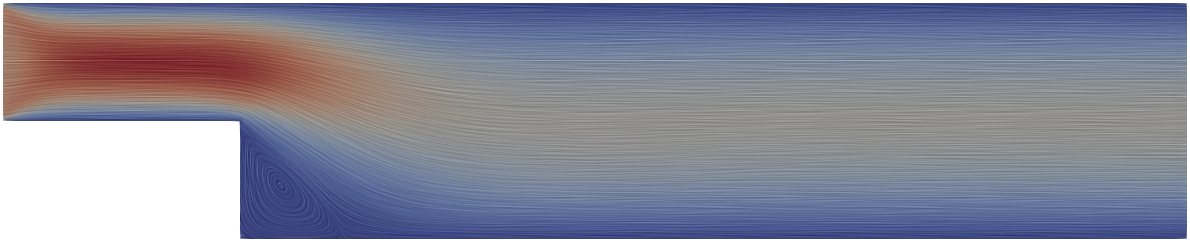}}\hfill
            \parbox{.13\linewidth}{\includegraphics[width=\linewidth]{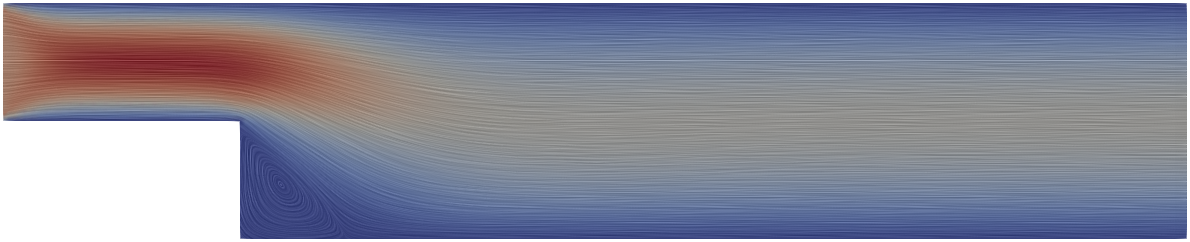}}\hfill
            \parbox{.13\linewidth}{\includegraphics[width=\linewidth]{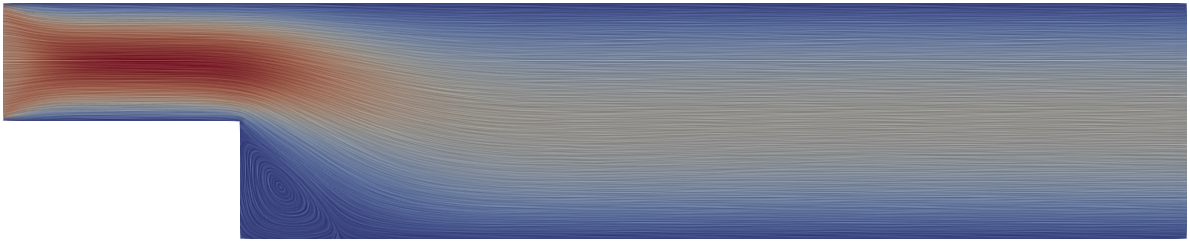}}\hfill
            \parbox{.13\linewidth}{\includegraphics[width=\linewidth]{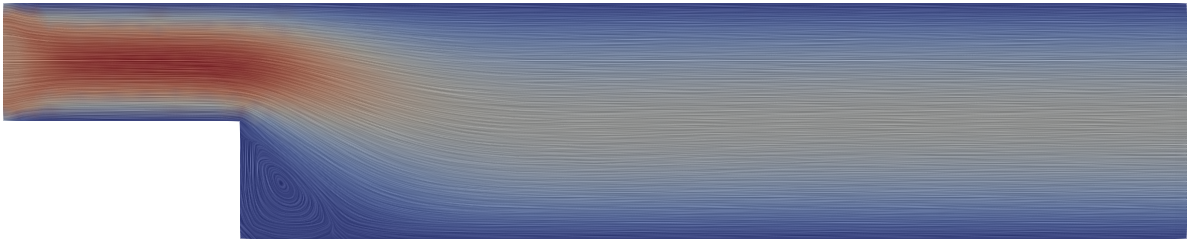}}\hfill
            \parbox{.13\linewidth}{\includegraphics[width=\linewidth]{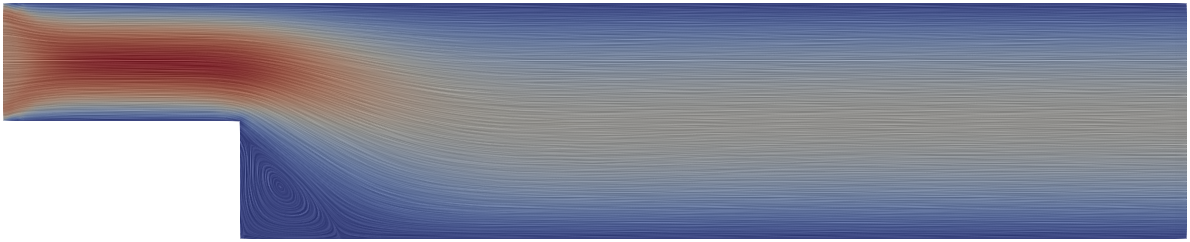}}\hfill
            \parbox{.13\linewidth}{~}\\
            \parbox{.13\linewidth}{\centering\tiny FE21-BDF3 6m, $e=5.32 \times 10^{-3}$}\hfill
            \parbox{.13\linewidth}{\centering\tiny FE11-SL 8m, $e=3.85 \times 10^{-3}$}\hfill
            \parbox{.13\linewidth}{\centering\tiny FE11-FLIP 8m, $e=5.41 \times 10^{-3}$}\hfill
            \parbox{.13\linewidth}{\centering\tiny FV-C 9m, $e=4.23 \times 10^{-3}$}\hfill
            \parbox{.13\linewidth}{\centering\tiny FE11-AB2AM2 10m, $e=1.26 \times 10^{-2}$}\hfill
            \parbox{.13\linewidth}{\centering\tiny FE21-AB2AM2 9m, $e=5.03 \times 10^{-3}$}\hfill
            \parbox{.13\linewidth}{~}
        }}
            %%%%%%%%%%%%%%%%
        \caption{Result of the open driven cavity at time $t = T = 6$ for different viscosities.}
        \label{fig:ocavity}
    \end{figure}

%% file: pics/vortex/plot-tri.tex
    \begin{figure}
        \centering\scriptsize
        \includegraphics[width=.25\linewidth]{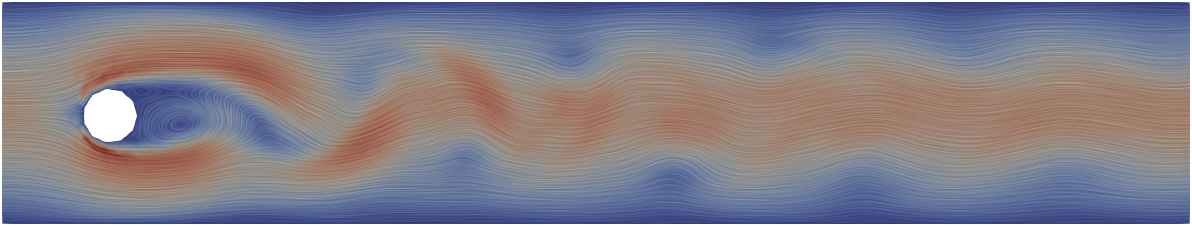}\hfill
        \includegraphics[width=.25\linewidth]{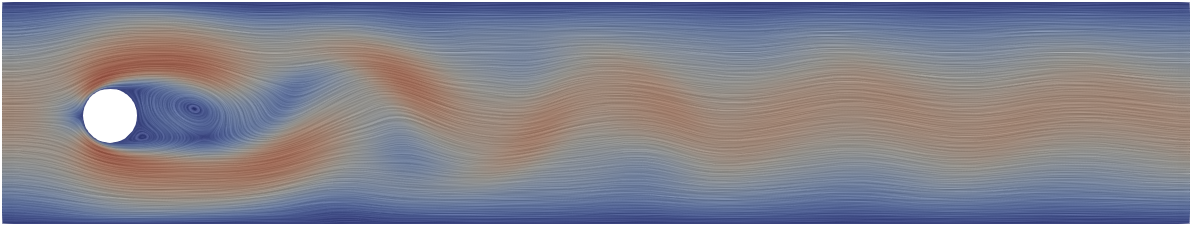}\hfill
        \includegraphics[width=.25\linewidth]{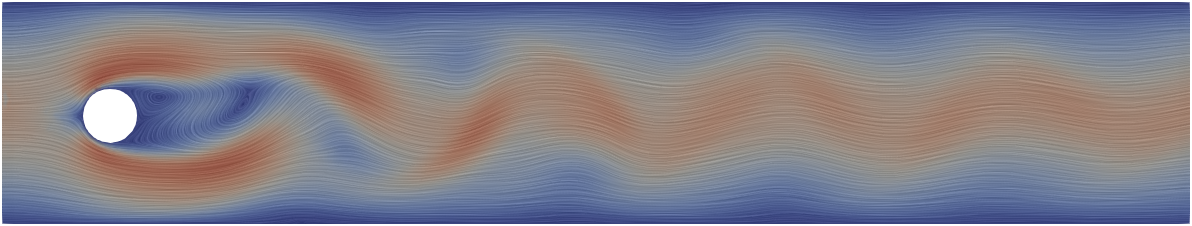}\\
        \parbox{.25\linewidth}{\centering FE21-BDF3 12m, $e=1.31 \times 10^{-1}$}\hfill
        \parbox{.25\linewidth}{\centering FE11-SL 11m, $e=4.01 \times 10^{-1}$}\hfill
        \parbox{.25\linewidth}{\centering FE11-FLIP 13m, $e=4.70 \times 10^{-1}$}\\
        \includegraphics[width=.25\linewidth]{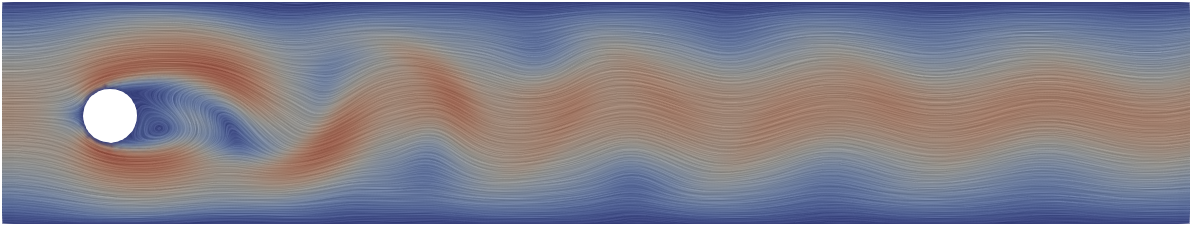}\hfill
        \includegraphics[width=.25\linewidth]{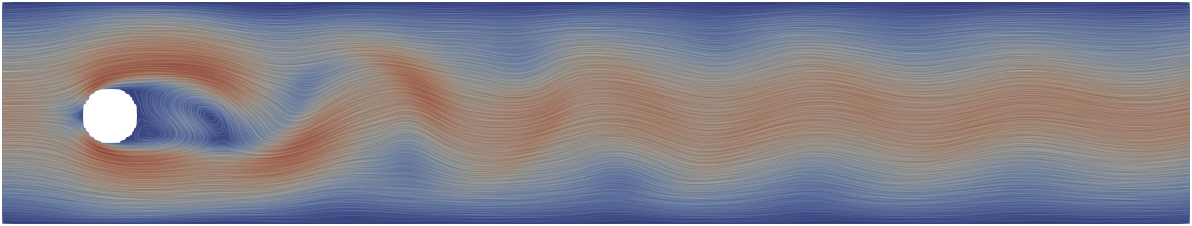}\hfill
        \includegraphics[width=.25\linewidth]{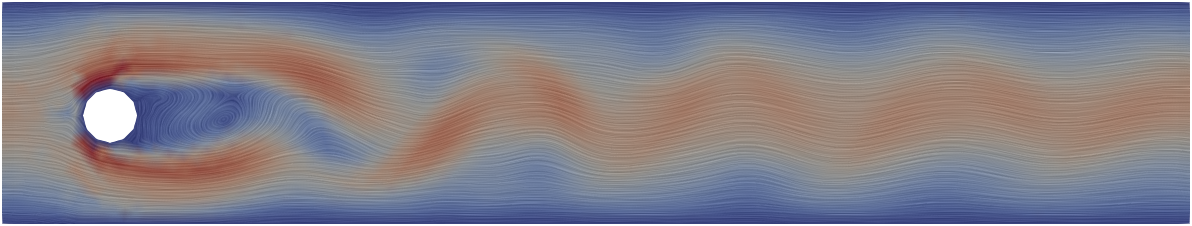}\hfill
        \includegraphics[width=.25\linewidth]{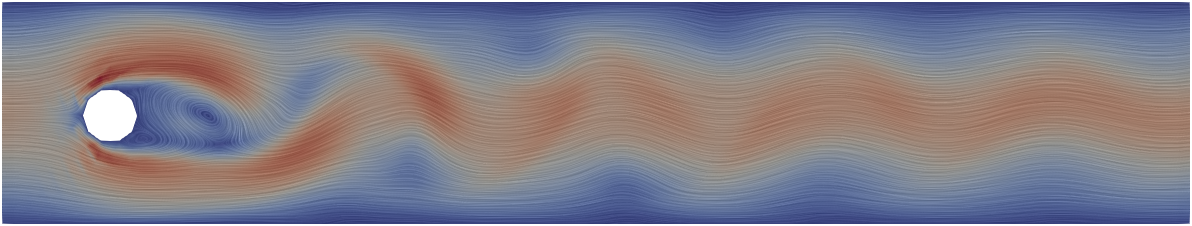}\\
        \parbox{.25\linewidth}{\centering FV-C 12m, $e={5.22 \times 10^{-2}}$}\hfill
        \parbox{.25\linewidth}{\centering FD-FLIP 12m, $e=1.27 \times 10^{-1}$}\hfill
        \parbox{.25\linewidth}{\centering FE11-AB2AM2 11m, $e=4.13 \times 10^{-1}$}\hfill
        \parbox{.25\linewidth}{\centering FE21-AB2AM2 10m, $e=4.73 \times 10^{-1}$}
        \caption{Result of the vortex street on a triangular mesh at time $t=T=6$.}
        \label{fig:vortex-tri}
    \end{figure}

%% file: pics/vortex/plot-quad.tex
    \begin{figure}
        \centering\scriptsize
        \includegraphics[width=.25\linewidth]{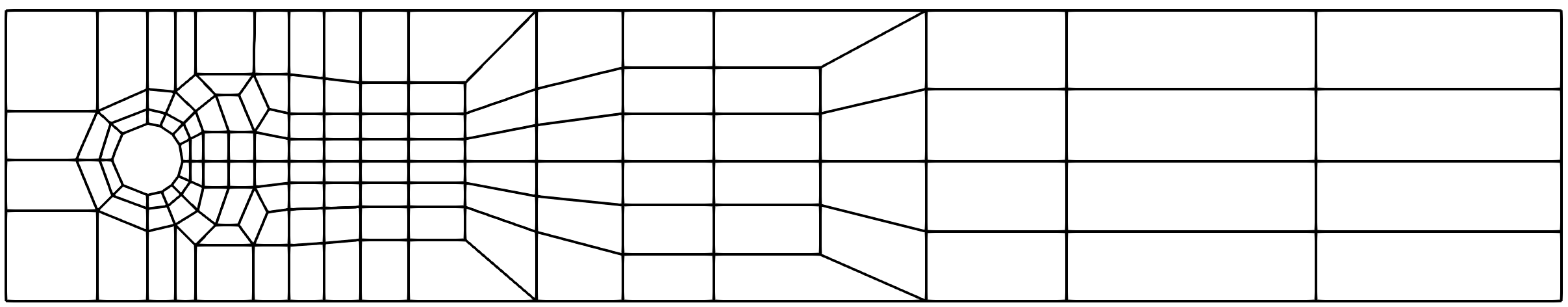}\\
        \includegraphics[width=.25\linewidth]{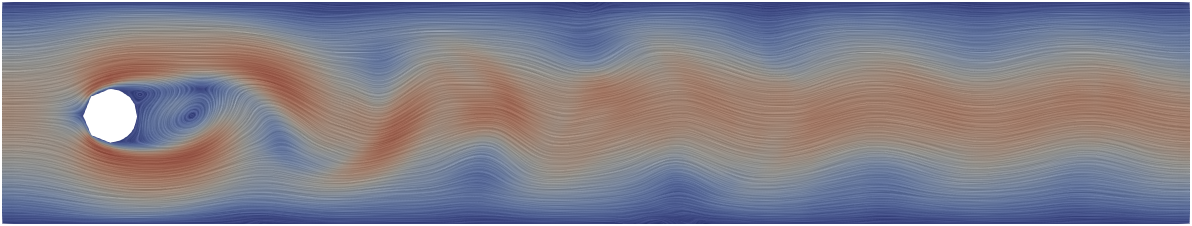}\hfill
        \includegraphics[width=.25\linewidth]{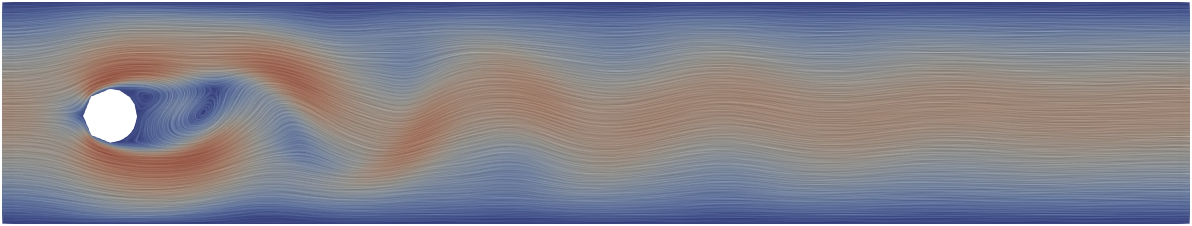}\hfill
        \includegraphics[width=.25\linewidth]{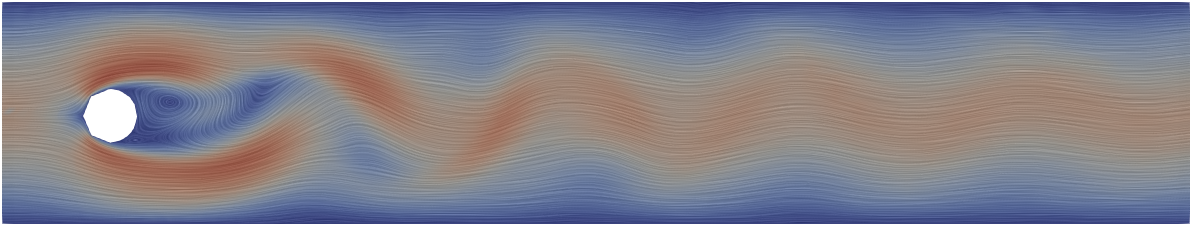}\\
        \parbox{.25\linewidth}{\centering FE21-BDF3 25m, $e=4.61 \times 10^{-2}$}\hfill
        \parbox{.25\linewidth}{\centering FE11-SL 7m, $e=4.34 \times 10^{-1}$}\hfill
        \parbox{.25\linewidth}{\centering FE11-FLIP 3m, $e=4.33 \times 10^{-1}$}\\
        \includegraphics[width=.25\linewidth]{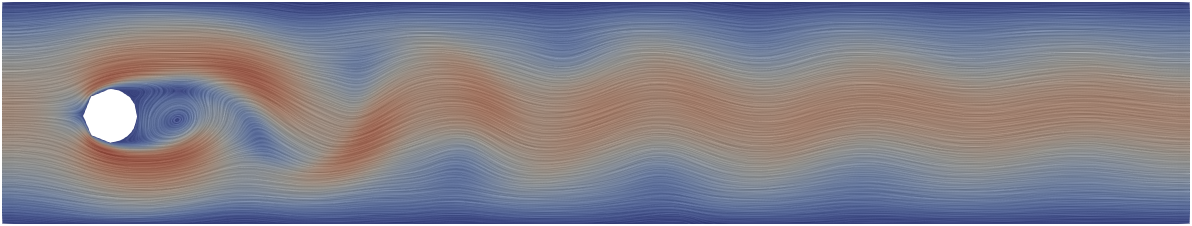}\hfill
        \includegraphics[width=.25\linewidth]{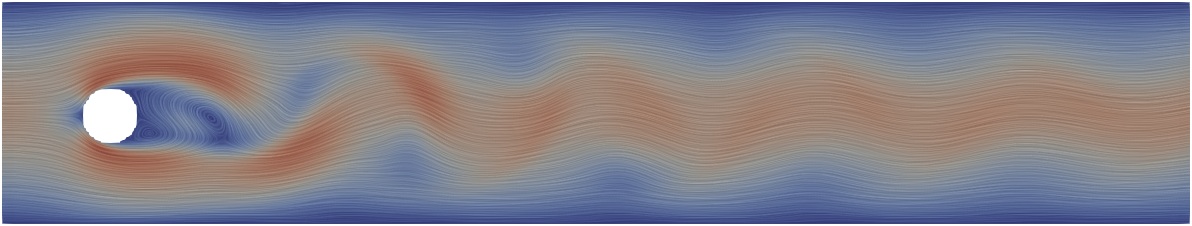}\hfill
        \includegraphics[width=.25\linewidth]{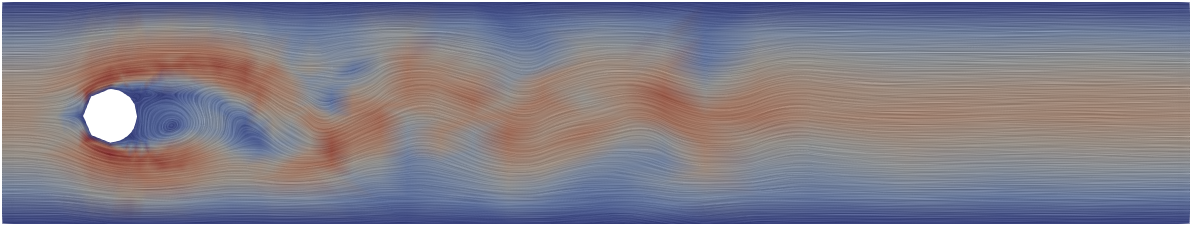}\hfill
        \includegraphics[width=.25\linewidth]{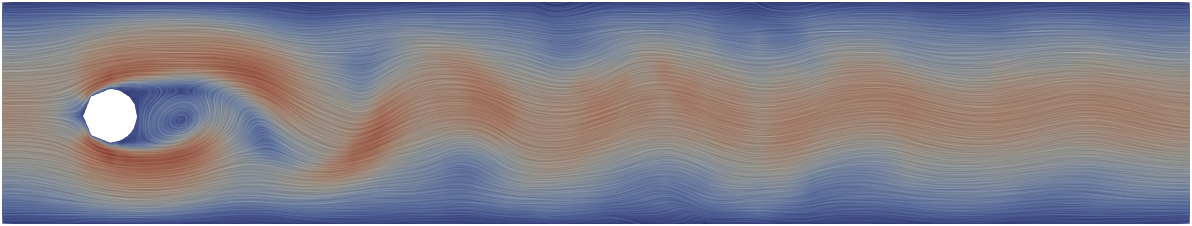}\\
        \parbox{.25\linewidth}{\centering FV-C 6m,\\$e=1.02 \times 10^{-1}$}\hfill
        \parbox{.25\linewidth}{\centering FD-FLIP 12m,\\$e=1.27 \times 10^{-1}$}\hfill
        \parbox{.25\linewidth}{\centering FE11-AB2AM2 12m,\\$e=4.39 \times 10^{-1}$}\hfill
        \parbox{.25\linewidth}{\centering FE21-AB2AM2 19m,\\$e=1.41 \times 10^{-1}$}
        \caption{Result of the vortex street on an adaptive quad-mesh at time $t=T=6$.}
        \label{fig:vortex-quad}
    \end{figure}

%% file: 06-largescale.tex
% \section{Manifactured Solutions}
\section{Large-Scale Benchmark}\label{sec:large-scale}

To quantify method errors on a large set of complex geometries, we use setups with known closed-form solutions: Taylor-Green vortex in 2D and the \emph{method of manufactured solutions} in 3D \cite{SALARI:2000:CVB}. In the latter case, we start from a target closed-form solution $u$, compute the corresponding right-hand side (which for our choice of $u$ is zero) by plugging it into the PDE, and evaluate $u$ on the boundary to obtain the Dirichet boundary conditions. 
For a complete description of our experimental setup, see Appendix \ref{app:setup}.%

The Taylor-Green vortex~\cite{taylor1937mechanism} in 2D is given by
\[
    \begin{aligned}
        \mathbf{u}_{\text{2D}}(x_1,x_2,t) & = (\cos(2\pi x_1)\sin(2\pi x_2), -\sin(2\pi x_1)\cos(2\pi x_2))e^{-8\nu \pi^2 t} \\
        p_{\text{2D}}(x_1,x_2,t)          & = \frac{-1}{4}(\cos(4\pi x_1)+\cos(4\pi x_2)) e^{-16\nu \pi^2 t}
    \end{aligned}
\]
We use the following manufactured solution in 3D, following \cite{Ethier1994}. It is one of the few cases for which exact fully three-dimensional solution of the incompressible NS equations  can be obtained without introducing artificial forcing terms. 
\[
    \begin{aligned}
        \mathbf{u}_{\text{3D}}(x_1,x_2, x_3, t)
                                     & = -e^{-\nu t} (e^x_1\sin(x_2+x_3)+e^x_3\cos(x_1+x_2),                        \\
                                     & e^x_2\sin(x_3+x_1)+e^x_1\cos(x_2+x_3),e^x_3\sin(x_1+x_2)+e^x_2\cos(x_3+x_1)) \\
        p_{\text{3D}}(x_1,x_2,x_3,t) & = -\frac{1}{2}e^{-2\nu t}(e^{2x_1}+e^{2x_2}+e^{2x_3}                         \\
                                     & +2\sin(x_1+x_2)\cos(x_3+x_1)e^{x_2+x_3}                                      \\
                                     & +2\sin(x_2+x_3)\cos(x_1+x_2)e^{x_3+x_1}                                      \\
                                     & +2\sin(x_3+x_1)\cos(x_2+x_3)e^{x_1+x_2})
    \end{aligned}
\]
with viscosity $\nu=0.01$, $T=0.1$. 

We evaluate the performance of all methods (except  FD-FLIP) on a large, diverse set of realistic domains, for two- and three-dimensional problems. We excluded FD-FLIP as it is designed to work on regular grids, and the error introduced by the implicit approximation of the boundary would lead to an unfair comparison.

For every domain, we generate the quad-/hex-meshes first, as they are more challenging to generate for automatic tools, and then create corresponding tri-/tet-meshes with a matching number of vertices.  For conciseness, we report only time versus $L^2$ error plots; the other metrics can be explored in our interactive plot at \url{https://cims.nyu.edu/gcl/fluids/}.

\subsection{2D Dataset} 
\begin{figure}
    \centering
    \includegraphics[width=.45\linewidth]{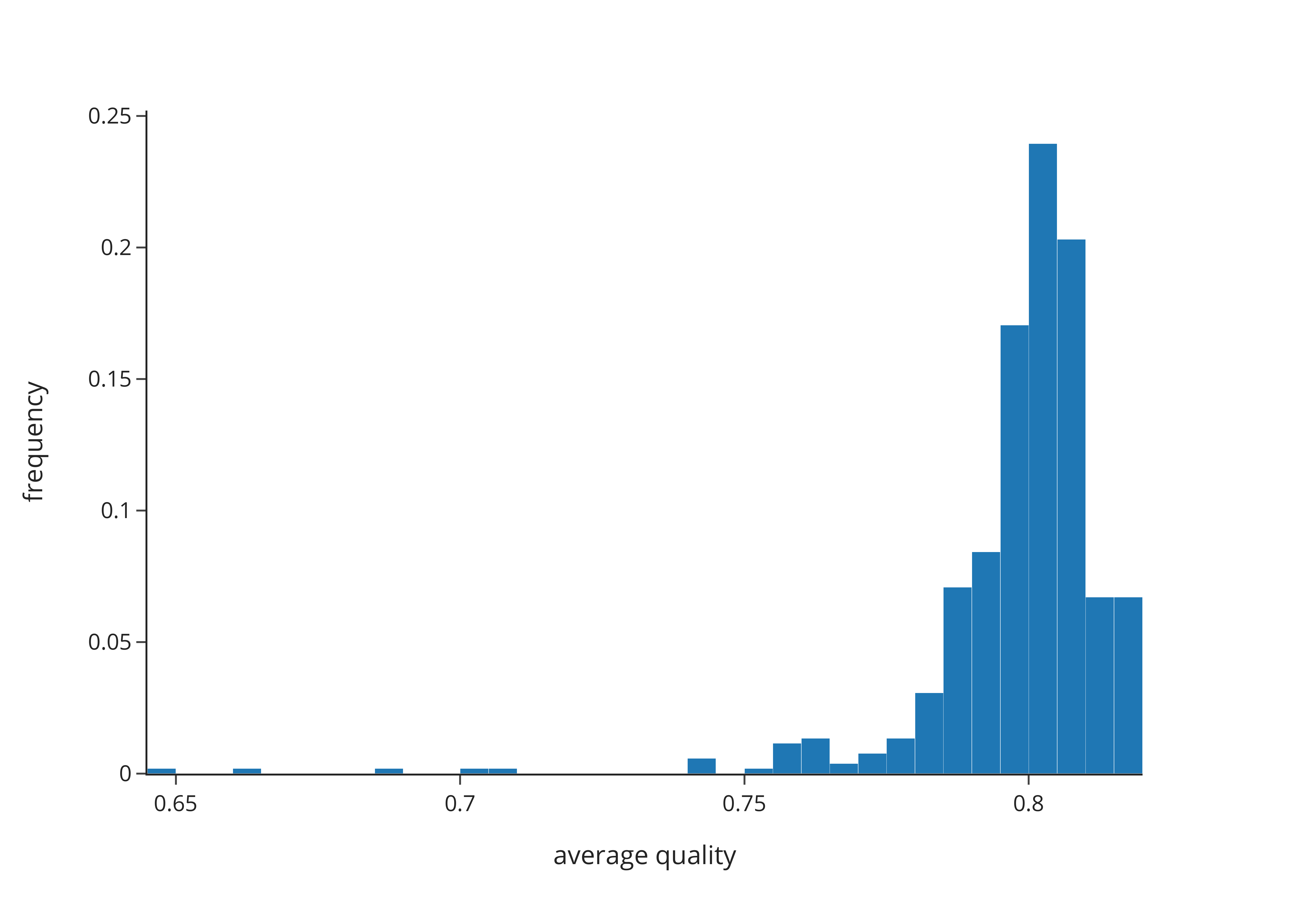}\hfill
    \includegraphics[width=.45\linewidth]{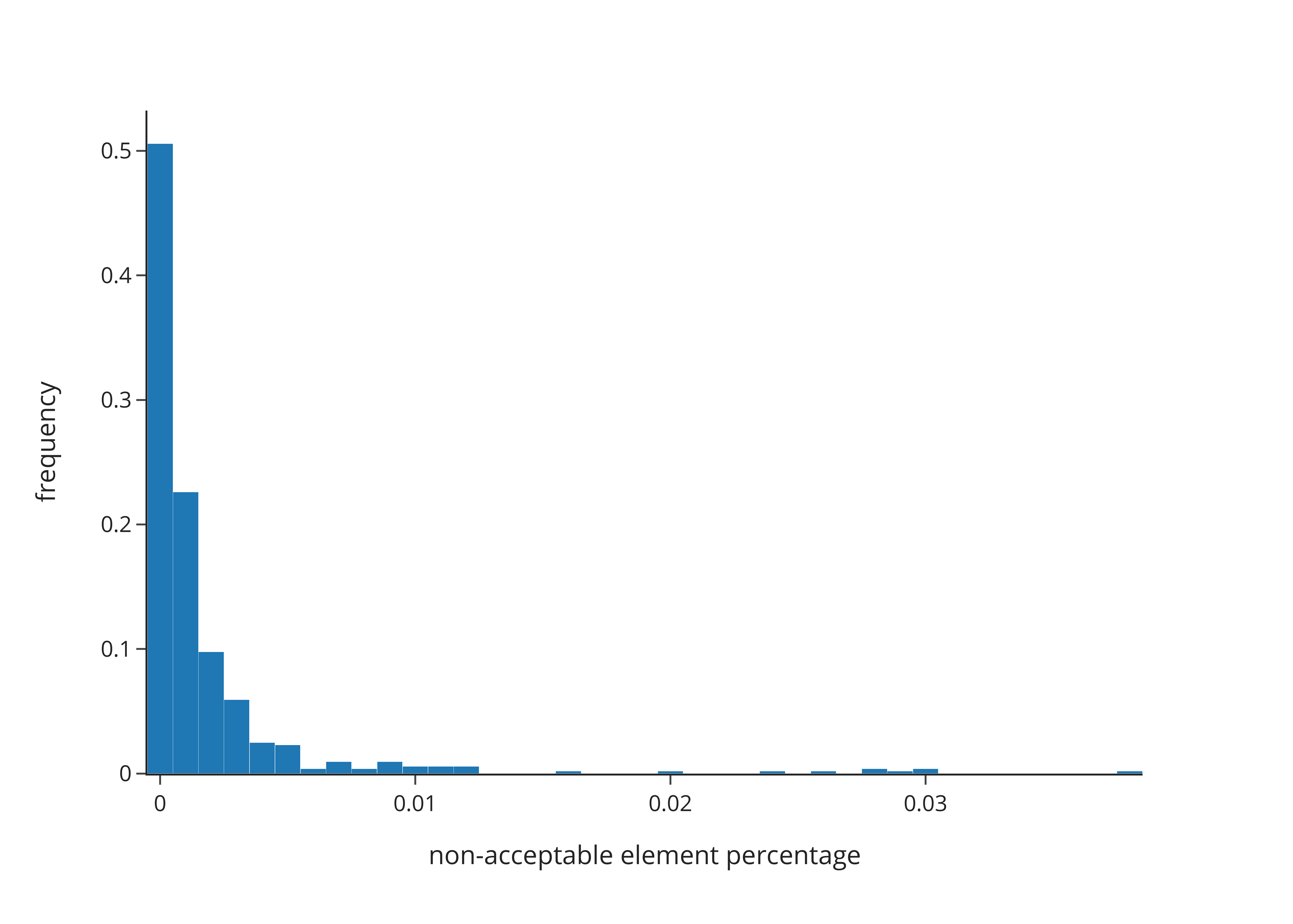}
    \parbox{\linewidth}{\centering Triangular meshes.}
    \centering
    \includegraphics[width=.45\linewidth]{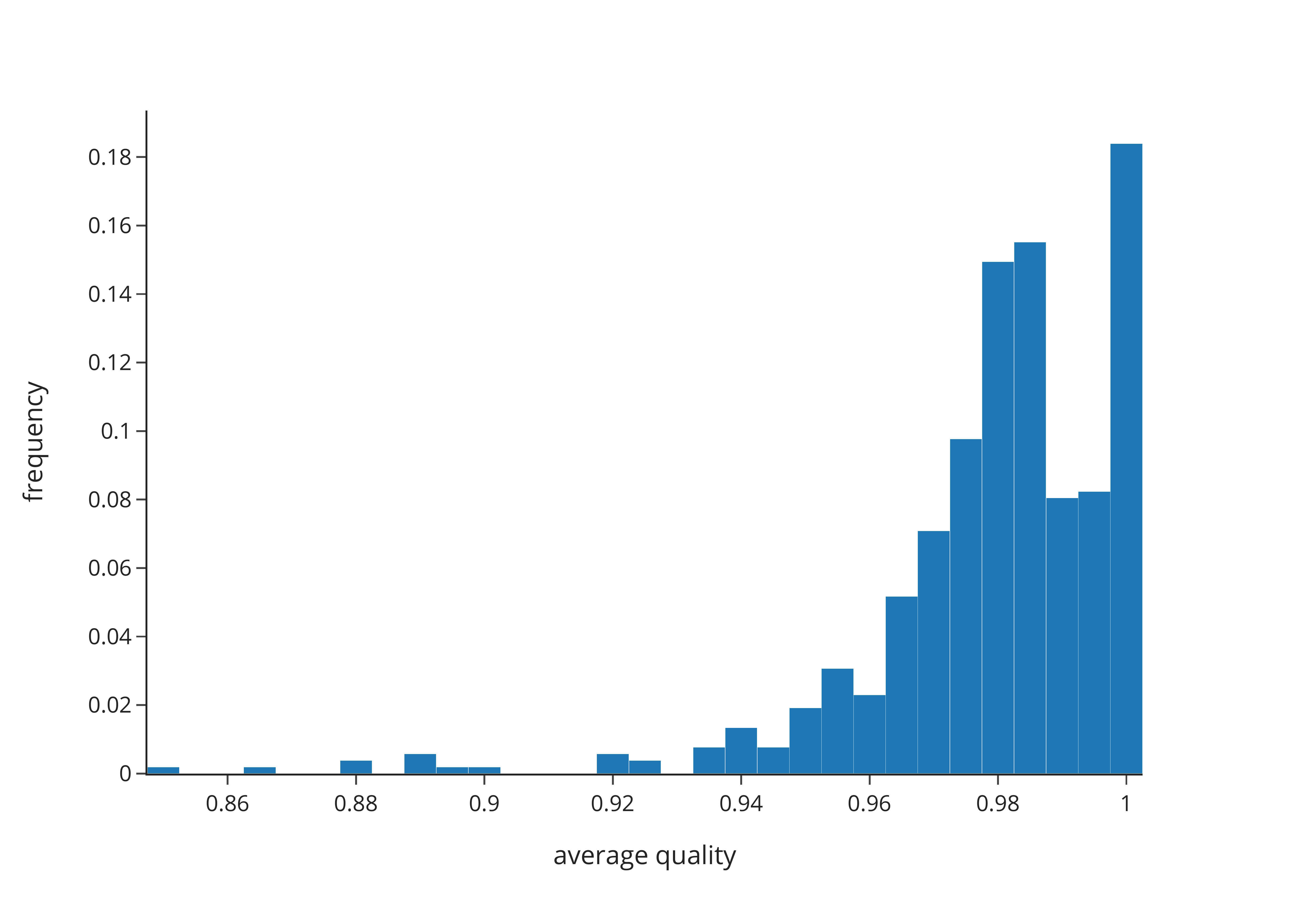}\hfill
    \includegraphics[width=.45\linewidth]{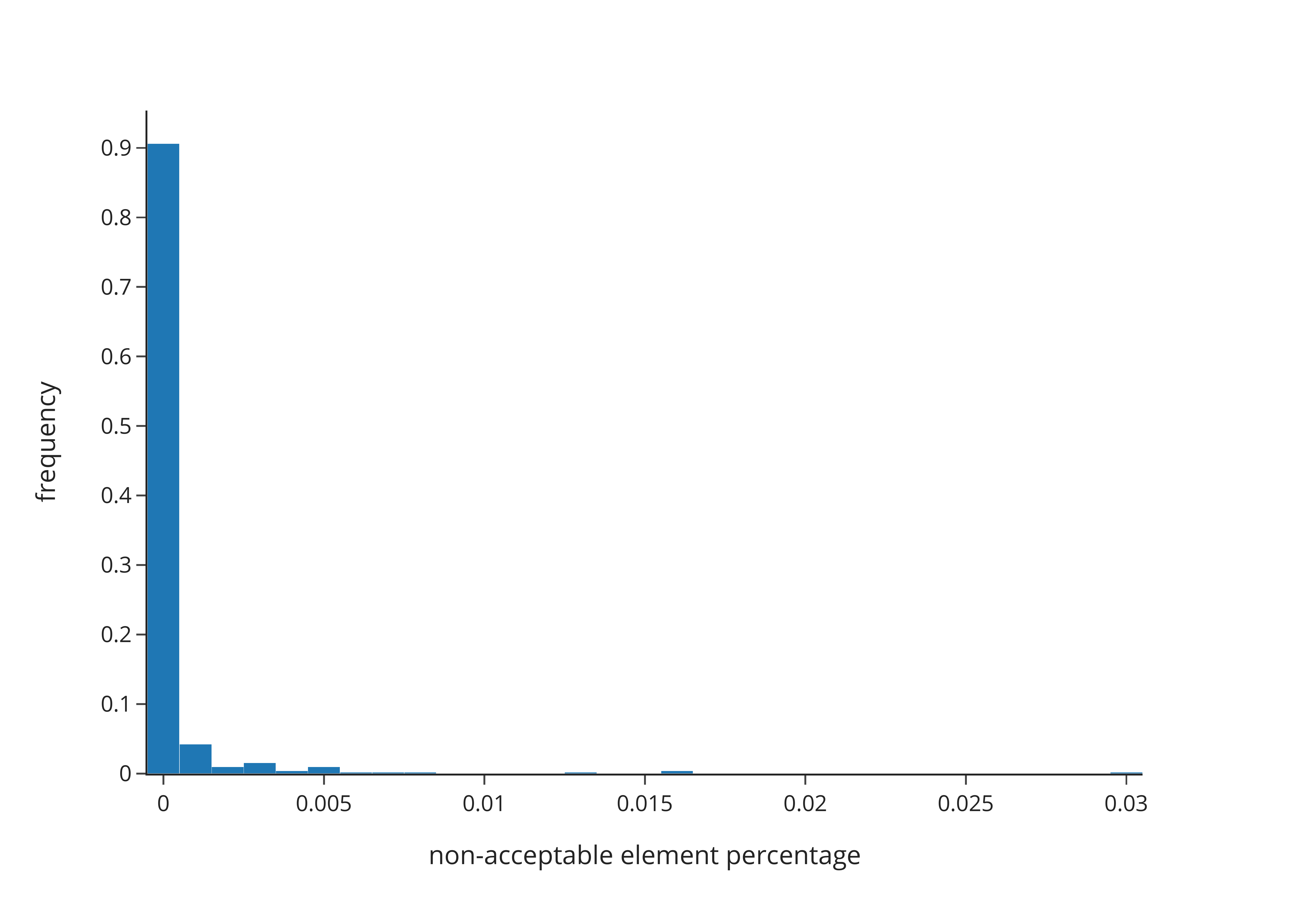}
    \parbox{\linewidth}{\centering Quadrilateral meshes.}
    \caption{Quality (left) and percentage of low quality elements (right) in our 2D dataset.}
    \label{fig:quality-2d}
\end{figure}

We create our 2D dataset using the 20k OpenClip dataset introduced in \cite{TriWild}. For each model, we create a quadrilateral mesh using the instant meshing algorithm \cite{Jakob2015Instant}, using the option to preserve and align to the mesh boundary. In Figure~\ref{fig:quality-2d}, left, we report the average element scaled Jacobian $q$~\cite[Section 4.9, Section 5.15]{quality2007} for quadrilaterals. To assess the quality of the meshes we also compute the percentage of low quality element (Figure~\ref{fig:quality-2d}, right); that is ratio of elements without $0.5< q < 2/\sqrt{3}$ for triangles and $0.3<q<1$ quadrilaterals over the total number of elements.

We then drop quadrilateral meshes with flipped elements, meshes with disconnected components, and non-manifold meshes. To keep the total running time acceptable, we drop the meshes that are too dense, which results in 522 quadrilateral meshes. For each quadrilateral mesh, we obtain a corresponding triangular mesh with the same number of vertices by splitting every quadrilateral into two triangles (we keep the shorter among the two diagonals). The meshes are normalized, so that the Reynolds number is the same for different meshes. For FE11-AB2AM2 and FE21-AB2AM2 we use $dt=0.1h^2$, while for other methods $dt=0.1h$.

\begin{figure}
    \centering
    \includegraphics[width=.45\linewidth]{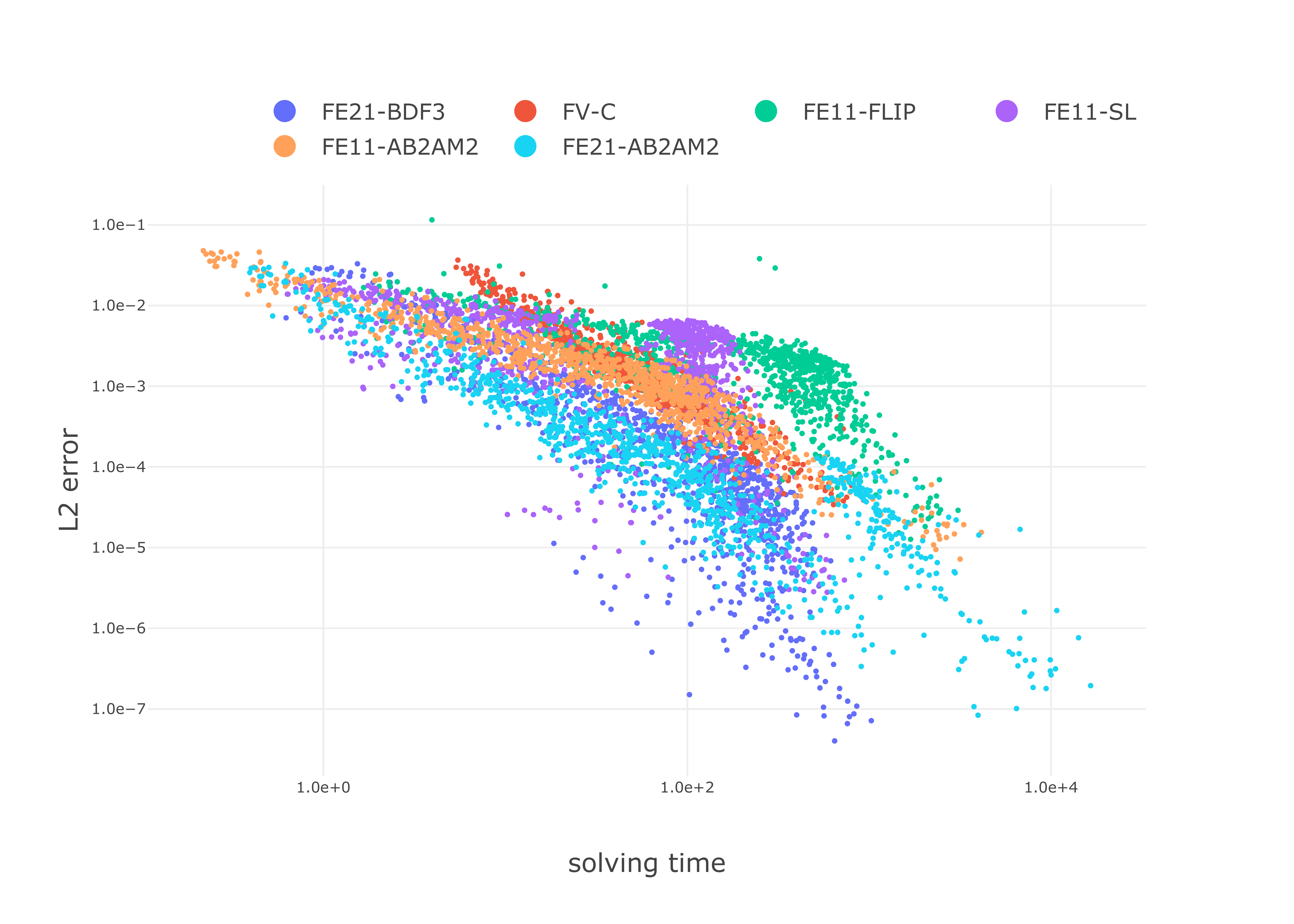}\hfill
    \includegraphics[width=.45\linewidth]{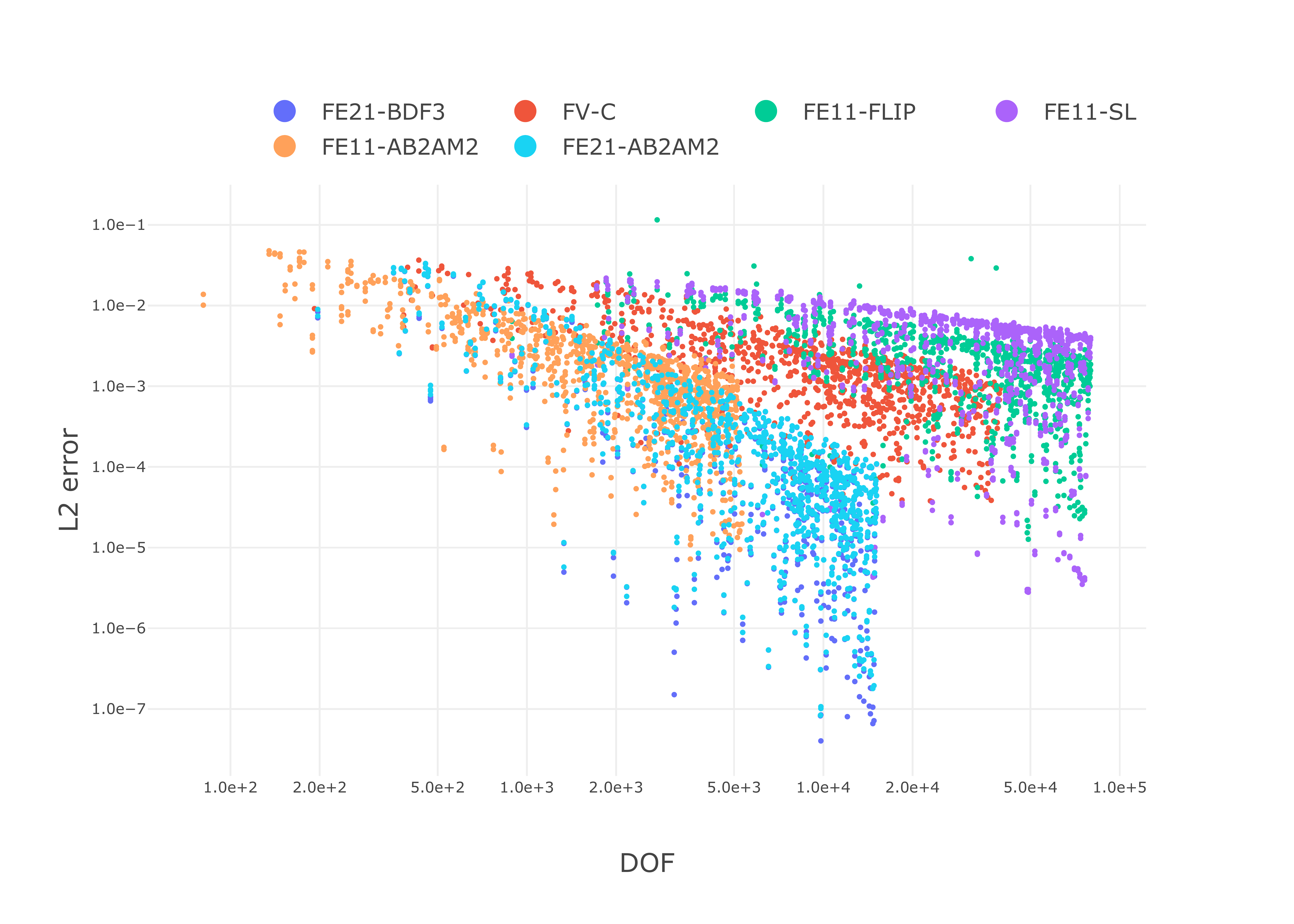}
    \caption{Comparisons among four methods on the 2D dataset.}
    \label{fig:error-time-2d-dataset}
\end{figure}

\begin{figure}
    \centering\scriptsize
    \includegraphics[width=.30\linewidth]{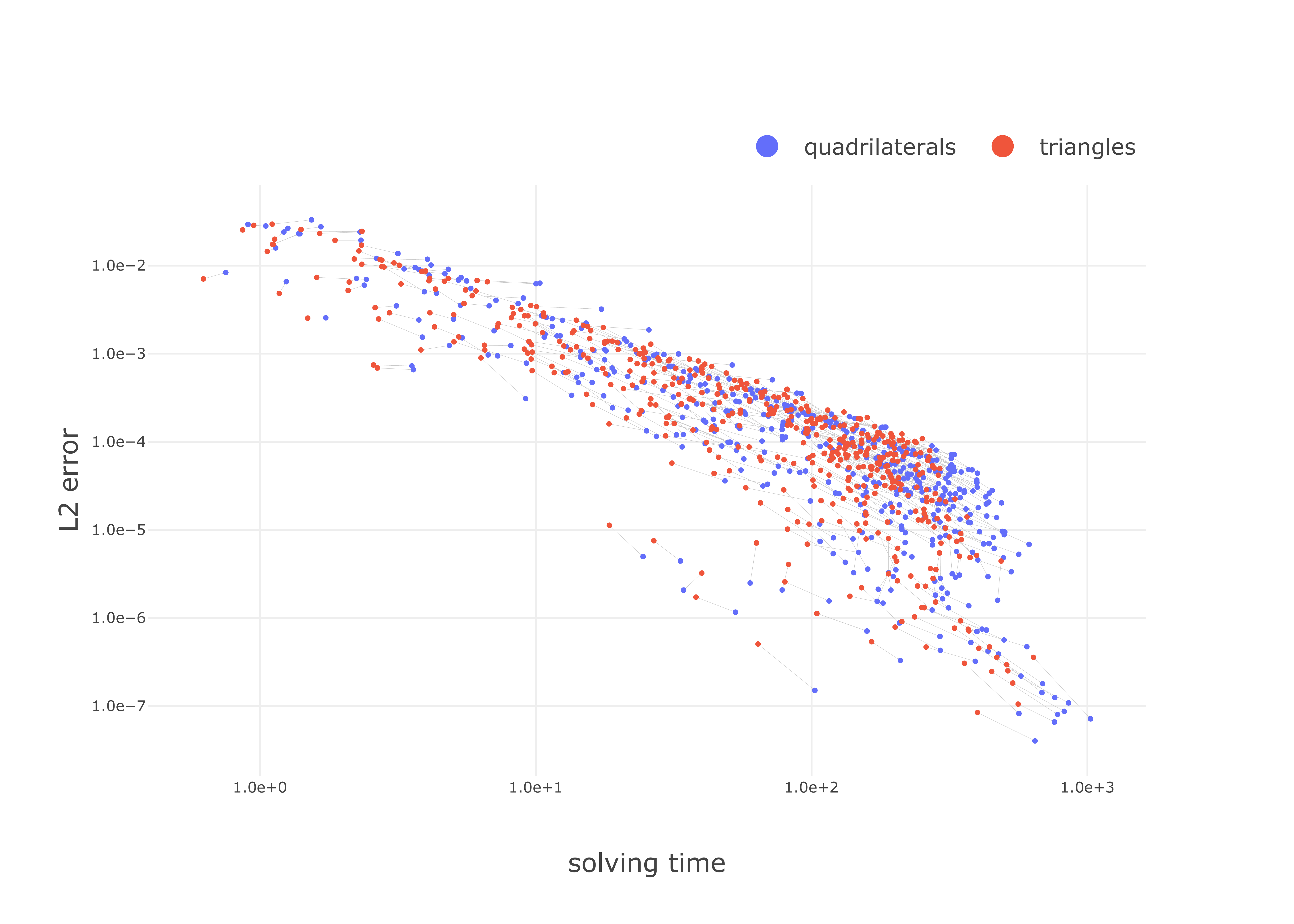}\hfill
    \includegraphics[width=.30\linewidth]{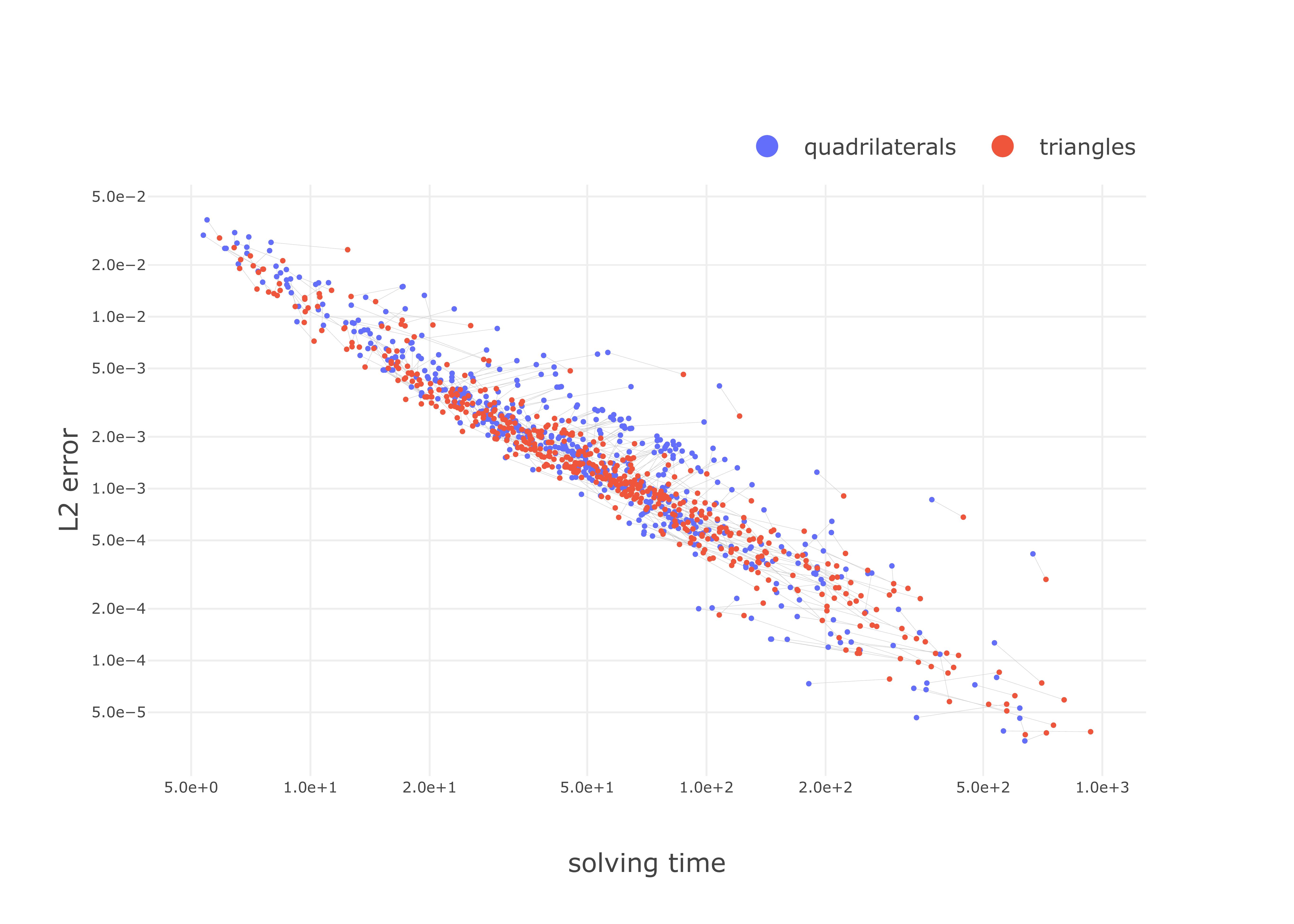}\hfill
    \includegraphics[width=.30\linewidth]{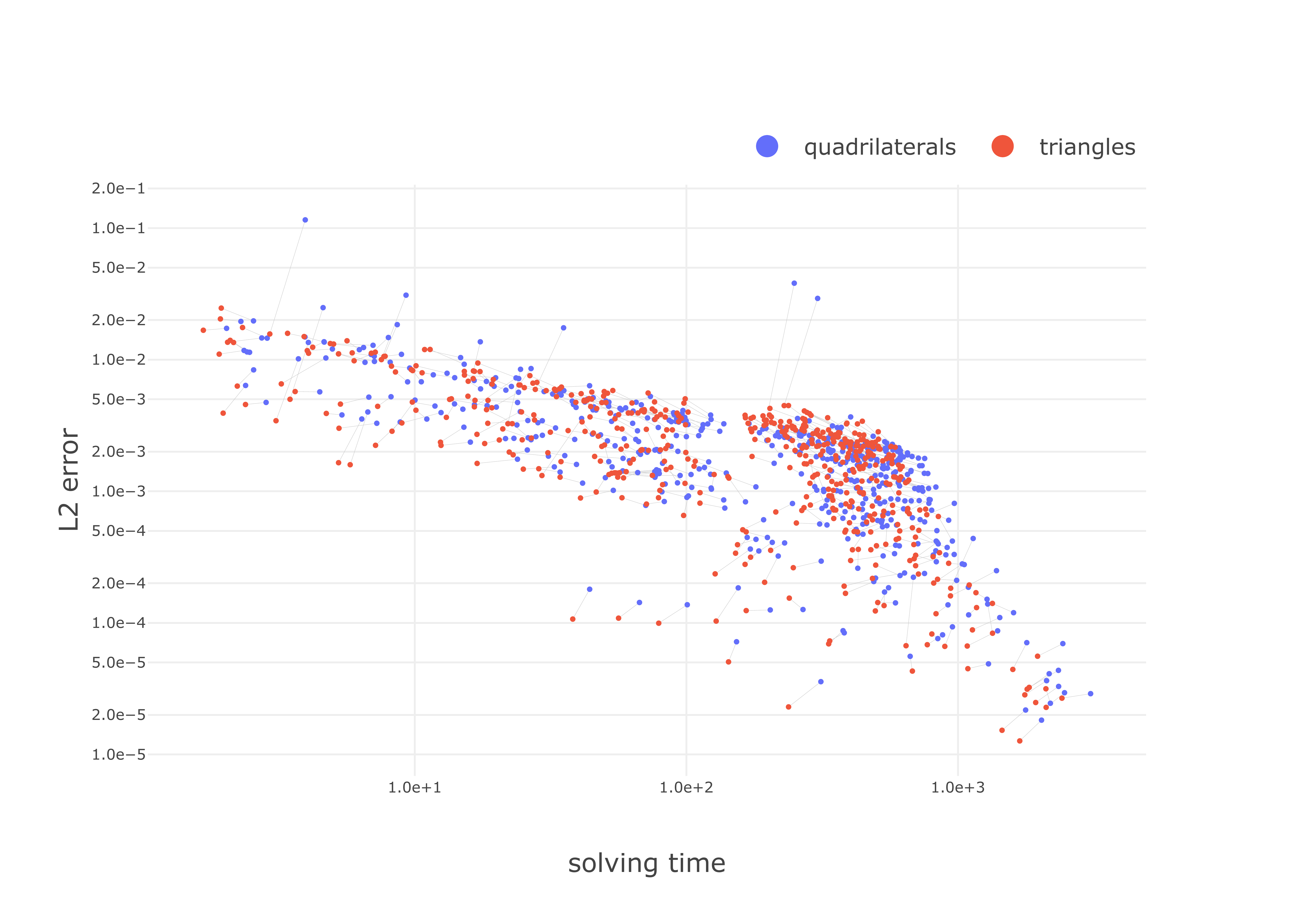}\\
    \parbox{.30\linewidth}{\centering FE21-BDF3}\hfill
    \parbox{.30\linewidth}{\centering FV-C}\hfill
    \parbox{.30\linewidth}{\centering FE11-FLIP}\\
    \includegraphics[width=.30\linewidth]{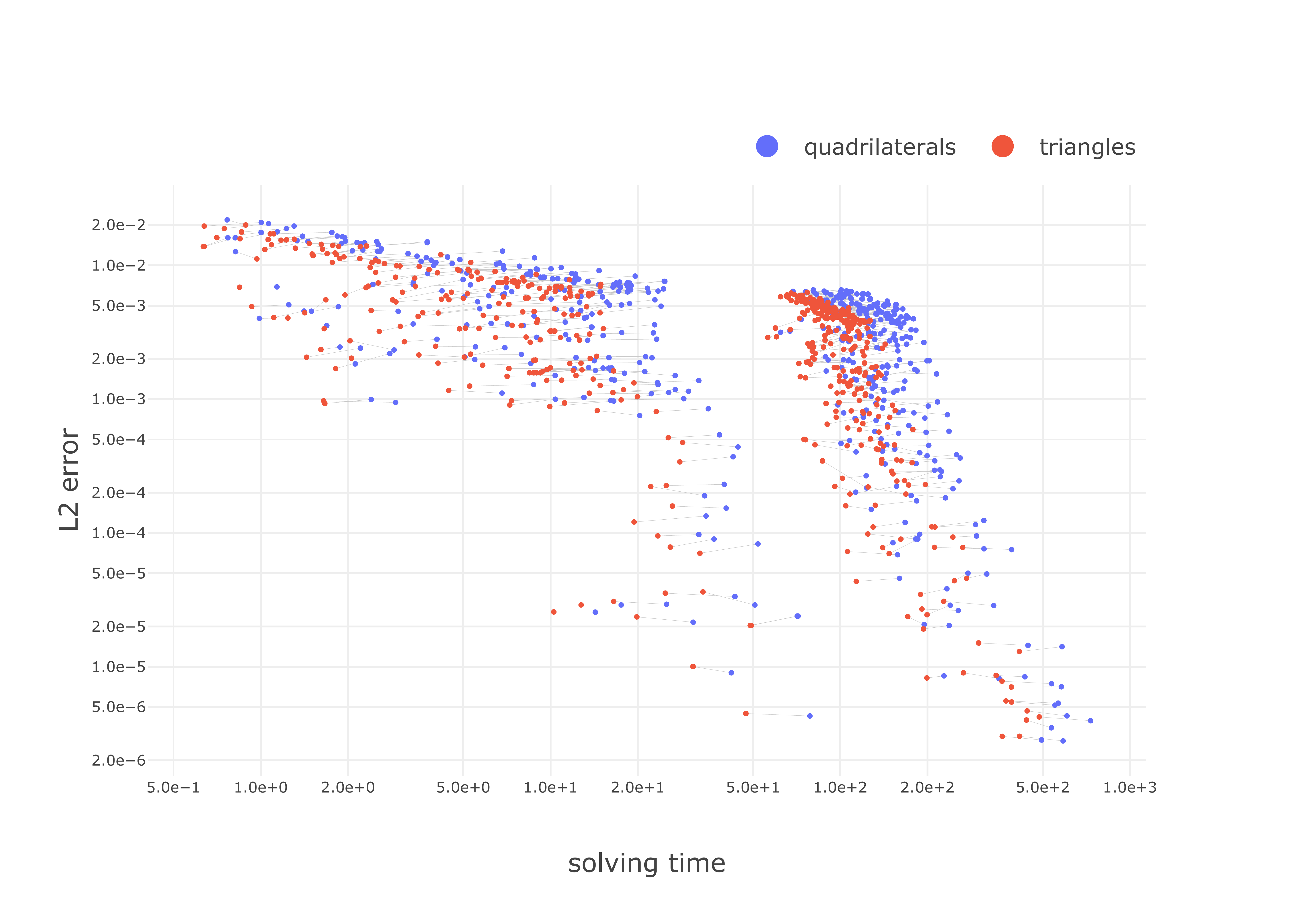}\hfill
    \includegraphics[width=.30\linewidth]{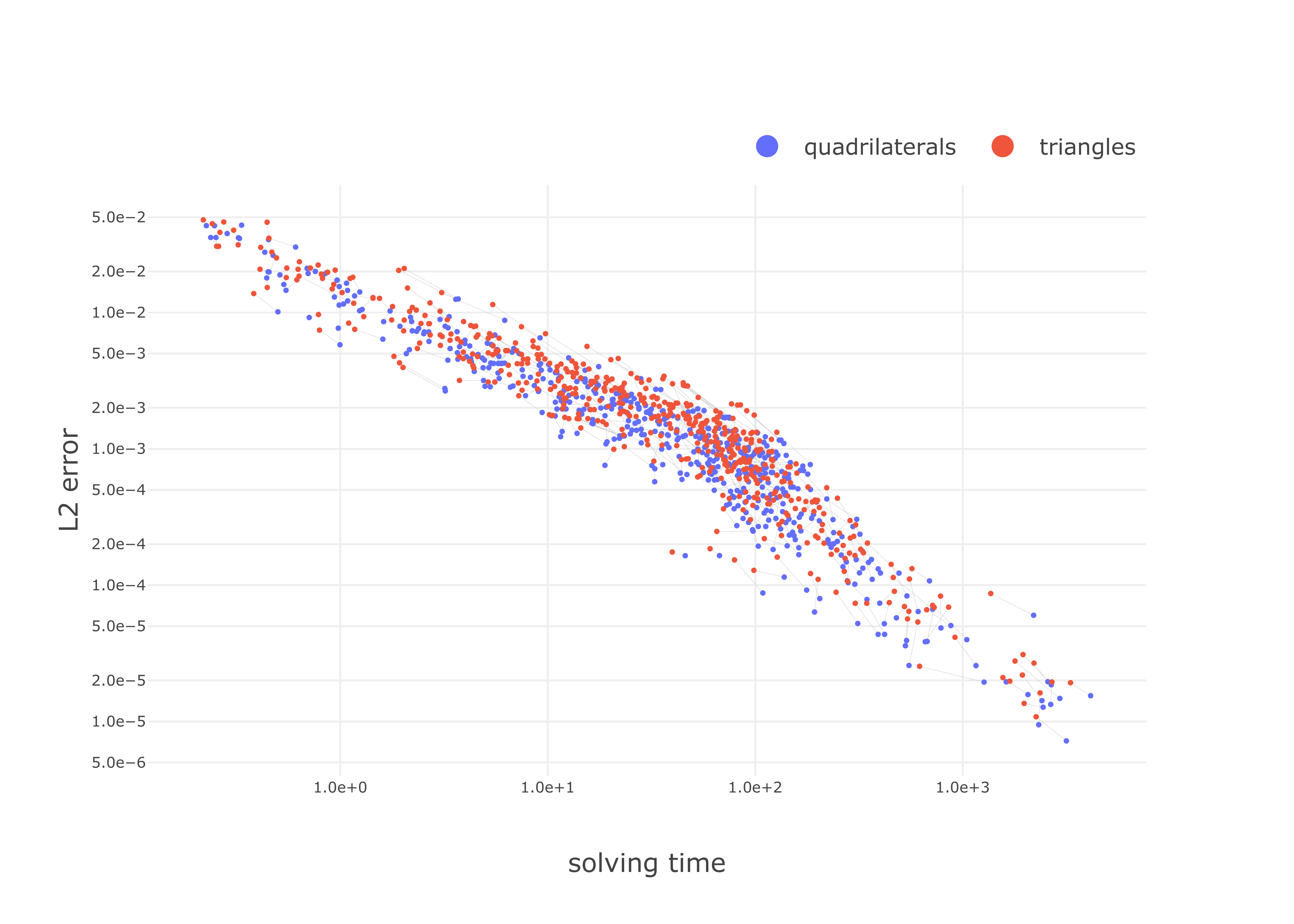}\hfill
    \includegraphics[width=.30\linewidth]{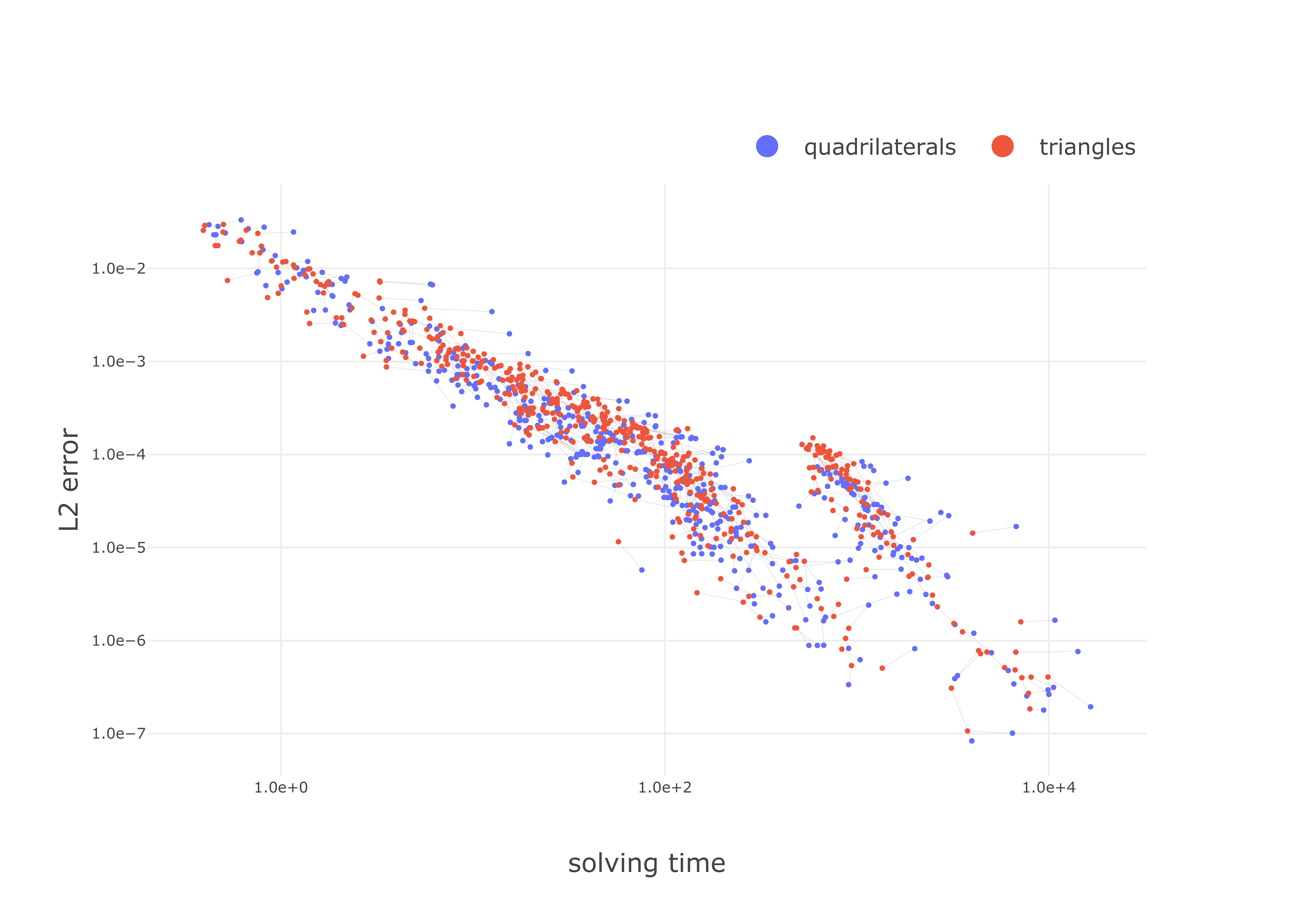}\par
    \parbox{.30\linewidth}{\centering FE11-SL}\hfill
    \parbox{.30\linewidth}{\centering FE11-AB2AM2}\hfill
    \parbox{.30\linewidth}{\centering FE21-AB2AM2}
    \caption{Comparisons between triangle and quad meshes on the 2D dataset.}
    \label{fig:error-time-2d-dataset-sl}
\end{figure}

Figure~\ref{fig:error-time-2d-dataset}, left, shows the results of the four different methods on the whole 2D dataset (triangle and quad meshes combined). The plot shows that FE21-BDF3 and FE21-AB2AM2 generally yield the lowest error for a given wall-clock time,  closely followed by FV-C for higher resolutions.
For every method, we compare the performance of triangle and quadrilateral meshes. Figure~\ref{fig:error-time-2d-dataset-sl} shows that all methods have similar performance for both mesh types.

For completeness, we also measure the performance using DOFs instead of solver time, as this is an indirect measure of problem complexity not affected by the choice of the specific numerical solver. The result does not show any major difference with respect to wall-clock time (Figure~\ref{fig:error-time-2d-dataset}, right).

\subsection{3D Dataset} 
\begin{figure}
    \centering
    \includegraphics[width=.45\linewidth]{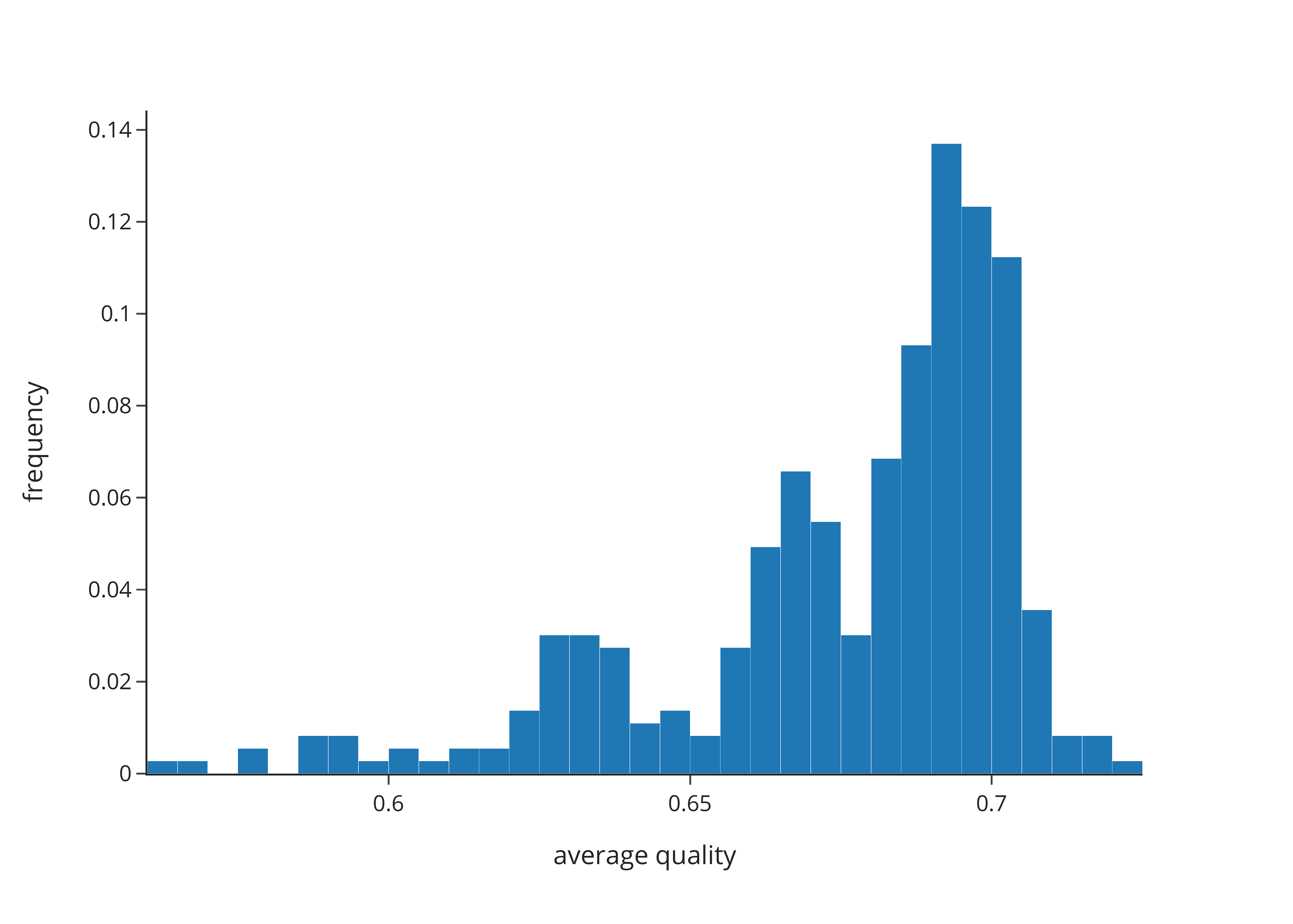}\hfill
    \includegraphics[width=.45\linewidth]{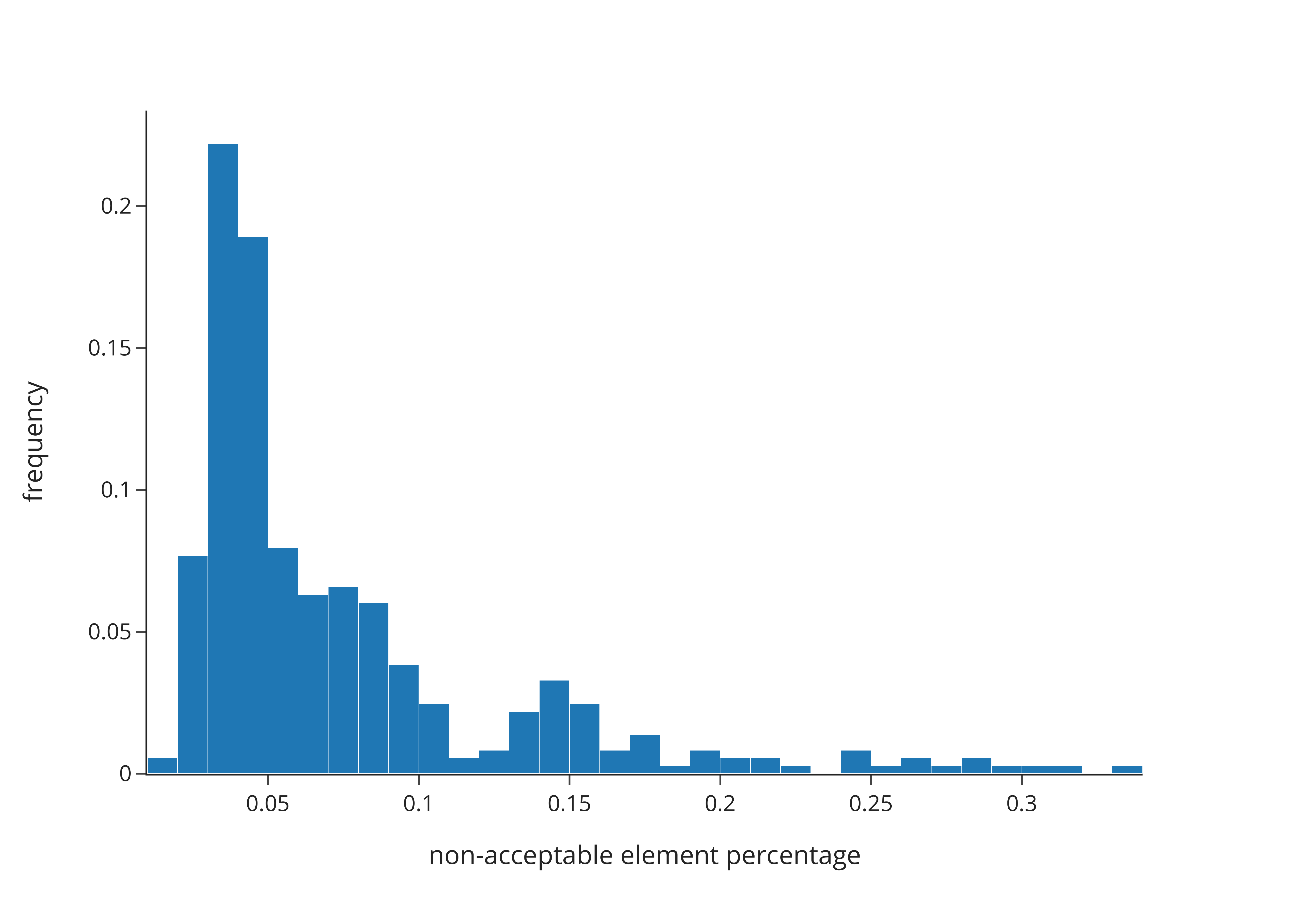}
    \parbox{\linewidth}{\centering Tetrahedral meshes.}
    \includegraphics[width=.45\linewidth]{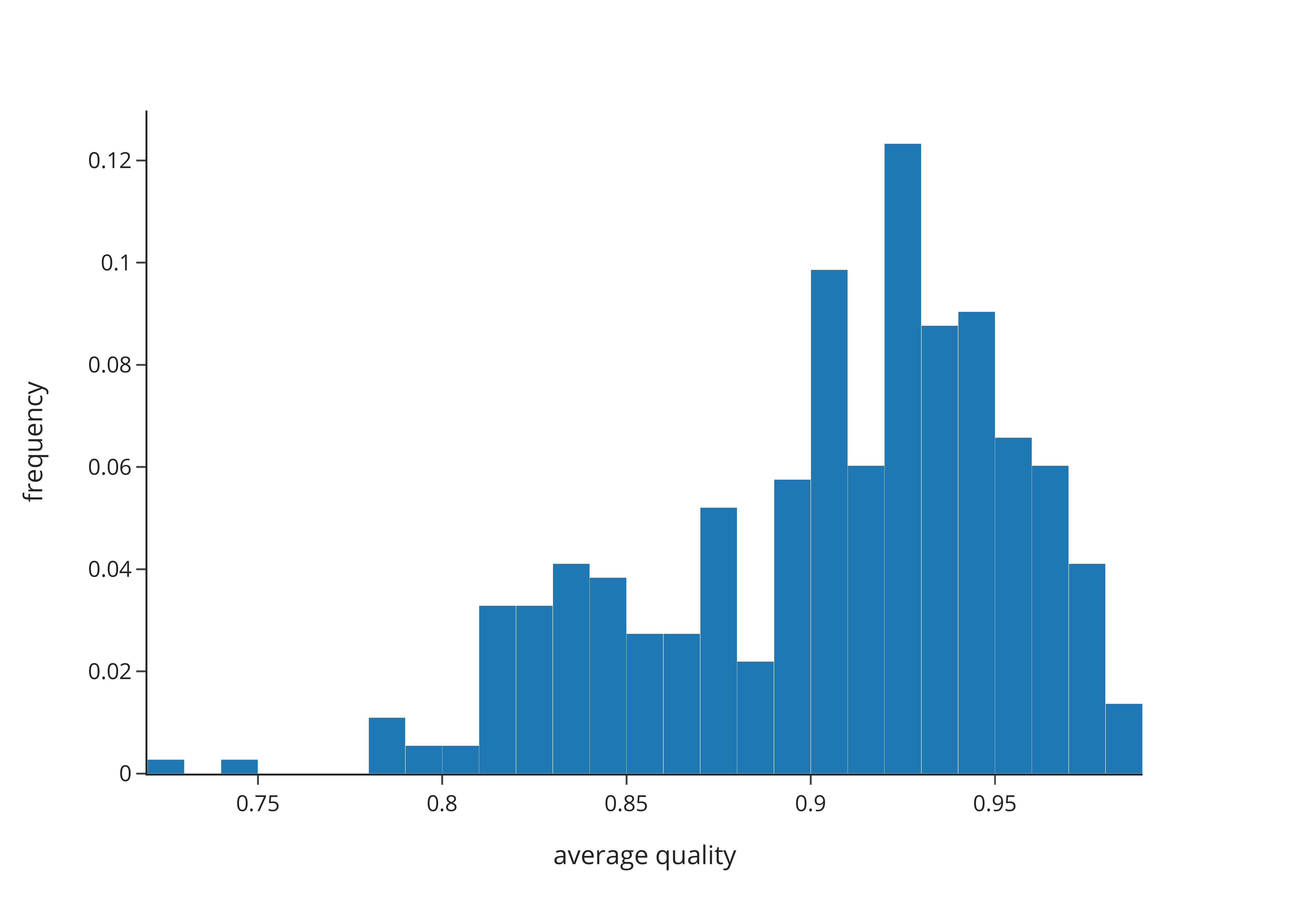}\hfill
    \includegraphics[width=.45\linewidth]{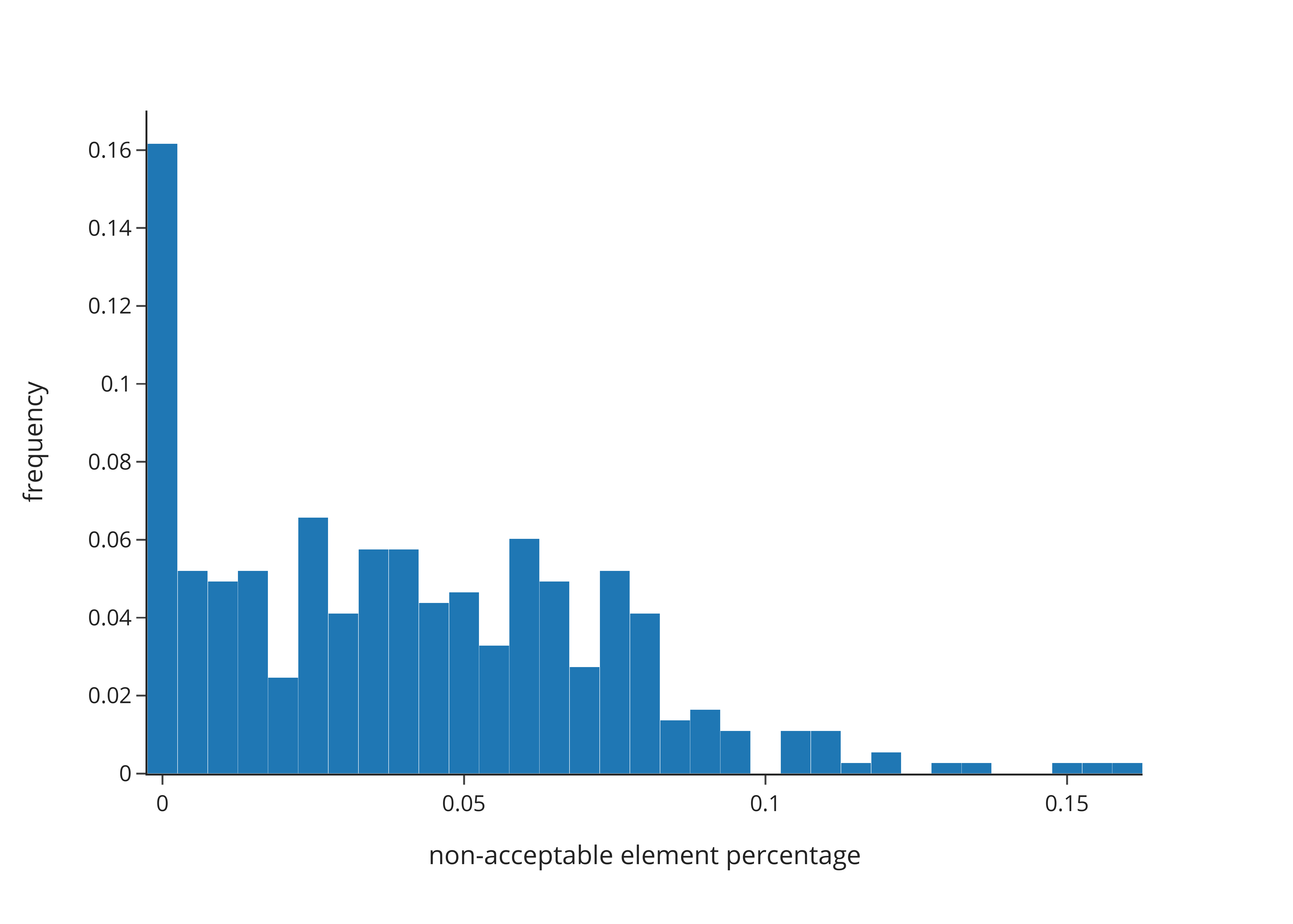}
    \parbox{\linewidth}{\centering Hexahedral meshes.}
    \caption{Quality (left) and percentage of low quality elements (right) in our 3D dataset.}
    \label{fig:quality-3d}
\end{figure}

We use hexahedral meshes from the dataset described in~\cite{tet-vs-hex}, excluding those requiring more than 150GB of memory for the linear system solve: for every mesh, we generate a tetrahedral mesh with similar mesh size. We opt for matching mesh size $h$ (differently from vertex count as in \cite{tet-vs-hex}) as for transient problems $dt$ is related to $h$, and the running time depends on $dt$. This leads to a dataset of 730 meshes. The meshes are normalized to fit the unit cube, so that the Reynolds number is consistent for different meshes. For FE11-AB2AM2 and FE21-AB2AM2 we use $dt=0.01h^2$, while for other methods $dt=0.1h$. As in 2D, in Figure~\ref{fig:quality-3d} left, we measure the scaled Jacobian $q$~\cite[Section 6.13, Section 7.11]{quality2007} of the meshes in our dataset. In Figure~\ref{fig:quality-3d} right, we report the percentage of low quality elements; an high quality element has $0.5<q<1$.

\begin{figure}
    \centering
    \includegraphics[width=.45\linewidth]{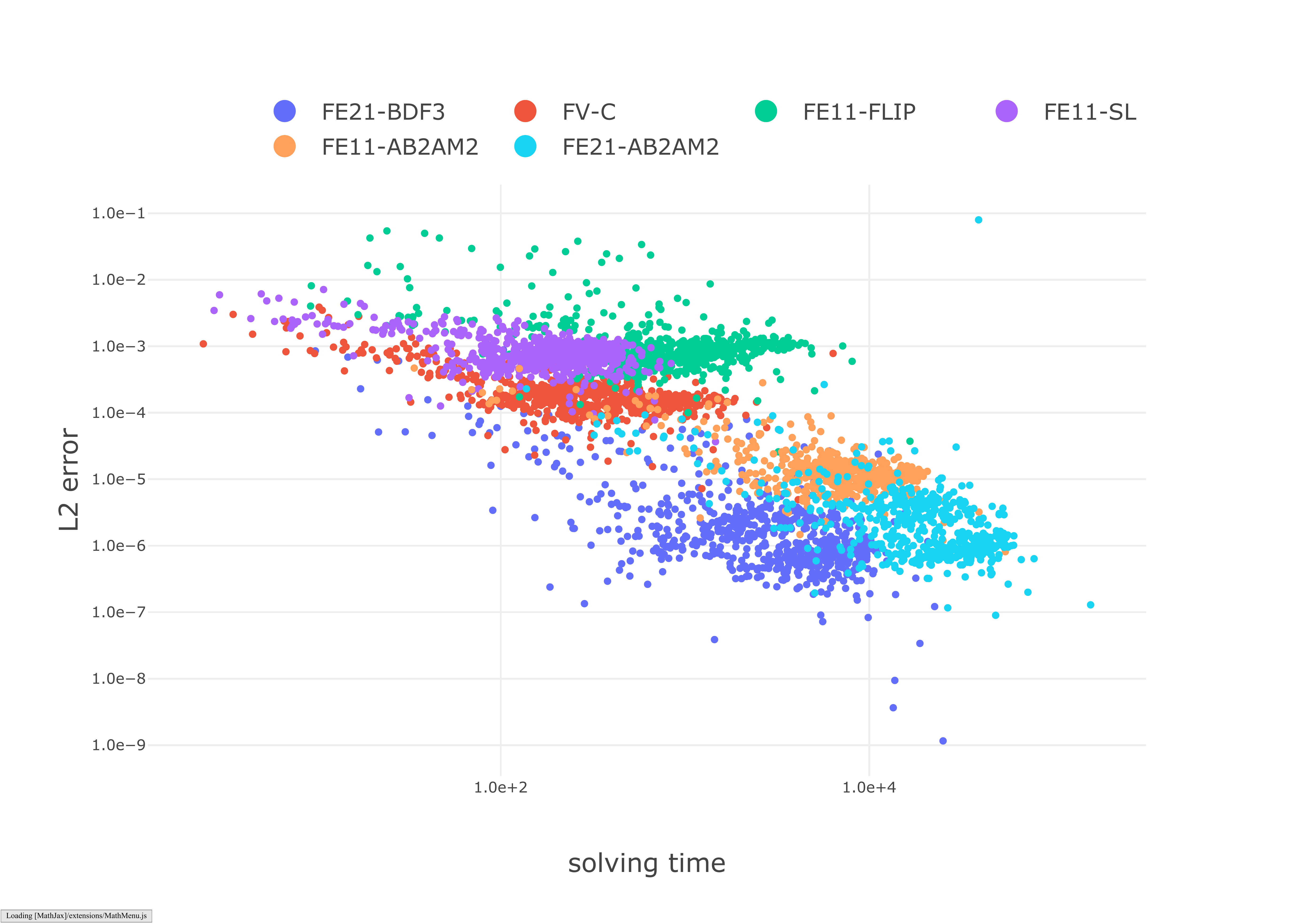}\hfill
    \includegraphics[width=.45\linewidth]{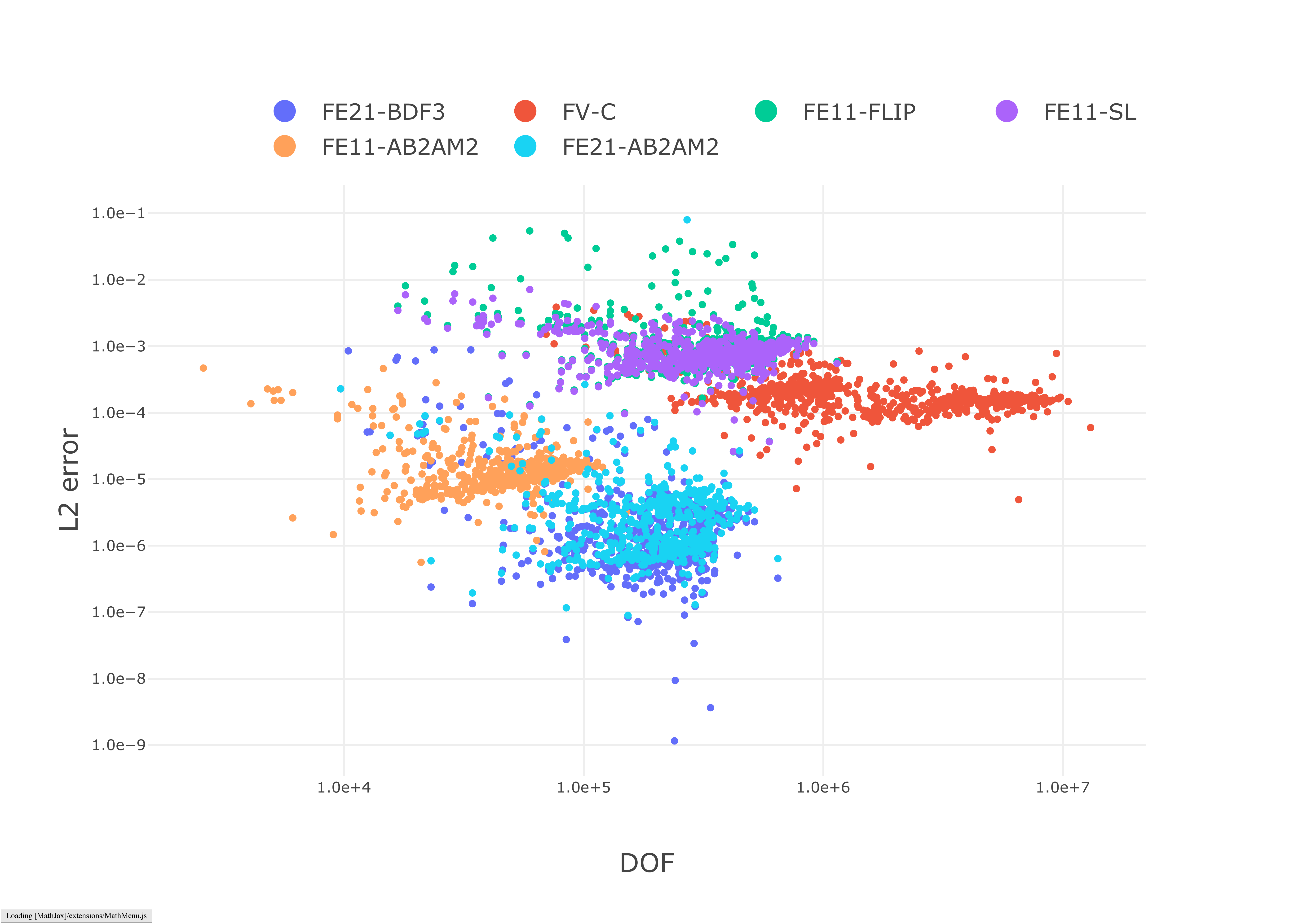}
    \caption{Comparisons among 4 methods on the 3D dataset.}
    \label{fig:error-time-3d-dataset}
\end{figure}

Figure~\ref{fig:error-time-3d-dataset} shows that moving from 2D to 3D affects all methods in a similar way: FE21-BDF3 is still the most accurate method (both in terms of time and DOFs versus error), closely followed by FV-C. FE21-AB2AM2 becomes slower as it requires more time steps.

\begin{figure}
    \centering\scriptsize
    \includegraphics[width=.3\linewidth]{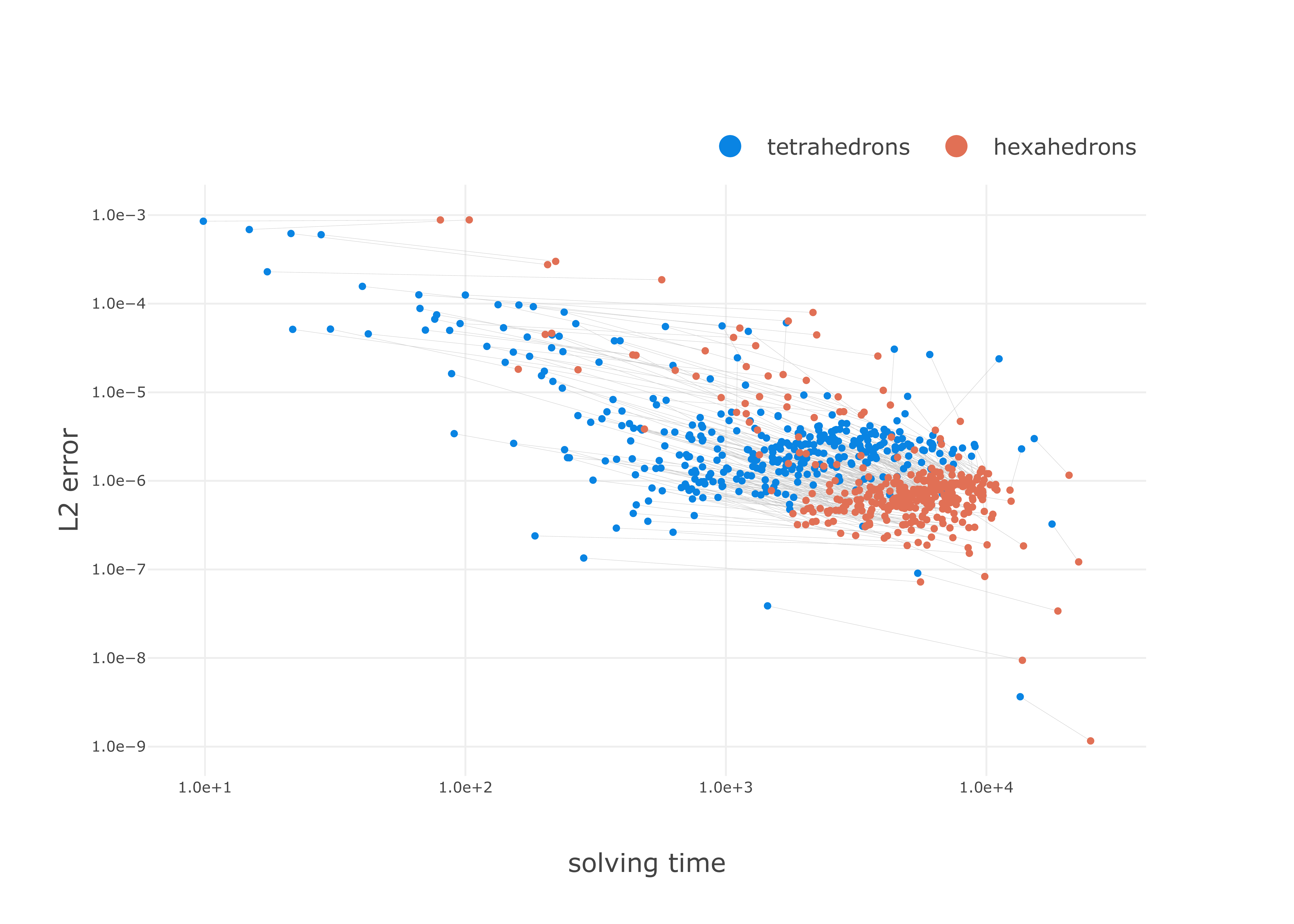}\hfill
    \includegraphics[width=.3\linewidth]{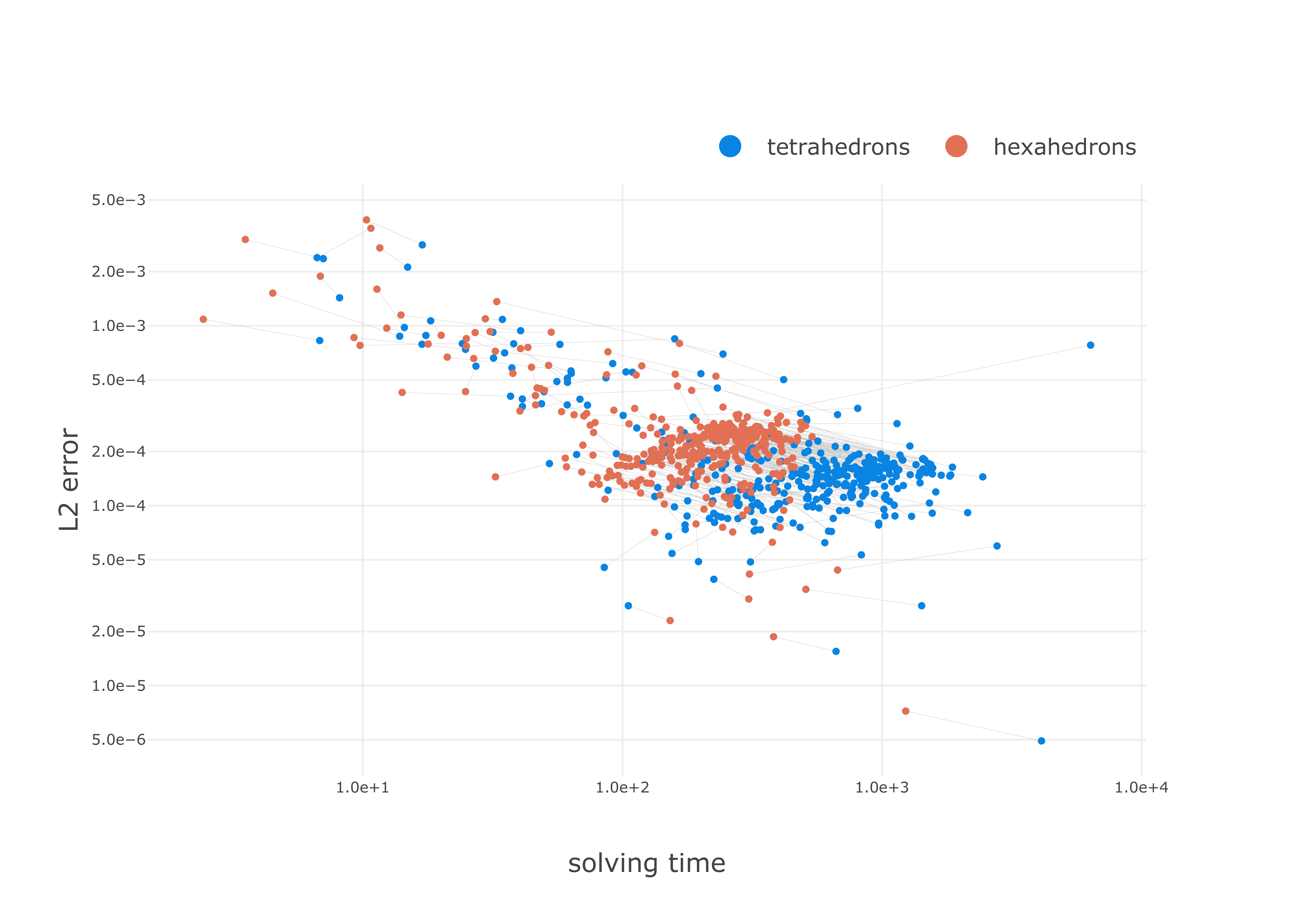}\hfill
    \includegraphics[width=.3\linewidth]{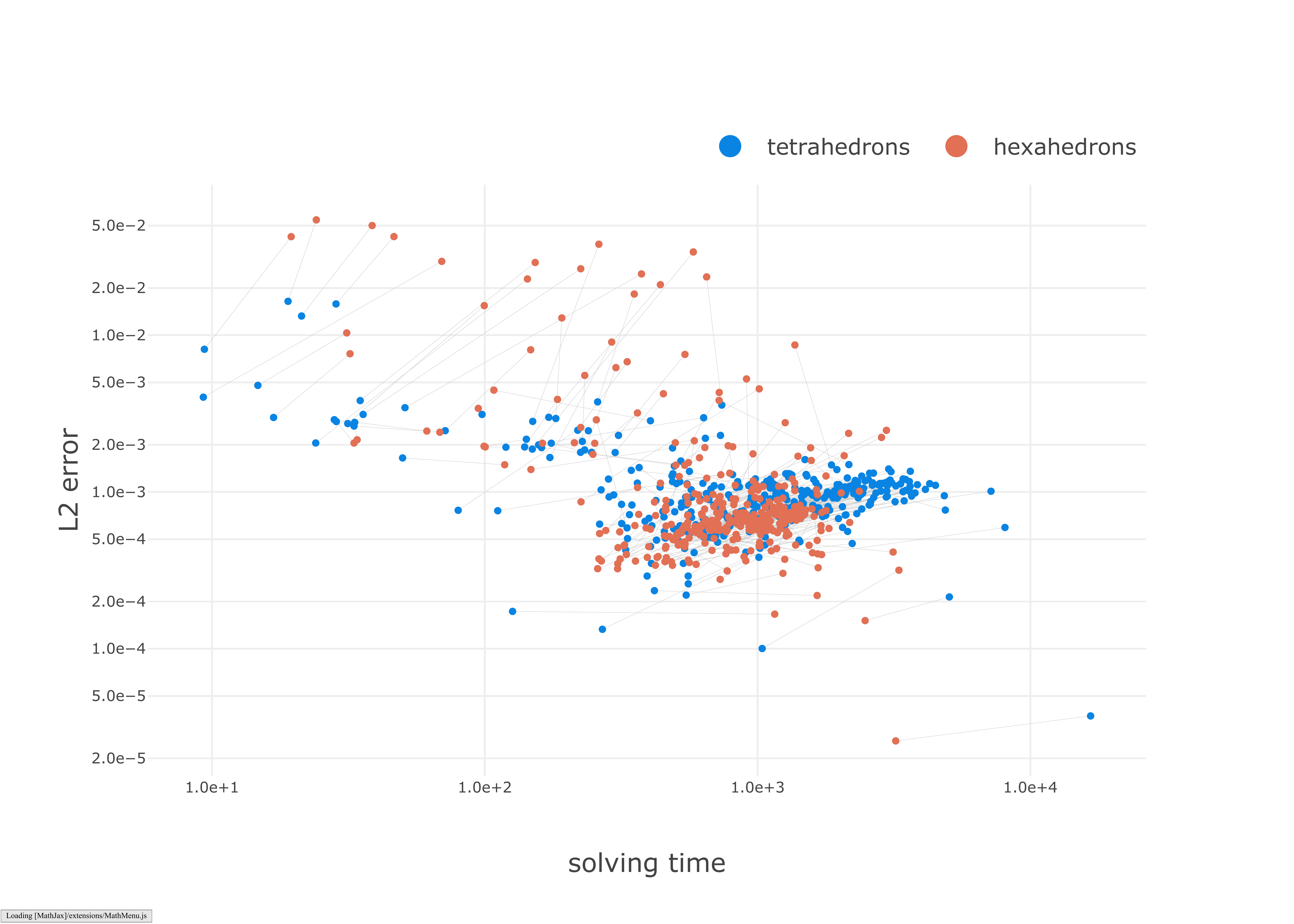}\\
    \parbox{.3\linewidth}{\centering FE21-BDF3}\hfill
    \parbox{.3\linewidth}{\centering FV-C}\hfill
    \parbox{.3\linewidth}{\centering FE11-FLIP}\\
    \includegraphics[width=.3\linewidth]{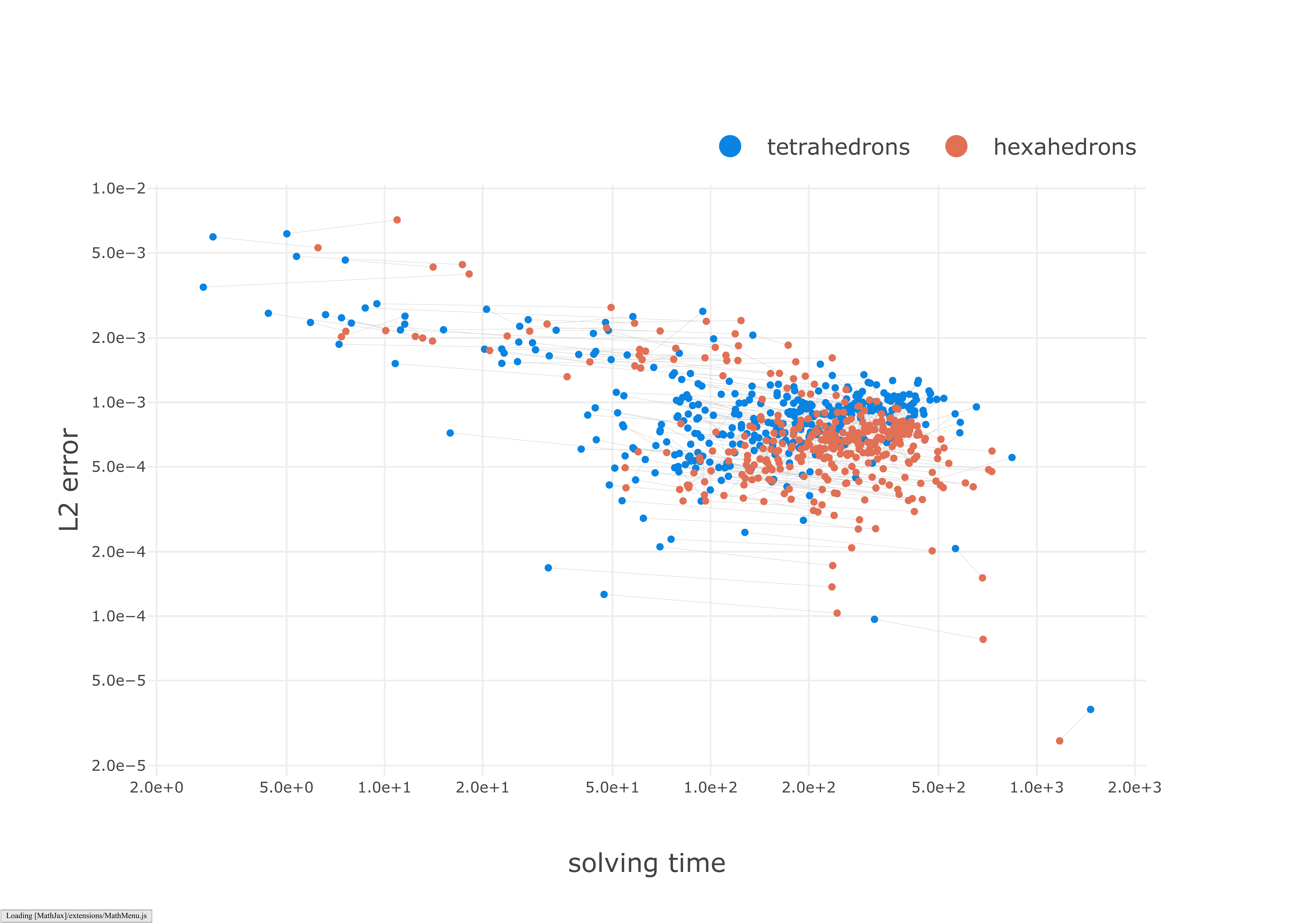}\hfill
    \includegraphics[width=.3\linewidth]{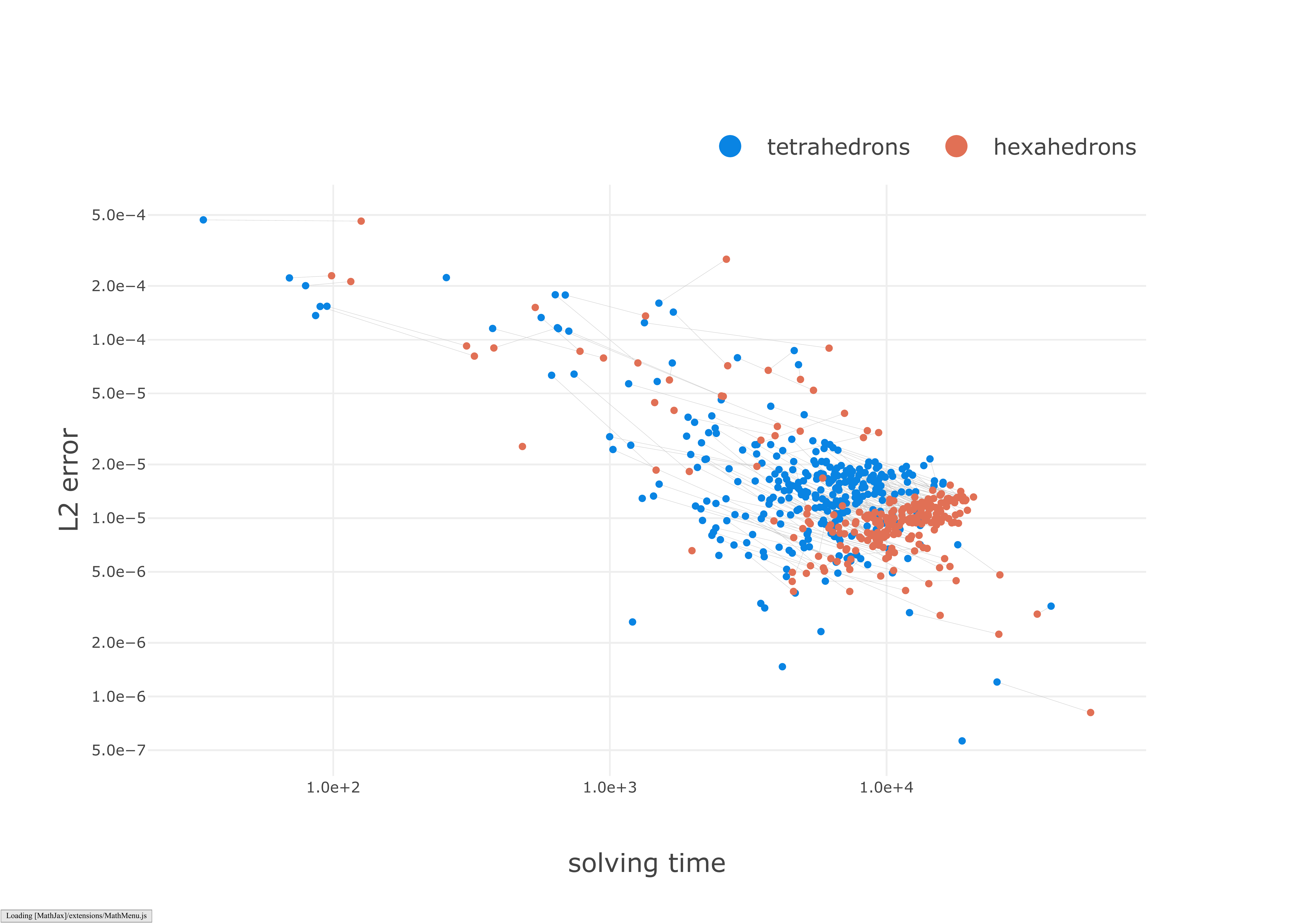}\hfill
    \includegraphics[width=.3\linewidth]{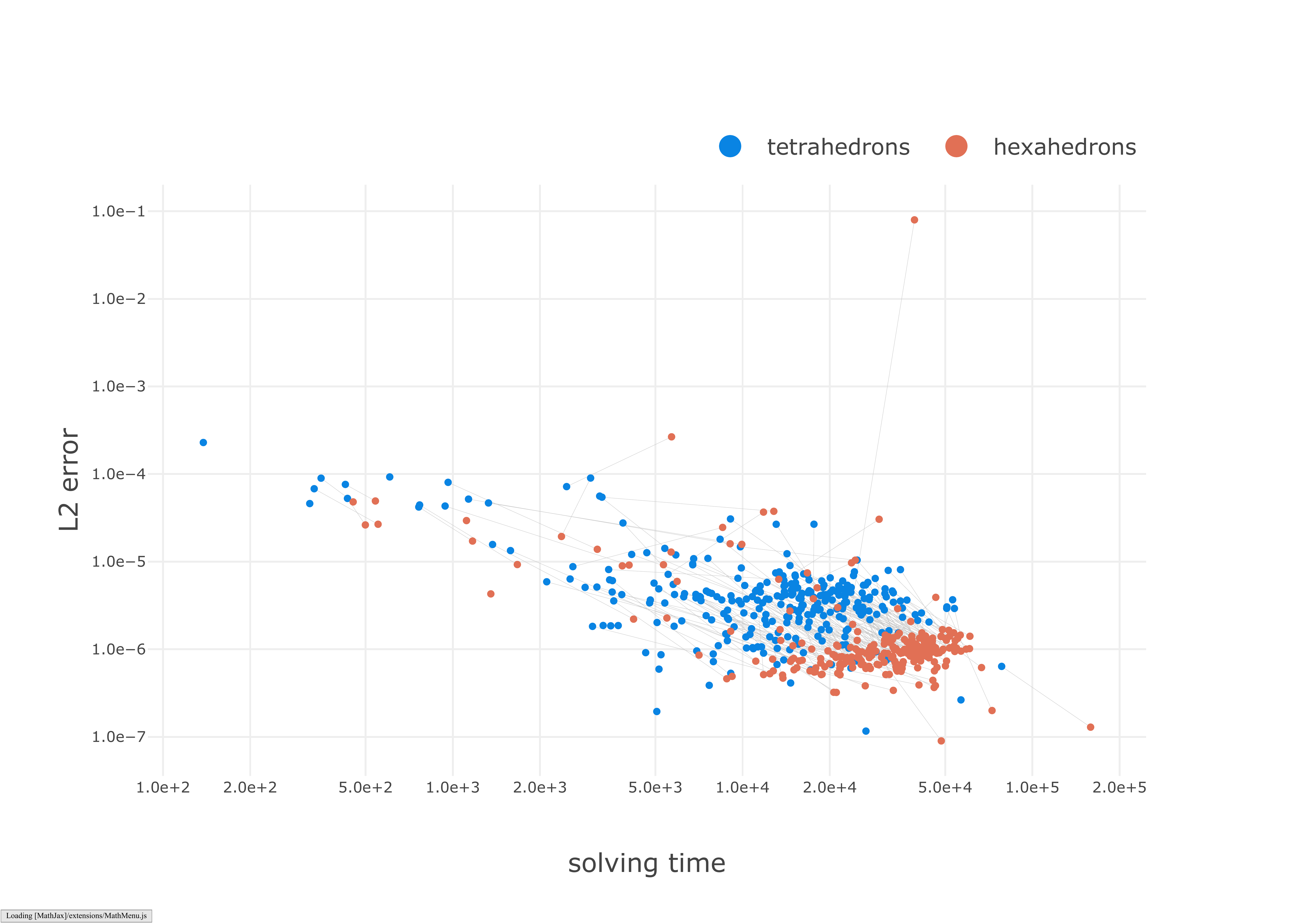}\\
    \parbox{.3\linewidth}{\centering FE11-SL}\hfill
    \parbox{.3\linewidth}{\centering FE11-AB2AM2}\hfill
    \parbox{.3\linewidth}{\centering FE21-AB2AM2}\\
    \caption{Comparisons between tetrahedral and hexahedral meshes on our 3D dataset.}
    \label{fig:error-time-3d-dataset-sl}
\end{figure}

In three dimensions, the difference between tetrahedral and hexahedral meshes is minor with the peculiar case of FE11-FLIP for which tetrahedral elements outperform hexahedral ones (Figure~\ref{fig:error-time-3d-dataset-sl}). This is likely caused by the requirement of FE11-FLIP to invert the trilinear geometric map several times for each timestep, which we compute with Newton's method.

%% file: 07-conclusions.tex
\section{Concluding Remarks}

We introduced a benchmark to quantitatively evaluate approaches for solving the incompressible Navier-Stokes equations for moderate Reynolds numbers. We included both test cases with closed-form or fabricated solutions, and more realistic scenarios for which an analytic solution is unknown. Overall, our dataset contains 1044 2D test cases and 730 3D test cases.

We believe that our work will enable future research on improving numerical methods for incompressible Navier-Stokes equations. In particular:
\begin{enumerate}
    \item New approaches can be easily tested against representative methods for the most popular  spatial and time discretizations.
    \item Scalability of new methods can be evaluated in a statistically meaningful way, using a common quantity of interest, such as running time, as reference. This applies to both problems with closed-form solutions on simple domains, and more realistic setups in complex geometries.
    \item Our preliminary study on using point-wise residuals of the strong form to evaluate the quality of methods in a discretization-agnostic way shows that discretization artefacts are an important factor for pointwise quantities, especially for low-order methods  (Appendix~\ref{app:residual}), confirming that extracting accurate pointwise velocities and pressures from these solutions is an important problem on its own.
\end{enumerate}

\subsection{Venues for Future Work}

Overall, we could not identify an existing solver allowing to solve larger-scale problems with high-order convergence rates (both on the interior and on the boundary of the domain), which has potential for solving  Navier-Stokes equations more efficiently, for a given target accuracy, especially for higher viscosities and in smoother domains.
This could be achieved by extending existing methods in different ways:

\paragraph{Linear System Solver for FE21-BDF3} The FE21-BDF3 algorithm performs consistently well in all our tests, with the exception of struggling to scale to large systems. Developing a preconditioner for this system to address the scaling issues is a promising direction for future work.

\paragraph{Implicit High-Order Splitting} The approaches based on high-order splitting perform very similarly to FE21-BDF3, while allowing to timestep problems with a large number of DOFs. However, since they are based on explicit time integration, they require small time steps, also making them impractical for large problems. We are not aware of high-order  implicit-splitting time integration schemes for this type of discretizations, making this an interesting venue for future work.

\paragraph{Boundary Conditions and Accuracy} Different method impose boundary conditions in different ways, with a significant impact on the accuracy close to the domain boundary. Improving boundary condition formulations may be a computationally inexpensive approach to achieving better performance.

\paragraph{Benchmark Extension} Our benchmark is restricted to incompressible, laminar flows. We believe extending our experimental approach to compressible flows, boundary layer meshes, and turbulent regimes are all important and exciting directions for future work.

\subsection{Open-Source Release} 
The code, data, and scripts to reproduce all our results will be released with an open source licence after publication.

%% file: 09-appendix.tex
\section[L2 Residual]{$L^2$ Residual}\label{app:residual}
$L^2$ norm of the pointwise PDE residual can be, in principle, used to measure the accuracy of different numerical methods in the cases when no analytic solution known, as it directly checks how well the original PDE is satisfied. 
The complication in direct residual evaluation is that it requires approximation of first and second derivatives of the solution, which is a non-trivial problem for commonly used low-order methods. 

The $L^2$ norm of the residual can be defined as
\begin{equation*}
    R^2(u,p)=
    % \sqrt{
    \int_0^T\int_{\Omega}\left\|\displaystyle\rho\frac{\partial u}{\partial t} + \rho(u \cdot \nabla) u - \nu \Delta u + \nabla p\right\|^2 
    % }
\end{equation*}
To compute this residual for a given numerical solution $u_h, p_h$, we need to evaluate the derivatives and integrate them over both time and space.   To treat all methods uniformly we sample velocities and pressure on vertices and compute a cubic spline approximation to the solution.

% The derivatives of the interpolant are considered as the derivatives of the numerical solution and used to compute the integral.

\begin{figure}
    \centering\scriptsize
    \includegraphics[width=.48\linewidth]{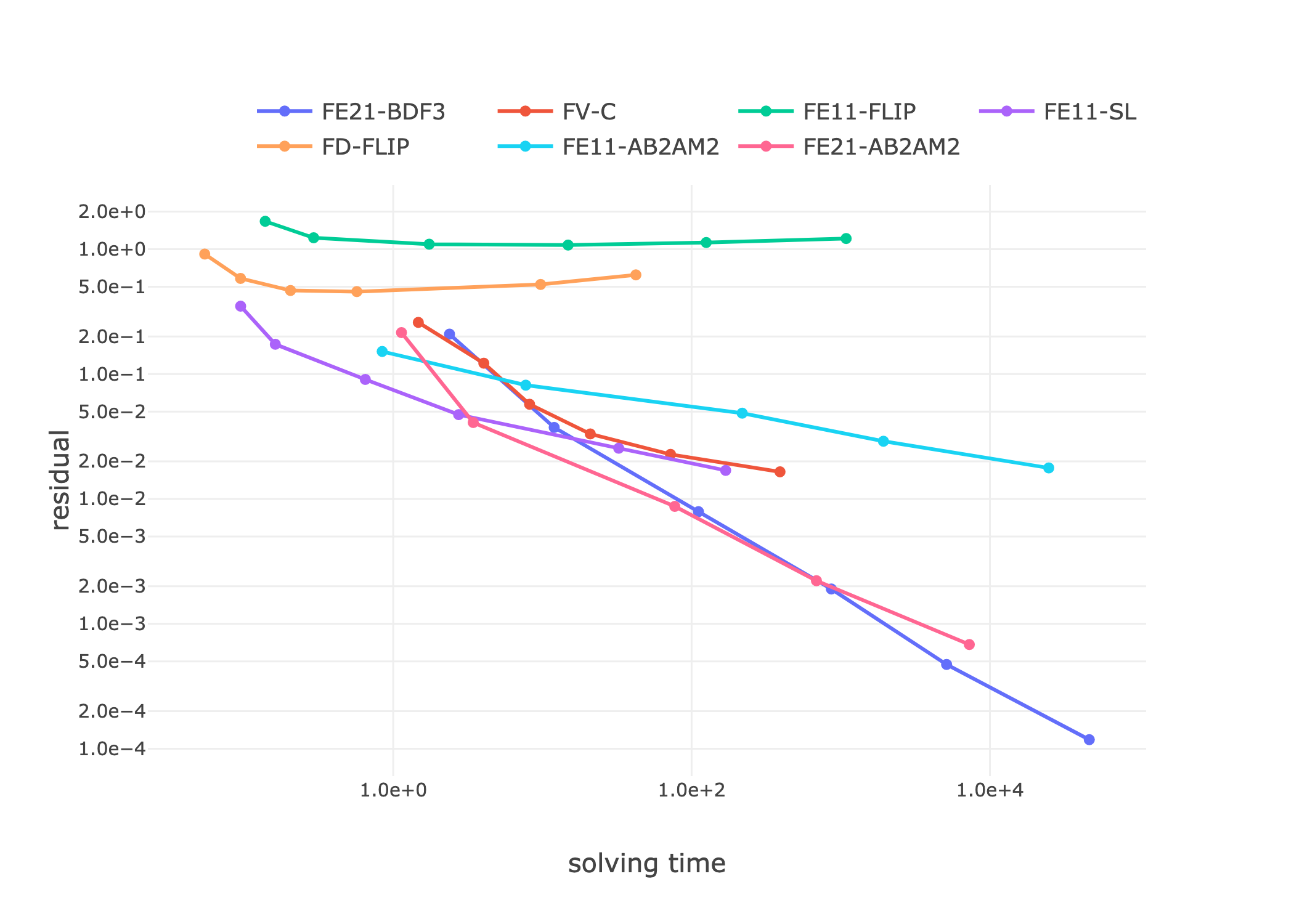}\hfill
    \includegraphics[width=.48\linewidth]{pics/2d-plots/pics/tgv-2d-quad-time-error}\\
    \parbox{.48\linewidth}{\centering Residual}\hfill
    \parbox{.48\linewidth}{\centering $L^2$ error}
    \caption{Convergence plots of residual and $L^2$ error against wall-clock time.}
    \label{fig:residual-time}
\end{figure}

\begin{figure}
    \centering\scriptsize
    \includegraphics[width=.48\linewidth]{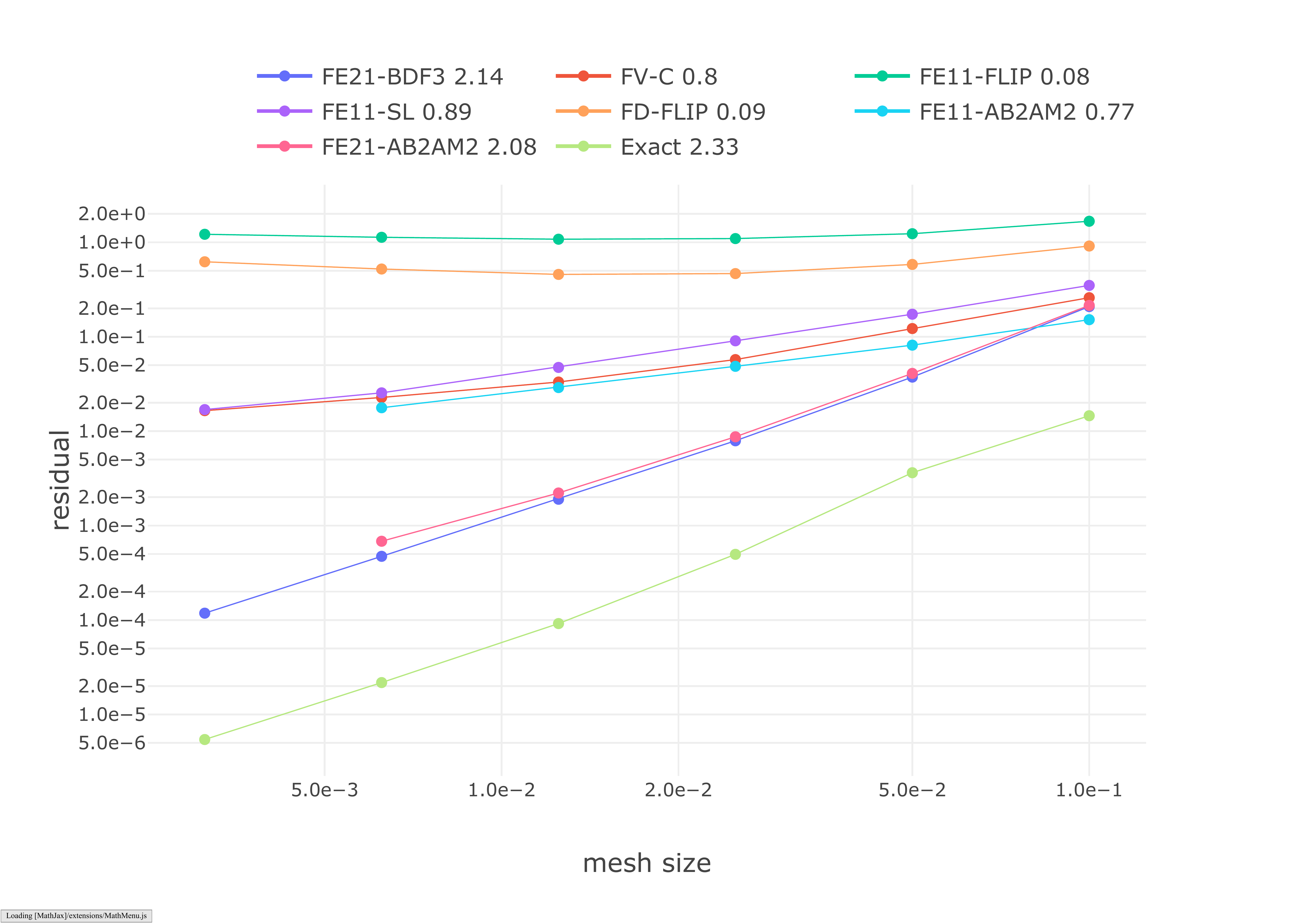}\hfill
    \includegraphics[width=.48\linewidth]{pics/2d-plots/pics/tgv-2d-quad-h-error}\\
    \parbox{.48\linewidth}{\centering Residual}\hfill
    \parbox{.48\linewidth}{\centering $L^2$ error}
    \caption{Convergence plots of residual and $L^2$ error against mesh size.}
    \label{fig:residual-h}
\end{figure}

To evaluate how the derivative approximation affects $R$, we compute it for the Taylor-Green vortex and compare it against the $L^2$ error
(Figure~\ref{fig:residual-time}). To make sure that the derivative error is not dominant, we also plot the residual computed from sampled exact solution as reference (Figure~\ref{fig:residual-h}).

In Figure~\ref{fig:residual-h}, we note that all seven methods have different convergence rates between two plots. Besides, in Figure~\ref{fig:residual-time}, only FE21-BDF3 is consistent between the two plots.

The reason why the residual of both FE11-FLIP and FD-FLIP is not decreasing under refinement is, most likely, because these methods use randomly sampled particles. The random sampling introduces noise in the numerical solution, which is then magnified by our local spline-based scheme for derivative approximation. For other methods, this may be due to discrete solutions not yielding sufficiently accurate estimates of the derivatives needed for residual evaluation. 

%%%%%%%%%%%%%%%%%%%%%%%%%%%%%%%%%%%%%%%%%%%%%%%%%%%%%%%%%%%%%%%%%%%%%%%%%%%%
%%%%%%%%%%%%%%%%%%%%%%%%%%%%%%%%%%%%%%%%%%%%%%%%%%%%%%%%%%%%%%%%%%%%%%%%%%%%
%%%%%%%%%%%%%%%%%%%%%%%%%%%%%%%%%%%%%%%%%%%%%%%%%%%%%%%%%%%%%%%%%%%%%%%%%%%%
\section{Experimental Setup}\label{app:setup}

All experiments are run on  individual nodes of an HPC cluster each using two Intel Xeon Platinum 8268 24C 205W 2.9GHz Processors and 48 threads. We limit the runtime to 48 hours. All  methods considered in this study, with the exception of the FV method that uses Ansys Fluent~\cite{ansys2016ansys}, and FD-FLIP, are part of the PolyFEM library \cite{polyfem}, and use Pardiso~\cite{code:pardiso-a,code:pardiso-b,code:pardiso-c} direct solver.

\paragraph{FV-C}
We use the pressure-based solver and the laminar model in Ansys Fluent~\cite{ansys2016ansys}. The coupled scheme is used for the pressure-velocity coupling. For the spatial discretization, we use the least square cell-based scheme for gradients, the second-order scheme for pressure, and the second-order upwind scheme for momentum. We use the second-order implicit scheme provided in Fluent as the transient formulation. The convergence criteria is set to $10^{-6}$ for the continuity and velocity residual. Other parameters are set to default values.

\paragraph{FE21-BDF3}
The non-linear equation is solved via Newton's method, whose tolerance is set to $10^{-8}$. To ease convergence, we first use Picard iterations~\cite{Elman:2005} up to $10^{-3}$ tolerance.

\paragraph{FE11-SL, FE11-FLIP}
We briefly overview \cite{stam1999stable},
focusing on the specific discretization choices that we use in our comparison.
For time step $n$ with velocity $u^n$, the advection step aims at integrating
\begin{equation*}
    \frac{\partial u}{\partial t} = -(u \cdot \nabla) u
\end{equation*}
over time by $\Delta t$ to obtain an intermediate velocity $\bar{u}^n$, where semi-Lagrangian-type methods \cite{sawyer1963semi,maccormack2003effect,dupont2003back,qu2019efficient} or particle-in-cell methods \cite{harlow1962particle,brackbill1986flip,jiang2015affine,fu2017polynomial} are often applied.
% \DZ{while this is explained in graphics in this way, the split equations are not the correct equations for fluid, esp. if the same variable is used (u) these are confusing.  I would recommend following a more conventional route}
After advection,  body forces are integrated into the velocity field via the forward Euler scheme. The diffusion step resolves the viscosity terms (smoothing the current velocity field) and obtains $\tilde{u}^n$ via
\begin{equation*}
    \rho\frac{\tilde{u}^{n} - \hat{u}^{n}}{\Delta t} = \nu \Delta \tilde{u}^{n}.
\end{equation*}
Finally, the pressure projection step projects the velocity field back to a divergence-free state:
\begin{equation*}
    \frac{u^{n+1} - \tilde{u}^n}{\Delta t} = -\frac{1}{\rho} \nabla p^{n+1} \quad \text{ s.t. } \quad \nabla \cdot u^{n+1} = 0.
\end{equation*}

We use the $P_1$ or $Q_1$ basis for both velocity and pressure in the FEM discretization. The advection requires to find the element containing a given point and compute its local coordinates in that element. We use spatial hashing~\cite{spatialHash} to identify  a small set of candidate elements. We then compute the local coordinates with respect to each, and identify the element for which these are within the valid $[0, 1]$ range. % The computation of local coordinates is done by inverting the $P_1$ (triangular and tetrahedral meshes) and $Q_1$ (quadrilateral and hexahedral meshes) Lagrange basis.
For triangular and tetrahedral meshes, local coordinates are computed by solving a small linear system; for quadrilateral and hexahedral meshes, the bi-linear and tri-linear map is inverted using  Newton's method.

In the pressure projection step, we solve the Laplacian equation by FE to get the pressure field. We then compute the gradient of pressure at each node by averaging the gradient in all adjacent elements.

\paragraph{FD-FLIP}
FLIP is used for advection, where the specific ODE per FLIP particle is solved via RK-3. During the RK-3 solve, when velocity is queried at a point outside of the domain, we project the sample point to the closest point on the domain boundary to interpolate velocity.
% \DZ{do not understand what this says} 
If a FLIP particle goes outside of domain, we simply ignore its contribution. As FLIP particle resampling is performed at the beginning of each time step to ensure $2^d$ particles per cell ($d=2$ or $3$ is the dimension of space), a nearly uniform particle distribution is always maintained at the expense of introducing a small amount of random noise. After advection, Dirichlet boundary conditions are enforced by overwriting the boundary velocities.

Diffusion and pressure projection linear systems are solved using Pardiso~\cite{code:pardiso-a,code:pardiso-b,code:pardiso-c}. For diffusion, velocities are the unknowns; we enforce boundary conditions with linear equality constraints, requiring the boundary velocity DOF at the face centers and the boundary velocity interpolation at the edge centers are all equal to the prescribed values. 
% \DZ{how do we ensure this? be more explicit} 
On the other hand, boundary conditions in the pressure projection system  follows the standard scheme\ \cite{bridson2015fluid}: a volume fraction weighted gradient operator is applied to better resolve solid boundaries that are not grid-aligned.

\section{Neumann boundary conditions}
\label{app:neumann}

The zero Neumann boundary condition is enforced differently for different methods.

\paragraph{FE21-BDF3}
The zero Neumann boundary condition $\nu\frac{\partial u}{\partial n}+pn=0$ is naturally enforced on $\partial\Omega_N$ and no special treatment is needed.

\paragraph{FD-FLIP}
To enforce the zero Neumann boundary condition, $\frac{\partial u}{\partial n}$ on the boundary $\partial\Omega_N$ is constrained to be zero, which can be directly enforced in the diffusion step. In the pressure projection step, this is treated by setting the ghost pressure outside the domain at the boundary $\partial\Omega_N$ to zero.

\paragraph{FE11-SL, FE11-FLIP}
Similar to FD-FLIP, $\frac{\partial u}{\partial n}$ on the boundary $\partial\Omega_N$ is constrained to be zero in the diffusion step, which is naturally enforced under FE spatial discretizations. In the pressure projection step, pressure on the boundary $\partial\Omega_N$ is set to zero.

\paragraph{FE11-AB2AM2, FE21-AB2AM2}
Similar to FE11-SL and FE11-FLIP, $\frac{\partial u}{\partial n}$ on the boundary $\partial\Omega_N$ is constrained to zero in the predictor and corrector step of velocity. When we solve pressure based on the velocity, pressure on the boundary $\partial\Omega_N$ is set to zero.

\paragraph{FV-C}
In Ansys Fluent, pressure outlet boundary \cite{ansysUserGuide} conditions are used to define the static pressure at flow outlets. We observed that FE with zero Neumann boundary and FV with zero pressure outlet boundary converge to the same result in the test problems. Thus, in all our experiments, we use this observation when imposing zero Neumann on FV simulations.